\documentclass[a4paper,fleqn,usenatbib]{mnras}

\usepackage[T1]{fontenc}
\usepackage{ae,aecompl}

\usepackage{graphicx}	
\usepackage{amsmath}	
\usepackage{amssymb}	
\usepackage{threeparttable} 

\title[The 1989 and 2015 outbursts of V404 Cygni]{The 1989 and 2015 outbursts of V404 Cygni: a global study of wind-related optical features}

\author[Mata S\'{a}nchez et al.] {D. Mata S\'{a}nchez$^{1,2,3,4}$\thanks{E-mail:matasanchez.astronomy@gmail.com}, T. Mu\~{n}oz-Darias$^{1,2}$, J. Casares$^{1,2,5}$, P. A. Charles$^{6,5}$, 
 \newauthor 
M. Armas Padilla$^{1,2}$, J. A. Fern\'{a}ndez-Ontiveros$^{1,2}$, F. Jim\'enez-Ibarra$^{1,2}$, P. G. Jonker$^{3,7}$, 
\newauthor
M. Linares$^{8,9}$, M.A.P. Torres$^{1,2,3}$, A. W. Shaw$^{10}$, P. Rodr\'{i}guez-Gil$^{1,2}$, T. van Grunsven$^{3,7}$,
\newauthor
 P. Blay$^{11}$, M. D. Caballero-Garc\'{i}a$^{12}$, A. Castro-Tirado$^{13,14}$, P. Chinchilla$^{1,2}$,
\newauthor
C. Farina$^{1,15}$, A. Ferragamo$^{1,2}$, F. Lopez-Martinez$^{15}$, J.~A. Rubi\~no-Martin$^{1,2}$,
\newauthor
L. Su\'{a}rez-Andr\'{e}s$^{1,2}$.
\\
$^{1}$Instituto de Astrof\'{i}sica de Canarias (IAC), E-38205 La Laguna, Tenerife, Spain\\
$^{2}$Departamento de astrof\'isica, Univ. de La Laguna, E-38206 La Laguna, Tenerife, Spain\\
$^{3}$SRON, Netherlands Institute for Space Research, Sorbonnelaan 2, NL-3584CA Utrecht, The Netherlands
\\
$^{4}$Jodrell Bank Centre for Astrophysics, School of Physics and Astronomy, The University of Manchester, Manchester M13 9PL, UK
\\
$^{5}$Department of Physics, Astrophysics, University of Oxford, Denys Wilkinson Building, Keble Road, Oxford OX1 3RH, UK
\\
$^{6}$Department of Physics and Astronomy, University of Southampton, Southampton SO17 1BJ, UK
\\
$^{7}$Department of Astrophysics/IMAPP, Radboud University, P.O. Box 9010, NL-6500GL Nijmegen, The Netherlands
\\
$^{8}$Departament de F{\'i}sica, EEBE, Universitat Polit{\`e}cnica de Catalunya, c/ Eduard Maristany 10, E-08019 Barcelona, Spain
\\
$^{9}$Institute of Space Studies of Catalonia (IEEC), E-08034 Barcelona, Spain
\\
$^{10}$Department of Physics, University of Alberta, CCIS 4-181, Edmonton, AB T6G 2E1, Canada
\\
$^{11}$Universidad Internacional de Valencia (VIU), 46002 Valencia, Spain
\\
$^{12}$Astronomical Institute, Academy of Sciences of the Czech Republic,
Boc\^{n}\'i II 1401, CZ-141 00 Prague, Czech Republic
\\
$^{13}$Instituto de Astrof\'isica de Andaluc\'ia (CSIC), E-18008 Granada, Spain
\\
$^{14}$Unidad Asociada Departamento de Ingenier\'ia de Sistemas y Autom\'atica,
Universidad de M\'alaga, E-29071 M\'alaga, Spain
\\
$^{15}$Isaac Newton Group of Telescopes, Apto. 321, E-38700 Santa Cruz de la Palma, Canary Islands, Spain
\\
}

\date{Accepted 2018 August 30. Received 2018 August 23; in original form 2018 May 25}

\pubyear{2018}

\begin{document}
\label{firstpage}
\pagerange{\pageref{firstpage}--\pageref{lastpage}}
\maketitle

\begin{abstract}
The black hole transient V404 Cygni exhibited a bright outburst in June 2015 that was intensively followed over a wide range of wavelengths. Our team obtained high time resolution optical spectroscopy ($\rm \sim 90$ s), which included a detailed coverage of the most active phase of the event. We present a database consisting of 651 optical spectra obtained during this event, that we combine with 58 spectra gathered during the fainter December 2015 sequel outburst, as well as with 57 spectra from the 1989 event. We previously reported the discovery of wind-related features (P-Cygni and broad-wing line profiles) during both 2015 outbursts. Here, we build diagnostic diagrams that enable us to study the evolution of typical emission line parameters, such as line fluxes and equivalent widths, and develop a technique to systematically detect outflow signatures. We find that these are present throughout the outburst, even at very low optical fluxes, and that both types of outflow features are observed simultaneously in some spectra, confirming the idea of a common origin. We also show that the nebular phases depict loop patterns in many diagnostic diagrams, while P-Cygni profiles are highly variable on time-scales of minutes. The comparison between the three outbursts reveals that the spectra obtained during June and December 2015 share many similarities, while those from 1989 exhibit narrower emission lines and lower wind terminal velocities. The diagnostic diagrams presented in this work have been produced using standard measurement techniques and thus may be applied to other active low-mass X-ray binaries.

\end{abstract}

\begin{keywords}
accretion, accretion discs -- X-rays: binaries -- stars: winds, outflows -- 
ISM: jets,outflows
\end{keywords}



\section{Introduction}

Low-mass X-ray binaries (LMXBs) are formed by a compact object, either a black hole (BH) or a neutron star, and a companion star typically lighter than the Sun $\lesssim 1\, \rm M_{\odot}$. The Roche lobe overflow of the companion leads to the transfer of matter onto the compact object via an accretion disc, which reaches temperatures of $\rm \sim 10^7\, K$ and produces strong X-ray emission (for a review, see e.g., \citealt{Charles2010}). The majority of LMXBs harbouring BHs are transient systems, which spend most of their lives in a faint, quiescent state ($\rm L_X\sim 10^{31-34}\, erg \, s^{-1}$, e.g., \citealt{Belloni2011}, \citealt{Casares2014b}, \citealt{Armas-Padilla2014b}) but exhibit occasional outbursts where their X-ray luminosity increases up to $\rm L_X \sim 10^{38-39}\,  erg\, s^{-1}$ (e.g., \citealt{CorralSantana2016}).

V404 Cygni (V404 Cyg hereafter) is an X-ray transient harbouring the first unambiguosly confirmed stellar-mass BH. It has a long orbital period ($P_{\rm orb}=6.5 \, \rm d$; \citealt*{Casares1992}) which allows the accumulation of material in a large accretion disc.  From its radio parallax, V404 Cyg is known to be relatively close ($2.39\pm 0.14\, \rm kpc$, \citealt{Miller-Jones2009}; consistent with Gaia DR2 optical parallax, see \citealt{Gaia2018} and \citealt{Gandhi2018}). This, combined with the extreme X-ray luminosities produced during the outburst (sometimes likely exceeding the Eddington limit, see \citealt{Motta2017b}), resulted in some of the brightest BH outburst events ever observed \citep{CorralSantana2016}. Several outbursts of V404 Cyg have been recorded over the last century: 1938 \citep{Wagner1989}, 1956 \citep{Richter1989}, 1989 \citep{Makino1989} and 2015 (June 2015, 2015J hereafter, \citealt{Barthelmy2015a}), with the latter followed by a shorter outburst only 6 months later (December 2015, 2015D hereafter; \citealt*{Barthelmy2015b}, \citealt{Lipunov2015}, \citealt{Motta2015} and \citealt{Hardy2016}). The 1938 event was historically known but had been misclassified as a classical nova \citep{Duerbeck1988}, while the 1956 outburst was only discovered after the 1989 X-ray outburst, by inspection of earlier photographic plates \citep{Richter1989}. Therefore, only three events (1989, 2015J and 2015D) had enough observing coverage to study their temporal evolution spectroscopically. 

Several works have reported on the 2015J outburst, taking advantage of the technological improvement as well as the faster response attained in the last decade. The photometric light-curves obtained in different optical bands allowed \citet{Kimura2016} to find optical oscillations on timescales of a hundred of seconds to a few hours, occurring at mass-accretion rates ten times lower than previously thought. On the other hand, the X-ray analysis performed by \citet{Motta2017b} revealed the presence of high-absorption and fast spectral variability which they propose is produced by a cold and clumpy outflow arising in an inner slim disc. In previous work (\citealt{MunozDarias2016}, MD16 hereafter), we presented the analysis of an optical spectroscopic database that allowed us to detect outflows in the form of P-Cygni (P-Cyg hereafter) profiles as well as the so-called nebular phase during the outburst decline. This nebular phase is characterised by unprecedentedly broad emission lines and high Balmer decrements, both indicators of an expanding, cooling nebula. All of these novel results have revealed the likely influence of outflowing matter in the outburst evolution of LMXBs (MD16, see \citealt{Soria2017} for HLX-1 and \citealt{Fender2016} for a review).

In this work, we exploit the full spectroscopic optical database gathered by our team during the three latest outbursts of V404 Cyg using a variety of telescopes and instruments: 2015J (651 spectra, MD16), 2015D (58 spectra, \citealt{MunozDarias2017}) and 1989 (57 spectra, \citealt{Casares1991}). We analyse this database in a systematic way aiming at: i) studying the evolution of line parameters during the different outburst phases, ii) developing new techniques to detect weak outflow signatures in the spectra and iii) comparing the resulting diagnostic diagrams of each outburst, aiming to extrapolate these results to other active LMXBs.

\section{Observations}
\label{observations}

\subsection{The June 2015 outburst}

Given that the most extensive database was acquired during the 2015J outburst (651 optical spectra), we will first analyse this event independently. The observation log is shown in Tables\,\ref{tab:summer1} and \ref{tab:summer2}. Following MD16, we established MJD=57190 (June 17, 00:00 UT) as the reference time for the 2015J outburst. It is worth noting that the X-ray trigger occurred a day earlier (MJD= 57188.77, day -1.2 in our temporal reference frame). In this work, we add new spectra obtained during days 0 and -1, as well as include a spectral set fortuitously obtained in day -1.8 (MJD=57188.21, i.e., half a day before the X-ray trigger reported in \citealt{Bernardini2016}). 

The telescopes used in this campaign are located at the Roque de Los Muchachos Observatory (La Palma, Spain): Gran Telescopio Canarias (GTC) equipped with the Optical System for Imaging and low-Intermediate-Resolution Integrated Spectroscopy (OSIRIS); William Herschel Telescope (WHT) with the Intermediate dispersion Spectrograph and Imaging System (ISIS) and the low-resolution long-slit spectrograph ACAM attached; Nordic Optical Telescope (NOT), equipped with the Andalucia Faint Object Spectrograph and Camera (ALFOSC) and the FIbre-fed Echelle Spectrograph (FIES); and the Isaac Newton Telescope (INT) with the Intermediate Dispersion Spectrograph (IDS) attached.

This heterogeneous database has been reduced using semiautomatic routines developed by our team and based on \textsc{iraf}\footnote{IRAF is distributed by the National Optical Astronomy Observatories, operated by the Association of Universities for Research in Astronomy, Inc., under contract with the National Science Foundation.}, \textsc{python} and \textsc{molly}\footnote{\textsc{molly} software developed by T. R. Marsh.}. Each spectrum has been de-biased and flat-field corrected. For those exposures with two objects in the slit (when possible, the slit was rotated to include a field star), a low-order spline function correction was applied along the spatial direction to account for large scale variability in the illumination. Individual arc spectra were extracted from the two-dimensional images at each object position (i.e., two different positions when the comparison star was present in the slit). Interpolated values from arcs obtained before and after the observation were used to obtain a precise wavelength calibration (when not available, the nearest arc was selected). The wavelength calibration was refined by comparing the observed position of sky emission lines with their corresponding rest wavelengths, from which we derived sub-pixel velocity drifts that were subsequently corrected. Finally, we moved all the spectra to the solar system barycenter by removing the velocity of the Earth relative to the source at each epoch of the observation.

To obtain a reliable flux calibration, the spectra were initially calibrated relative to the comparison star included in the slit. The bright field star USNO-B1.0 1238-00435227 was employed during the brightest phases of the outburst, while USNO-A2.0 1200-15039207 was used for the longer exposures, when the flux of V404 Cyg had dropped significantly. The spectra of both field stars were carefully flux-calibrated relative to the spectrophotometric standard star Wolf 1346 \citep{Koen2010}, and this absolute flux calibration was propagated to the whole database. A total of 622 spectra (out of 651) were observed under this configuration and therefore flux-calibrated. The 4 spectra obtained in day -1.8 (were no comparison star is present in the slit) were calibrated using the nearby flux standard star BD +25 4655 \citep{Oke1990}, which was observed at a similar airmass. The remaining 25 spectra were only considered when performing measurements on normalised spectra. 

All the flux-calibrated spectra were finally de-reddened using \textsc{molly} routines and a reddening value of $E(B-V)=1.3$ (\citealt*{Chen1997}, obtained during quiescence). Every flux-related measurement quoted in this work has been extracted from these final, de-reddened spectra. We note, however, that additional absorption components (e.g., outflows) are present in the outburst data. 

We finally report that, due to the remarkable variability of V404, the H$\rm \alpha$ line saturated the instrument detector in the following exposures: day -0.9, GTC/OSIRIS R1000B (1 spectrum); and day 10, GTC/OSIRIS R1000B and R2500R (9 spectra). For such cases, we extract the spectra from the wings of the bidimensional spatial profile rather than from its core, in an attempt to avoid saturation. While this method produces spectra with no clear signs of saturation (Gaussian-like H$\rm \alpha$ lines), we note that measurements derived from them must be treated with caution. They have been properly marked in the figures.

\subsection{The December 2015 and 1989 outbursts}

We also present the analysis of a total of 58 optical spectra covering the 2015D outburst, only 6 months after the main event. Both the telescopes used in this campaign and the reduction process are very similar to those of the 2015J event. The observation log is shown in Tab. \ref{tab:winter}. We established MJD=57379 (2015 December 23, 00:00 UT) as the reference time for 2015D. To complete the picture of V404 Cyg evolution during different outbursts we added 57 optical spectra obtained during the 1989 event whose reduction is detailed in \citet{Casares1991}. The corresponding observation log is shown in Tab. \ref{tab:1989}. We established MJD=47667 (1989 May 21, 00:00 UT) as the reference date for the 1989 event. 

\section{Analysis and results}

\subsection{Methods}
\label{methods}

The shape and strength of the emission lines present in the spectra vary significantly along the outburst. To study their evolution, and despite the different wavelength coverage of each instrument, we aimed at measuring parameters in a uniform way, as follows:

\begin{itemize}

\item Equivalent widths (EWs) and line fluxes (only in flux-calibrated spectra): contrary to the usual convention, we define EW as positive for emission features. 

\item Line flux ratios: from the line flux measurements we derive i) the ionisation ratio ($I_{\rm ratio}$), defined as the flux ratio between the He {\sc ii}--$\lambda$4686 and H$\rm \beta$ lines; and ii) the Balmer decrement ($\rm BD$), obtained from the flux ratio between the H$\rm \alpha$ and H$\rm \beta$ lines.

\item De-reddened $r'$ magnitude: only the SDSS (Sloan Digital Sky Survey, \citealt{Abazajian2009}) $r'$-band is fully covered in a sufficient number of spectra to trace the outburst evolution. We multiplied every spectrum that covers the wavelength range $\sim 5400-7100${~\AA} by the SDSS filter transmission profile of the $r'$-band filter. Given that the spectra are calibrated in flux density, we obtain the $r'_{obs}$ flux density folding the spectrum through the filter transmission curve.

The correction between the SDSS $r'$ magnitude and $r'_{\rm AB}$ is negligible (less than 0.01, see \citealt*{Bohlin2001}). Therefore, we can directly apply the conversion between SDSS AB magnitudes and physical fluxes, where the zero-point flux density is defined to be $3631 \, \rm Jy$ in every filter. This finally results in $r'_{\rm AB}$ (i.e., $r'_{obs}$) values for each flux calibrated spectrum. The quiescent observed magnitudes of V404 Cyg are: $V=18.42\pm 0.02$, $B=20.63\pm 0.05$ \citep{Casares1993}, which correspond to de-reddened magnitudes of $V=14.26$ and $B=15.17$ (assuming the interstellar reddening law $A_V=3.2\, E(B-V)=4.16$ from \citealt{Schild1977}). This implies a de-reddened $r'=14.0\pm 0.3$ quiescent magnitude \citep{Jester2005}, where the uncertainty accounts for the range of optical variability observed in quiescence \citep{Pavlenko1996}. We note that all the de-reddened $r'$ magnitudes presented in this work are related to the observed values as: $r'= r'_{obs}-3.6$.

\item Gaussian fit: we fitted every emission line present in the spectra with a single Gaussian with height, full-width-at-half-maximum (FWHM) and velocity offset as free parameters. For every spectrum, the same fitting mask was applied to each particular line. For those lines with strong P-Cyg profiles (blue absorption and red emission, e.g., \ion{He}{\sc i}--$\lambda$5876) and/or extended wings, the applied mask was always wide enough to include them. We note that the FWHM measurements shown in this work have already been deconvolved from the instrumental resolution, measured on nearby sky lines for each instrumental configuration.

\item Search for outflows: as shown in MD16, the most prominent P-Cyg profiles are observed in \ion{He}{\sc i}--$\lambda$5876, while $\rm H\alpha$ (the most intense line in the whole spectrum) sits on extended wings ($\rm \pm 3000\, km\, s^{-1}$) during the nebular phase. In order to search for outflows in these lines, we build diagnostic diagrams by subtracting the Gaussian fit described above from the original spectra, keeping the continuum normalised. We then measured the EW of the residual profile in blue and red velocity bands avoiding the line core (-4000 to -1000 $\rm km \,s^{-1}$ and 1000 to 4000 $\rm km \,s^{-1}$ respectively) to search for outflows. These are called EW blue and red excesses, respectively ($\rm EW_{b+}$ and $\rm EW_{r+}$ hereafter).

\end{itemize}

\subsection{June 2015 outburst}
\label{2015J_methods}

We will start by comparing the optical ($r'$) and X-ray emission to then detail the spectral evolution day by day. We will produce diagnostic diagrams by using different parameters measured in the H, He \textsc{i} and He \textsc{ii} emission lines. This will allow us to develop techniques to detect outflow features in the spectra and to obtain information about the ionisation and nebular contribution to the emitted light.

\subsubsection{X-ray-optical flux correlation}

\begin{figure*}
\includegraphics[keepaspectratio, trim=0cm 0cm 0cm 0cm, clip=true, width=\textwidth]{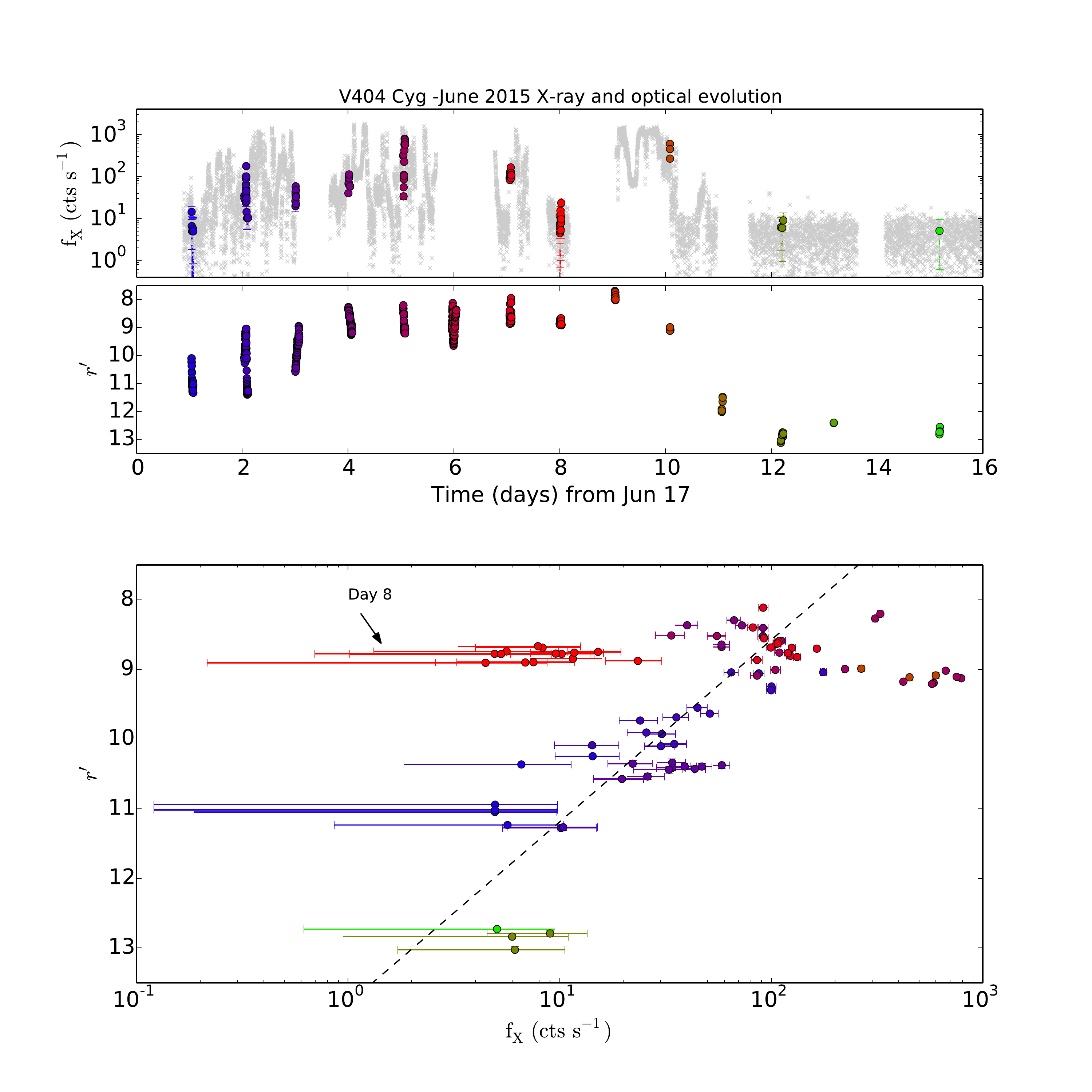} \caption{Top panel: time vs X-ray flux in counts per second of \textit{INTEGRAL} (light-grey crosses). Coloured circles highlight those observations simultaneous (within $\sim 90\, \rm s$) with our spectroscopic database. Middle panel: $r'$ vs time obtained from our de-reddened spectra. Bottom panel: X-ray flux in counts per second vs $r'$. The colour code denotes the observation day. The dashed line serves as a reference with slope defined by the correlation found by \citet{Paradijs1994}. Day 8 observations have been marked with an arrow.}
    \label{fig: xrayvsopt}
\end{figure*}

\begin{figure*}
\includegraphics[keepaspectratio, trim=1cm 0cm 4cm 0cm, clip=true, width=\textwidth]{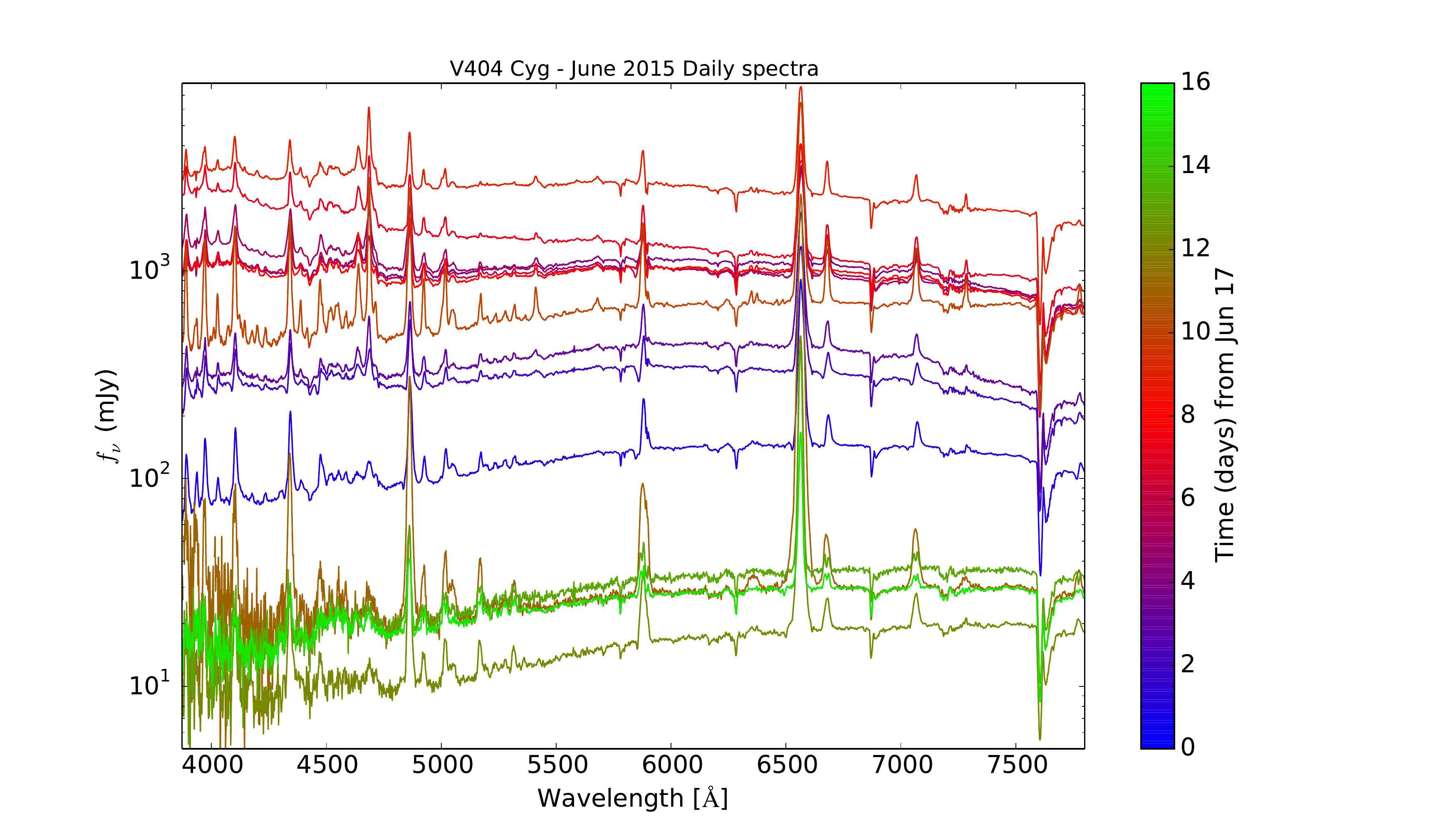} \caption{Day-averaged, flux-calibrated spectra, covering days 1 to 15 (except day 14). The colour bar denotes the observation day. The flux density is on a logarithmic scale so as to fully cover the large variability range.}
    \label{fig: spectra_fluxtot}
\end{figure*}

V404 Cyg was observed by the INTErnational Gamma-Ray Astrophysics Laboratory ( \textit{INTEGRAL}, \citealt{Winkler2003}) equipped with the Imager on-Board the \textit{INTEGRAL} Satellite (IBIS) during its 2015J outburst (\citealt{Kuulkers2015}, \citealt{Natalucci2015}) in hard X-rays ($25-200\, \rm keV$). We use this data, originally processed by \citet{Kuulkers2015} and later reduced by MD16, to obtain a light curve with a time resolution of $\sim 64~s$ (see MD16 for the data reduction details). We selected the X-ray detections simultaneous with our $r'$ magnitudes within $\sim 90\, \rm s$ (our best temporal resolution). In Fig. \ref{fig: xrayvsopt}, we show the temporal evolution of X-rays (top panel) and $r'$ (middle panel). By plotting optical emission against X-rays (Fig. \ref{fig: xrayvsopt}, bottom panel), we find that X-ray and optical emission roughly follow the same trend. A correlation between X-ray and optical emission in LMXBs has been observed by several authors when comparing the X-ray and optical fluxes of different LMXBs (e.g., \citealt{Paradijs1994}; dashed line in Fig. \ref{fig: xrayvsopt}, bottom panel). It is probably due to the reprocessing of the X-rays in the accretion disc, which is able to account for the main contribution to the optical luminosity.

However, we note that this correlation is not so well defined when inspecting day-by-day data. The most remarkable example is found on day 8, when observations deviate from the trend, having a higher optical emission than those observed in days 1--3 at similar X-ray fluxes. Studies on the X-ray to optical correlation during the 2015J outburst of V404 have revealed a non-trivial relationship, showing from null to 30 minutes lags when comparing flares in different epochs (e.g., \citealt{Rodriguez2015}, \citealt{Kimura2017}; see also Alfonso-Garzon et al. 2018 in prep.). These lags may be responsible for the deviation from the trend in our lower time-resolution observations.

\subsubsection{The overall spectral evolution}

\begin{figure*}
\includegraphics[keepaspectratio, trim=2cm 0cm 5cm 0cm, clip=true, width=\textwidth]{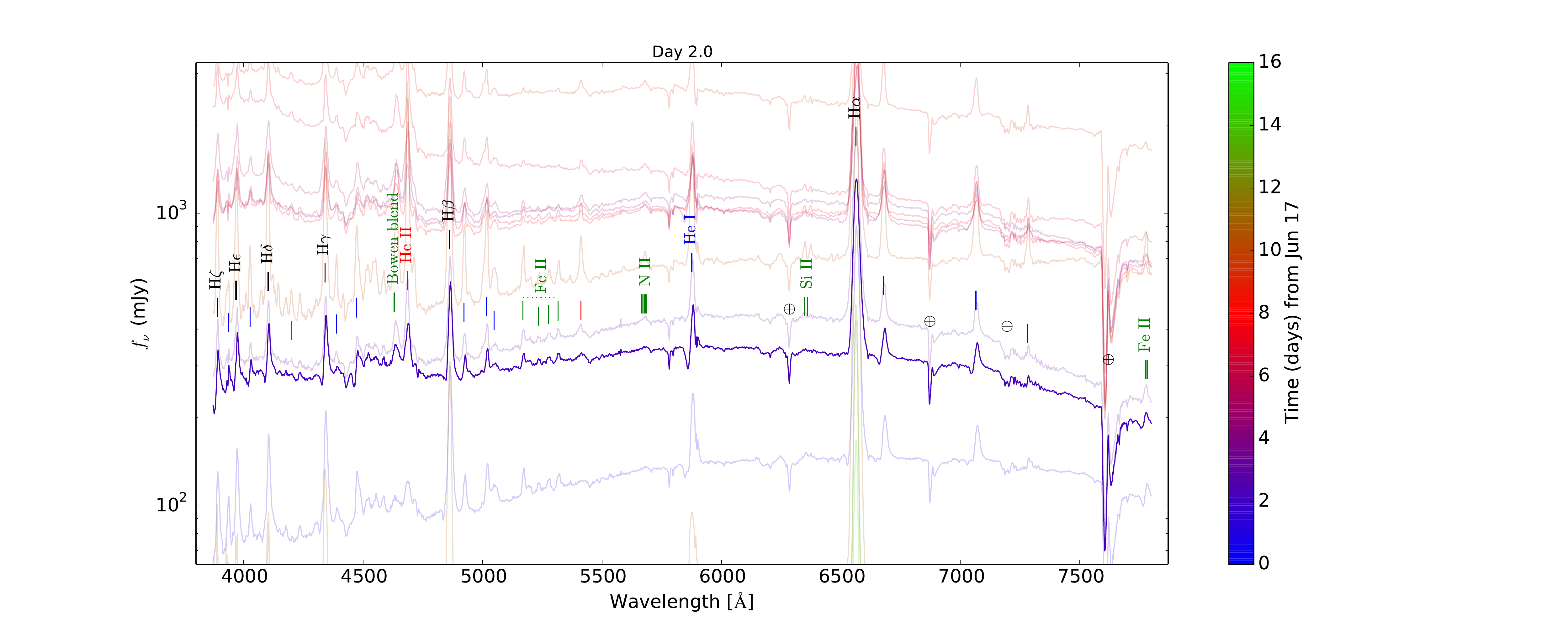} \caption{Same as Fig.\,\ref{fig: spectra_fluxtot} but focused on day 2, as it contains the best example of P-Cyg profiles. Emission line identifications are also colour encoded: black for H, blue for \ion{He}{\sc i}, red for He {\sc ii} and green for other species (Si {\sc ii}, N {\sc ii}, Fe {\sc ii} and the Bowen blend). Telluric absorption lines are marked as $\oplus$.}
    \label{fig: spectra_day2}
\end{figure*}

\begin{figure*}
\includegraphics[keepaspectratio, trim=2cm 0cm 5cm 0cm, clip=true, width=\textwidth]{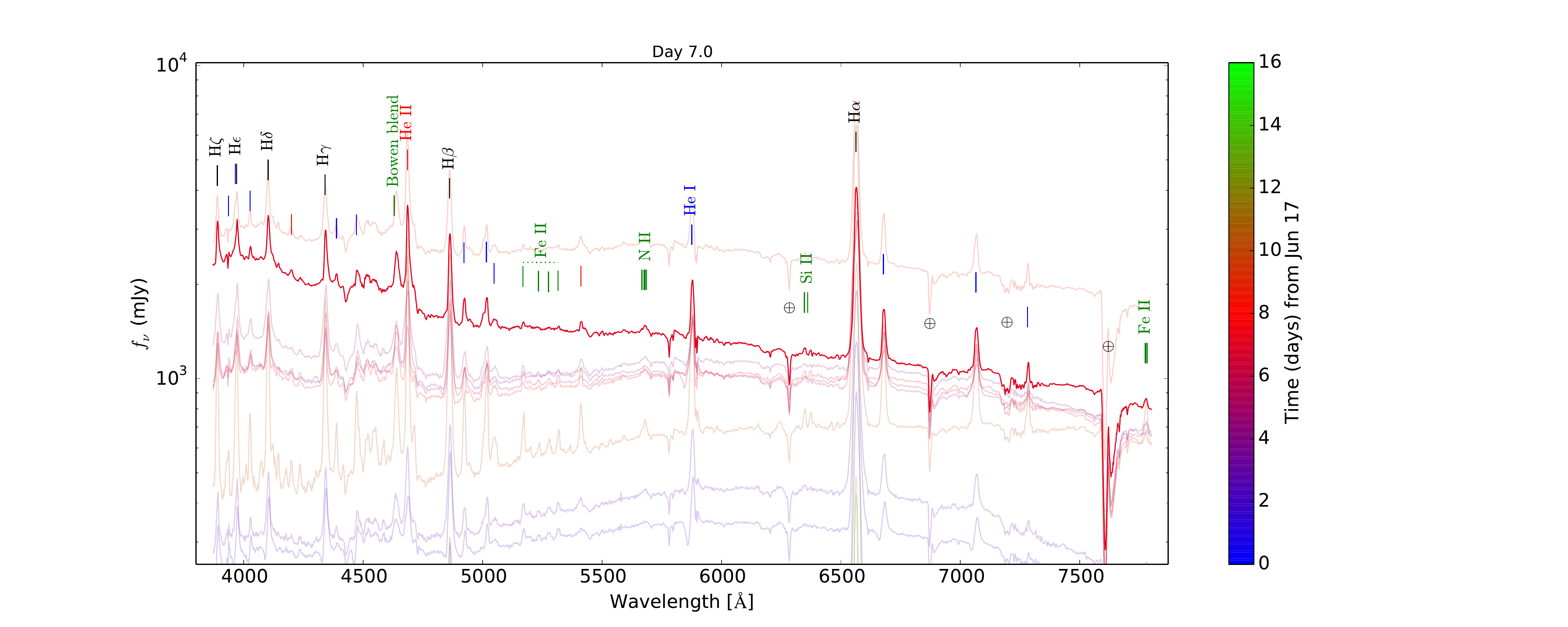} \caption{Same as Fig.\,\ref{fig: spectra_fluxtot} but focused on day 7.}
    \label{fig: spectra_day7}
\end{figure*}

\begin{figure*}
\includegraphics[keepaspectratio, trim=2cm 0cm 5cm 0cm, clip=true, width=\textwidth]{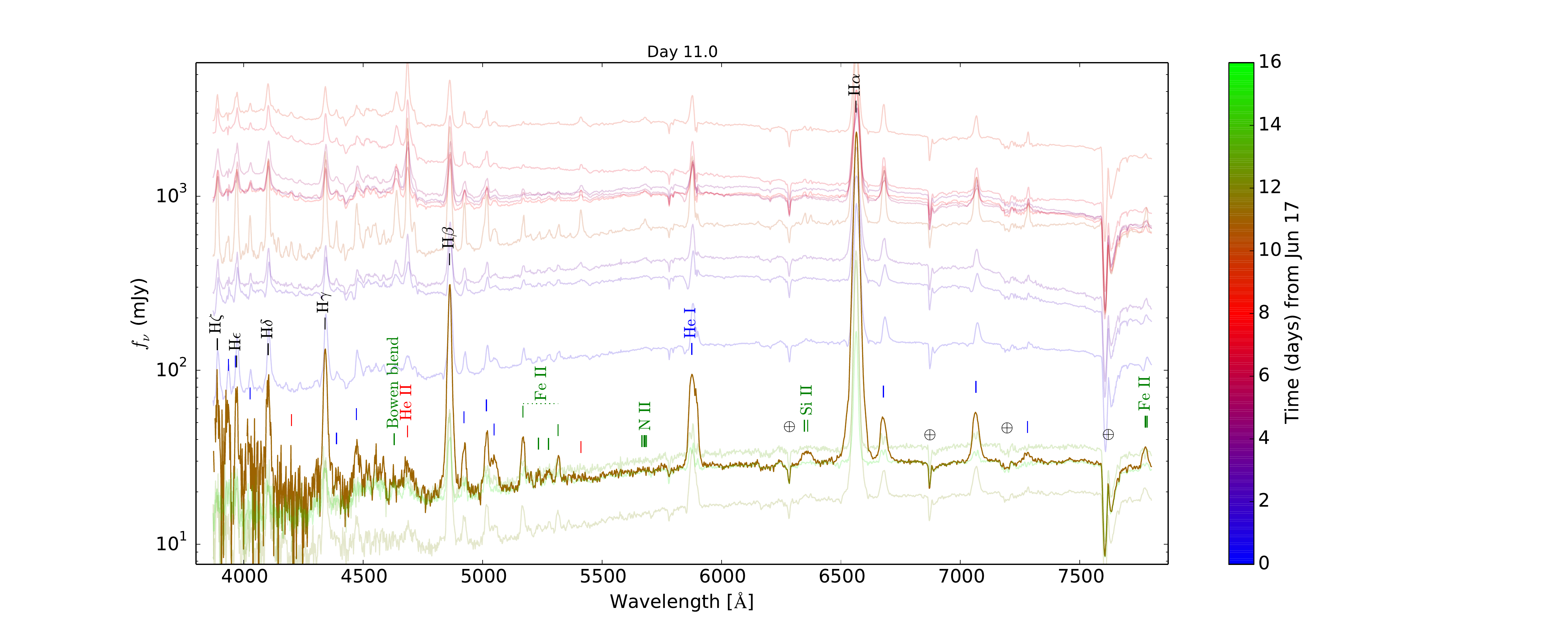} \caption{Same as Fig.\,\ref{fig: spectra_fluxtot} but focused on day 11, as the best example of nebular emission.}
    \label{fig: spectra_day11}
\end{figure*}

\begin{figure*}
\includegraphics[keepaspectratio, trim=2cm 0cm 5cm 0cm, clip=true, width=\textwidth]{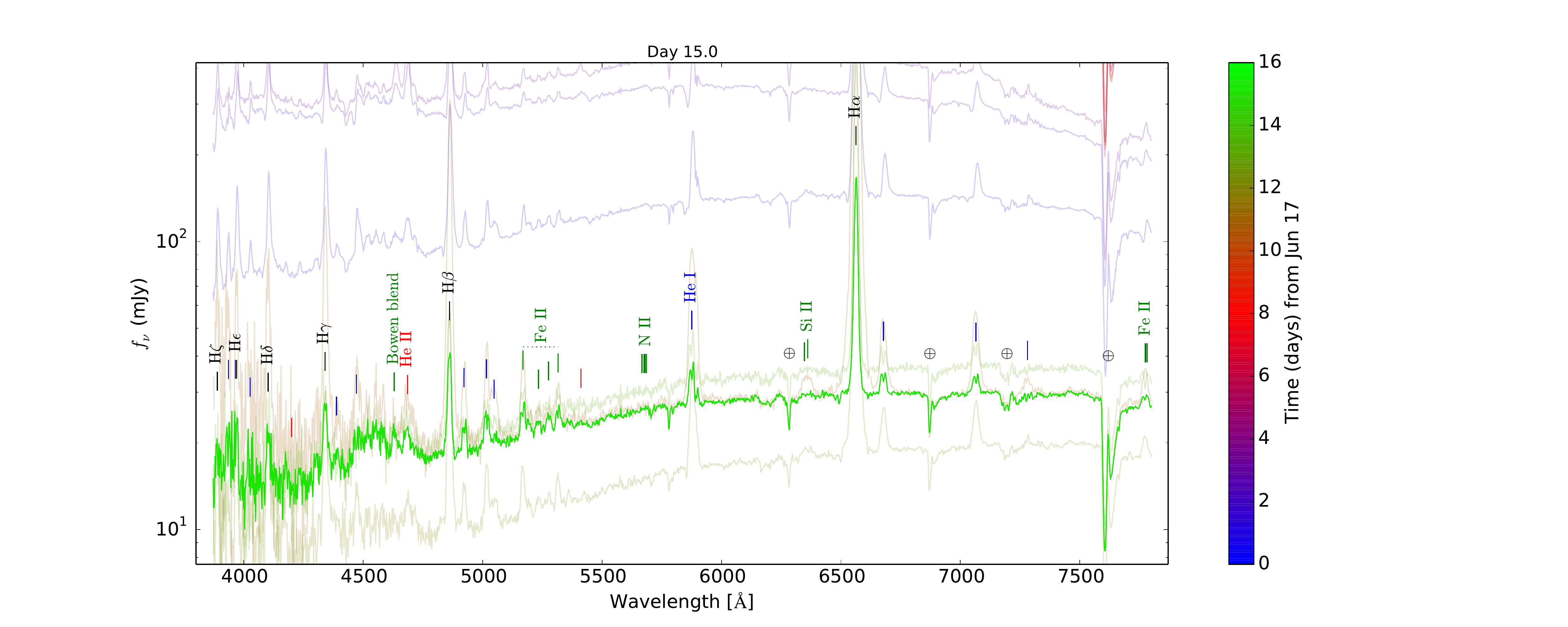} \caption{Same as Fig.\,\ref{fig: spectra_fluxtot} but focused on day 15, as the best example of double-peaked profiles (except for $\rm H\alpha$ and $\rm H\beta$).}
    \label{fig: spectra_day15}
\end{figure*}

The database of flux-calibrated spectra allowed us to follow the 2015J outburst from day 1 to 15 (except day 14, see Tab.\,\ref{tab:summer1}) with the highest time resolution data (GTC+OSIRIS, R1000B; one spectrum every 84 s from day 1 to 15). We found remarkable variations within our 0.5--2 h observing blocks. These data were completed with a handful of spectra obtained in days -2 to 0 (when possible, flux-calibrated). The evolution of the day-averaged spectra reveals the extreme variability of the system through the entire outburst (see Fig.\,\ref{fig: spectra_fluxtot}), containing dramatic changes in the line profiles. In this section, we present a concise overview of 2015J through a day-by-day summary of the outburst evolution, analysing the following key features (based on the parameters defined in Sec. \ref{methods}):

\begin{itemize} \item Outflows: classic P-Cyg profiles (Fig.\,\ref{fig: spectra_day2}, \ref{fig: spectra_day7}) are associated with the presence of winds. We detect these features in a particular spectrum when $\rm EW_{b+}<0$ and $\rm EW_{r+}>0$. We also analyse if they depend on the $I_{\rm ratio}$ (a good tracer of the outer disc irradiation; see e.g., \citealt{Groot2001}), $\rm BD$ and their terminal velocities ($v_{\rm T}$, the absolute value of the minimum velocity defined by the absorption feature).

On the other hand, optically thin expanding envelopes result in broad emission line wings during the nebular phase (Fig.\,\ref{fig: spectra_day11}). This phase is characterised by $\rm EW_{b+}>0$ and $\rm EW_{r+}>0$ (as a result of the broad line wings), but also high BDs, low $I_{\rm ratio}$ and high EWs (MD16, see also \citealt{Rahoui2017}).

 The continuous presence of these features in V404 Cyg spectra implies that \textit{an expanding outflow is present throughout the outburst}.

\item Optical continuum: it changes both in flux density (from $\rm 10$ to $\rm 3000\, mJy$) and spectral slope (see Fig.\,\ref{fig: spectra_day7}), as they might be associated with the variable contribution of different components of the X-ray binary to the optical spectrum (e.g., \citealt{Gandhi2016,Gandhi2017}).

\item Double-peaked emission lines (Fig.\,\ref{fig: spectra_day15}): emission lines have been seen to evolve from double-peaked (quiescence spectra) to single-peaked profiles (outburst spectra) in many LMXB outbursts. The origin of this effect is still controversial (see Sec. \ref{Disc_spec_evol}). We analyse such variations in V404 Cyg, mainly focusing on the optical brightness of the system ($r'$) when such changes occur.

\end{itemize}

\bigskip

The day-by-day summary of the outburst evolution follows (see also Tab.\,\ref{tab:summatory}; Fig. \ref{fig: paramd1-2} to \ref{fig: paramd9}):

\begin{itemize}
\item From days -2 to 0 we only have a handful of spectra obtained with different telescopes and instruments (see Tab. \ref{tab:summer1}, also Fig. \ref{fig: spectra_special}). Day -2 spectra were flux-calibrated using a standard star (no comparison star was placed in the slit), resulting in $r' = 13.23\pm 0.15$. The $\rm H\alpha$ $\rm EW$ is in the range $\rm 270-309${~\AA}, well above the quiescence level ($\rm EW =19.0\pm 5.2${~\AA}, \citealt{Casares2015}), revealing that the system was already becoming active by then ($\rm 0.6 \, d$ before the X-ray trigger). The high $\rm BD=5.4- 5.6$ together with the low $I_{\rm ratio}=0.05-0.19$ lead us to the conclusion that a nebular phase is likely present. The spectra from days -1 and 0 also reveal high values of the EW ($\rm \sim 458.8  \pm 1.1${~\AA} and $\rm 104.3 \pm 0.1${~\AA}, respectively). The lack of a flux-calibration hampered the determination of more parameters in these spectra.

\item On day 1, we observe a decay of more than 1 mag (i.e., $r' = 10.2-11.4$) in $\rm \sim 1\, h$, which suggests the presence of a flare before the observation started. The \ion{He}{\sc i}--$\lambda$5876 emission line (the most prolific in the exhibition of P-Cyg profiles during the outburst) exhibits blue-shifted absorption depths of $\rm 3-12 \%$ (referred to the continuum level), which become shallower as the continuum flux decreases. Their terminal velocities are in the range of $1500-2000 \, \rm km\, s^{-1}$. The $I_{\rm ratio}=0.1-0.7$ is relatively low, while $\rm BD=2.4-3.2$ and the EW of $\rm H\alpha$ varies between $\rm EW =65-195${~\AA}, increasing as $r'$ becomes fainter.

\item Day 2 is characterised by some of the deepest P-Cyg profiles in \ion{He}{\sc i}--$\lambda$5876 (blue absorption depth of $\rm 8-30 \%$). This observing block covers a full flare, including its rise, peak (shallower P-Cyg) and decay ($r' = 9.0-11.3$). The ionisation ratio varies during the event between $I_{\rm ratio}=0.03-2.5$, while $\rm BD$ varies dramatically between $0.9-5.1$. The initial rise ($\rm \sim 1 \, mag$ in $\rm \sim 1\, h$) is slower than the decay, which occurs over similar time scales than on the previous day ($\rm \sim 1\, h$) despite the drop in flux being $\sim 3$ times larger. After this decay, the deepest P-Cyg observed in \ion{He}{\sc i}--$\lambda$5876 during the outburst are seen simultaneously with the lowest $I_{\rm ratio}$ values. Indeed, the comparison of the $I_{\rm ratio}$ with the P-Cyg profiles' depth during the outburst reveals that these quantities are anti-correlated, with deeper absorptions for lower $I_{\rm ratio}$ values (see Fig.\,\ref{fig: IratiovsBD}, left panel). As noted in MD16, this probably implies that the ionising SED affecting the outer disc is also ionising the optical wind, thus preventing the observation of neutral P-Cyg absorptions at high $I_{\rm ratio}$. We also note that, simultaneously with the deep P-Cyg in \ion{He}{\sc i}--$\lambda$5876, a short nebular phase is observed in $\rm H\alpha$. It is characterised by an increase of both the $\rm H\alpha$ EW ($40-270 ${~\AA}, see Sec. \ref{halpha-evol}) and the BD (see Fig.\,\ref{fig: IratiovsBD}, right panel).

\item Day 3 exhibits a slow rise in the continuum flux ($r' = 10.6-8.9$ in $\sim 1.75 \rm \, h$), where the P-Cyg in \ion{He}{\sc i}--$\lambda$5876 gradually disappears (no clear detection when $\rm r' < 9.5$). At the same time, both He {\sc ii}--$\lambda$4686 and the Bowen blend ($\sim 4640$ \AA) become stronger, suggesting a higher ionisation of the disc material. Emission lines also exhibit broad wings at fainter stages, extending up to $\rm 2500 \, km\, s^{-1}$ in $\rm H\alpha$ ($\rm EW \sim 200${~\AA}). It becomes narrower (its wings narrow down to $\rm 1500 \, km\, s^{-1}$) as the flux steadily increases and returns to $\rm EW \sim 55${~\AA}, similar to values observed in previous nights. Indeed, excess on both red and blue $\rm H\alpha$ emission is measured in these data (see Sec. \ref{outflows}), confirming a faint nebular phase. Lower values of the ionisation ratio ($I_{\rm ratio}=0.2-1.7$) as well as a higher BD ($2.5-4.3$) being measured during this epoch also supports this interpretation.

\item Day 4 is a brighter version of day 1, exhibiting a slow decay of $r' = 8.2-9.3$ in $\rm \sim 2\, h $. The wings of $\rm H\alpha$ ($\rm EW =35-92${~\AA}, lower for higher fluxes) are not particularly broad (similar to day 1), which suggests that a necessary condition for observing a nebular phase is a relatively fast drop in flux. The ionisation ratio is high $I_{\rm ratio}=1.0-4.7$ (reaching the maximum value of the outburst), while BD is in the range $2.4-4.0$.

\item Day 5 exhibits an initial ($\rm \sim 0.5\, h $) decay of $r'~=~8.2~-~9.2$, ending in a plateau. The flux drop is not fast nor deep enough for a faint nebular phase to be observed ($\rm EW =43-108${~\AA}, lower for higher fluxes). The ionisation ratio $I_{\rm ratio}=0.6-2.6$ and $\rm BD=1.5-2.2$ are consistent with the observed weak P-Cyg in \ion{He}{\sc i}--$\lambda$5876 ($\rm EW_{b+} < 2.5${~\AA}).

\item Day 6 starts with a flare, similar but brighter ($r' = 8.1-9.7$) than that observed in day 2, and followed by a faster decay ($\rm \sim 0.25\, h$). After this event, a slow rise analogous to that of day 3 is observed, starting with a nebular phase in $\rm H\alpha$ which slowly fades as the flux increases ($\rm EW =53-177${~\AA}). Deep P-Cyg profiles in \ion{He}{\sc i}--$\lambda$5876 (up to $<20\%$) are observed throughout the day, being weaker at higher fluxes. Their terminal velocities are also higher than on previous days ($\sim 3000 \, \rm km\, s^{-1}$). The ionisation ratio is $I_{\rm ratio}=0.2-1.3$ while $\rm BD=1.9-3.3$. 

\item Day 7 exhibits the fastest ($\rm < 5 \, min$) but also weakest ($r' = 7.9-8.9$) flares. This night is also characterised by a continuum shape becoming significantly redder at the flares' peak (see Sec. \ref{Disc_spec_evol}); there are flux density variations of $\rm 880-1800\,  mJy$ in the $\rm H\alpha$ continuum, and $\rm 2400-2800\,  mJy$ near $\rm H_\gamma$ ($\rm 4340${~\AA}). Shallow P-Cygs in \ion{He}{\sc i}--$\lambda$5876 are detected through visual inspection. $\rm H\alpha$ EW excesses (both in the red and blue halves) are present, suggesting a faint remnant of the nebular contribution. The ionisation ratio is $I_{\rm ratio}=1.6-2.6$ and the Balmer decrement is $\rm BD=1.5-2.6$.

\item Day 8 shows a $\rm 1 \, h$ plateau at $r' \sim 8.6-8.9$, with no significant spectral changes. The null excess in the $\rm H\alpha$ line suggests that neither nebular phase nor P-Cyg are present. The most remarkable feature is the presence of high ionisation ratios $I_{\rm ratio}\sim 1.8-2.2$, combined with $\rm BD=3.5-4.3$.

\item Day 9 corresponds to the peak of the outburst at $r' \sim 7.7-8.0$, but only $\sim 0.5 \rm \, h$ of data were taken. Neither flares nor nebular phases are observed, while high ionisation features like the Bowen blend and He {\sc ii}--$\lambda$4686 prevail ($I_{\rm ratio}=1.3-2.8$). The $\rm BD=2.0-2.8$ and EW of $\rm H\alpha$ ($\rm EW =50-71${~\AA}), combined with no blue excess and small red excess does not reveal any interesting feature. Regarding the \ion{He}{\sc i}--$\lambda$5876 line, we only detect marginal P-Cyg ($\rm EW_{b+} \gtrsim - 0.9${~\AA}) with terminal velocities of $\sim 2000 \, \rm km\, s^{-1}$.

\item Day 10 corresponds to a transition stage with $r' \sim 9.1$ at the begining of the main nebular phase event. Unfortunately, the short exposure time for these spectra (40 s) was still long enough to saturate the peak of the $\rm H\alpha$ line, but its broad wings confirm the nebular contribution. The ionisation ratio is $I_{\rm ratio}=0.9-1.2$, while $\rm BD= 2.7-2.9$ and $\rm H\alpha$ $\rm EW = 204-317${~\AA} (note that measurements on $\rm H\alpha$ must be treated with caution due to saturation, see Sec. \ref{observations}). Hereafter, we will consider the following thresholds as indicators of the main nebular phase: $\rm H\alpha$ $\rm EW \gtrsim 100${~\AA}, $\rm BD \gtrsim 5$, $I_{\rm ratio}<1.2$; all these combined with the simultaneous presence of blue and red excess in $\rm H\alpha$.

\item Day 11 shows the highest $\rm H\alpha$ EW values ($\rm EW =1472-1904${~\AA}) as well as the broadest wings (reaching up to $\rm \sim 3000 \, km\, s^{-1}$, the wind terminal velocity of day 6). This is the result of a strong drop in the system brightness ($r' \sim 11.5-12.0$) mainly due to the continuum flux ($\rm 20\,  mJy$), which is not immediately followed by the $\rm H\alpha$ line flux ($\rm 2000\,  mJy$ peak flux density). This extreme phenomenology is interpreted as caused by the presence of a cooling, expanding nebula. The ionisation ratio is extremely low $I_{\rm ratio}=0.06-0.12$, while $\rm BD=4.8-5.9$ reaches high values. Even at such low values of $I_{\rm ratio}$, we note that only shallow P-Cygs are found when inspecting the \ion{He}{\sc i}--$\lambda$5876 line residuals (Sec. \ref{outflows}). This contrasts with days 2 and 6, when deep P-Cyg were present at similar $I_{\rm ratio}<0.15$ values. Therefore, a low $I_{\rm ratio}$ is not a sufficient condition to observe P-Cyg features. This might be prevented by broad line wings associated with the nebula, which compensates for the blue absorption.

\item Day 12 still exhibits nebular phase features in $\rm H\alpha$ ($\rm BD=5.5-7.4$), but with lower $\rm EW =408-685${~\AA}, and narrower wings. On the other hand, the Bowen blend and He {\sc ii} high-excitation lines almost disappear at the lowest fluxes ($I_{\rm ratio}=0.03-0.3$). Nevertheless, narrow ($\rm 1500 \, km \, s^{-1}$ terminal velocity) but clear P-Cyg features are present in \ion{He}{\sc i}--$\lambda$5876, becoming deeper (up to $<15\%$) after a faint rebrightening of the system ($ r'= 13.1-12.8$). This shows that the wind is present even at very low luminosity outburst phases, in agreement with our results from the 2015D outburst (see \citealt{MunozDarias2017}).

\item Day 13 to 15: double-peaked profiles appear in several lines, starting with \ion{He}{\sc i} ($\rm 5876${~\AA}, $\rm 6678${~\AA}, $\rm 7065${~\AA}), and later also in Fe {\sc ii} (the ``forest" at $\rm 5150-5300${~\AA}), as the system starts returning to quiescence. Balmer lines remain single-peaked for a few more days because of filled in residual nebular emission. The ionisation ratio is $I_{\rm ratio}= 0.2$ and $\rm BD= 5.3$ on day 13; $I_{\rm ratio}=0.3-0.7$ and $\rm BD=3.0-3.6$ on day 15. The corresponding EWs of $\rm H\alpha$ are $\rm 183-189${~\AA} (on day 13) and $\rm 74-91${~\AA} (on day 15), with continuum fluxes of $r' \sim 12.4$ and $\sim 12.5-12.8$, respectively. It is interesting to note that double-peaked profiles are observed at brighter $r'$ magnitudes than on day 12. The arbitrary threshold of $\rm EW \sim 100${~\AA} and $\rm BD \sim 5$, as well as the presence of low blue and red excess in $\rm H\alpha$ ($\rm EW_{excess} \sim 1${~\AA}) on day 13 allow us to set the end of the nebular phase after this day. Despite double-peaked profiles and low continuum fluxes, the presence of He II shows that there is still accretion activity in the system by day 15 ($I_{\rm ratio}$ is similar to day 1).

\end{itemize}

\subsubsection{Evolution of $\rm H\alpha$}
\label{halpha-evol}

\begin{figure*}
\begin{center}
\includegraphics[keepaspectratio, width=0.5\textwidth]{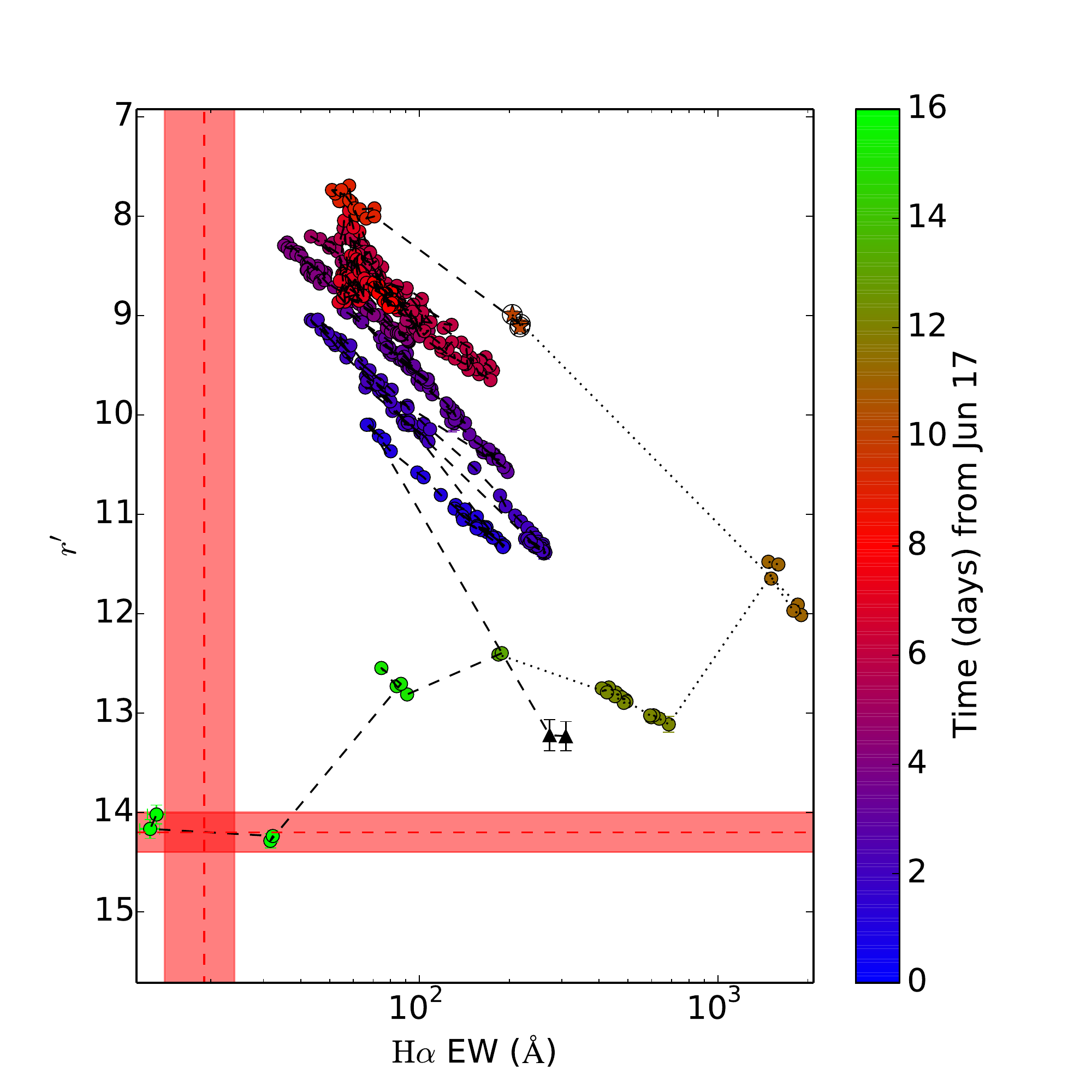}\includegraphics[keepaspectratio, width=0.5\textwidth]{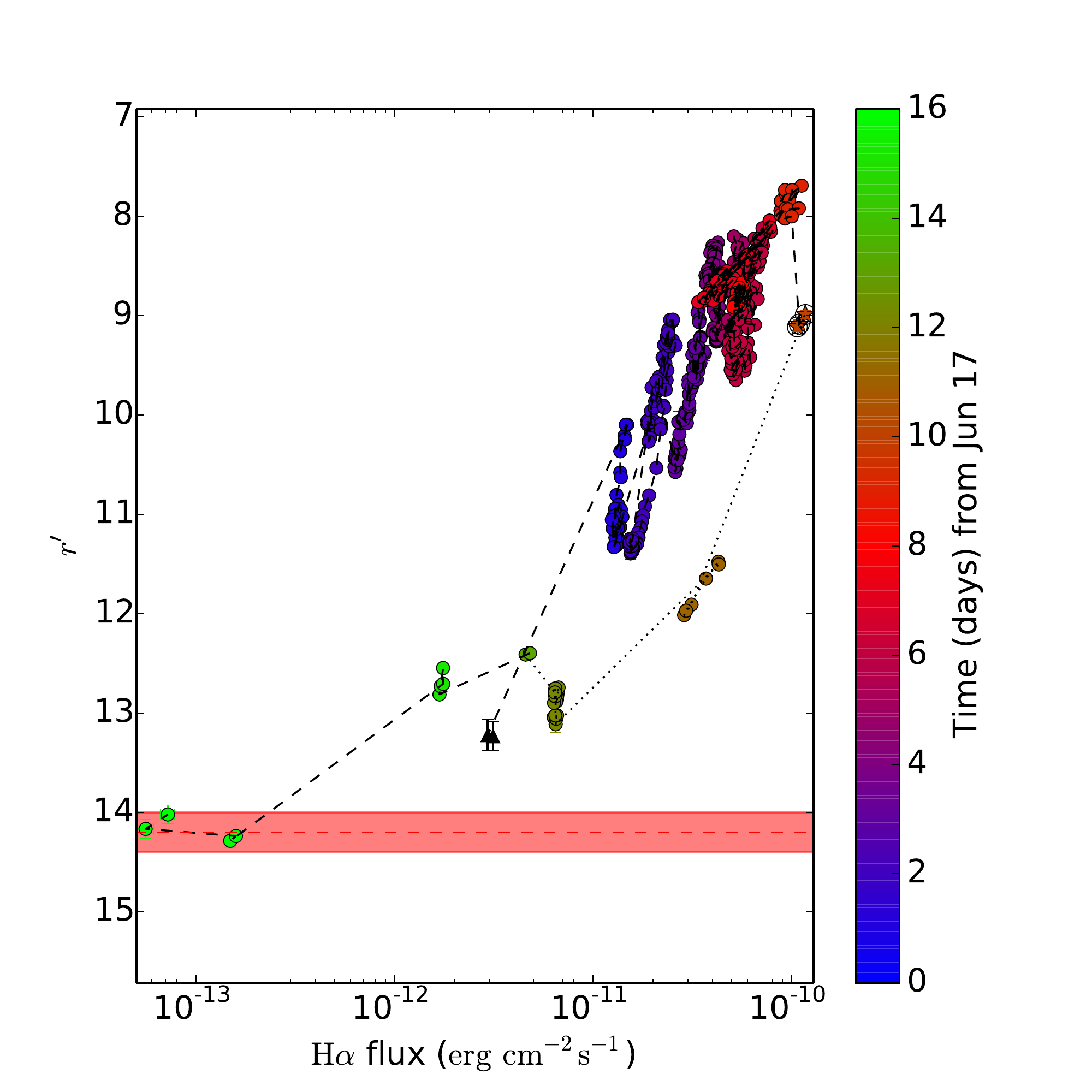}
\end{center}

    \caption{Left: $\rm H\alpha$ EW versus de-reddened $r'$. Right: $\rm H\alpha$ flux versus de-reddened $r'$. The quiescent magnitude is depicted as a red line, with a shaded region for the conservative 0.3 mag uncertainty. Colour bars refer to the observation date in days. From day 16 onwards, all the points have the same green colour. Encircled, star-shaped points refer to those where $\rm H\alpha$ saturated the detector. The black-dashed line connects the consecutive points to show the temporal evolution. Those corresponding to the nebular phase (days 10--13) have been connected with a black-dotted line. Black triangles refer to observations on day -1.8 (that is, $\rm\sim 13 \, h$ before the first X-ray detection of the outburst).}
    \label{fig: rvsHalpha_flux}
\end{figure*}

\bigskip

$\rm H\alpha$ is the most intense optical emission line in almost every LMXB spectrum. Therefore, its analysis provides valuable results that can be compared with those from many other LMXBs in outburst.

\bigskip

\textbf{EW evolution: the nebular loops}

\bigskip

We first examine the relationship between $\rm H\alpha$ EW and $r'$ (see left panel of Fig.\,\ref{fig: rvsHalpha_flux}). There is an initial rise of the continuum flux, starting from the earlier phases of the outburst to the peak luminosities (days 1 to 9) with typical $\rm EWs$ in the range $\sim 50$--$\rm 200${~\AA}. The decay to quiescence is preceded by the so-called nebular phase (days $10-13$), characterised by a dramatic drop in the continuum flux, as well as an intense and relatively narrow ($\rm FWHM < 1000\, km\, s^{-1}$) $\rm H\alpha$ line (peaking $\sim 100$ times above the continuum level) with extended wings (up to $\rm \pm \sim 3000\, km\, s^{-1}$). This results in extremely large EWs ($\rm 1000$--$\rm 2000${~\AA}, see Fig.\,\ref{fig: rvsHalpha_flux}). After this stage, the system slowly returns to quiescent levels (defined by $r' \sim 14$, $\rm EW=19.0 \pm 5.2$; \citealt{Casares2015}) over a timescale of $\rm \sim 40$ days (on day 41, both the EW and $r'$ magnitude are consistent with quiescence within $2\sigma $). This evolution draws a clockwise loop pattern in the EW-$r'$ diagram due to the presence of the nebular phase. Such a pattern is observed in the evolution of several parameters during the outburst, and will be referred to hereafter as the \textit{nebular loop}.

\begin{figure*}
\includegraphics[trim=0cm 0cm 0cm 0cm, clip=true, width=\textwidth]{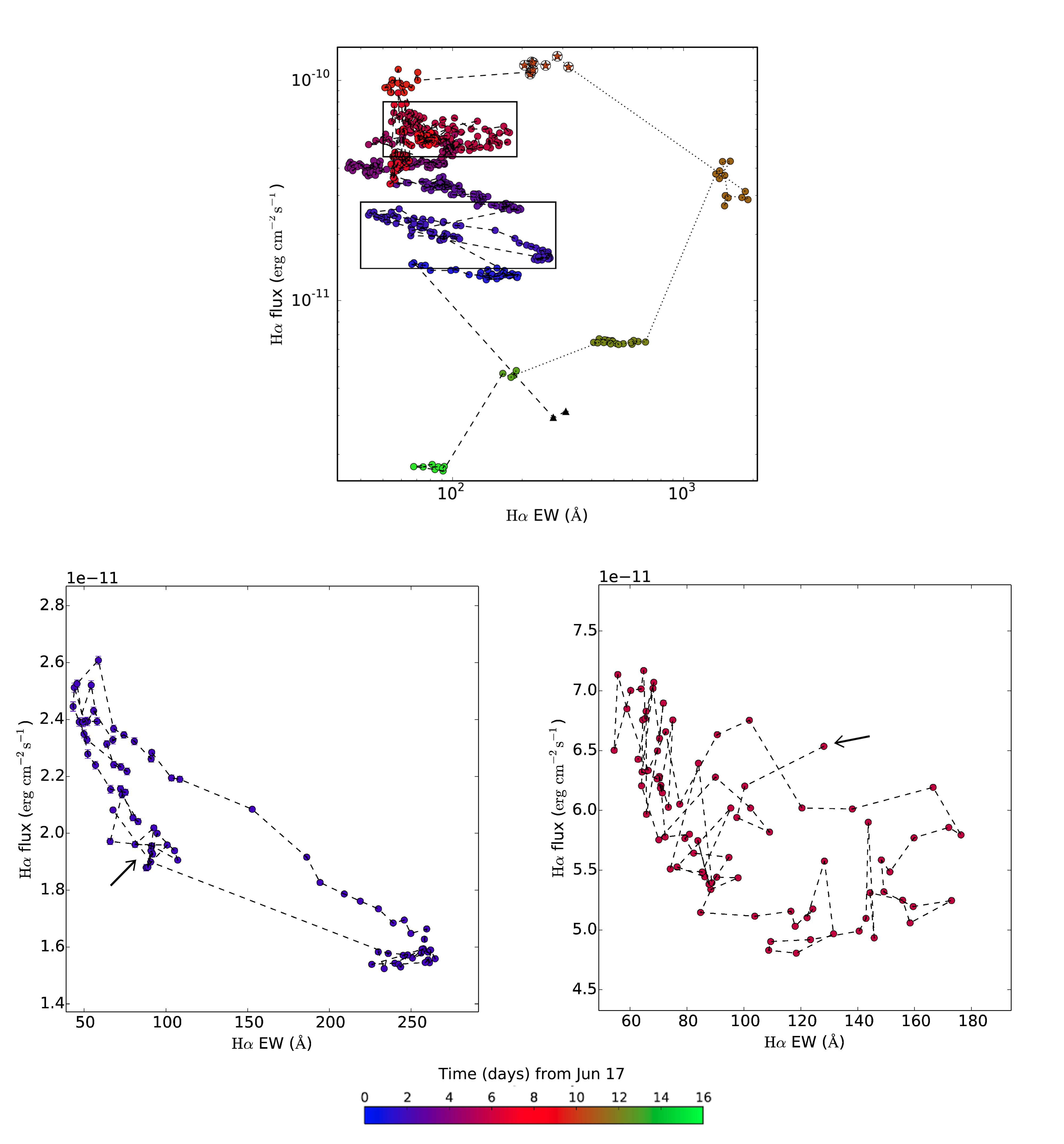}
    \caption{Top panel: EW versus flux for $\rm H\alpha$. The colour bar refers to the observation date in days. Only observations up to day 15 are depicted. As in Fig.\,\ref{fig: rvsHalpha_flux}, encircled, star-shaped points refer to those where $\rm H\alpha$ saturated the detector. Black triangles refer to observations before the first X-ray detection of the outburst (day -1.8). Bottom panels: zoom-in of top-left panel for days 2 (left) and day 6 (right), both marked by squares in the full diagram. The starting observation on each day has been marked with an arrow.}
    \label{fig: Halphaplots}
\end{figure*}

\bigskip

The $\rm H\alpha$ flux follows a similar nebular loop when plotted versus $r'$ (Fig.\,\ref{fig: rvsHalpha_flux}, right panel). The decay of the $\rm H\alpha$ flux during the nebular phase occurs at a slower rate than that of the continuum, exhibiting $\rm H\alpha$ fluxes similar to those observed at much brighter outburst phases ($\Delta r'\sim 3$). In fact, during the main nebular phase (day 11: largest EW values), the contribution of this single line produces $\sim 70\%$ of the integrated flux over the whole $r'$-band. This suggests that the main contribution to the emission line during this phase is not the accretion disc, but a cooling, expanding nebula (see MD16). 

A similar nebular loop is seen when plotting the EW versus the $\rm H\alpha$ line flux, as shown in Fig.\,\ref{fig: Halphaplots} (top panel). When looking at this relation on a day-by-day basis, we also find smaller clockwise nebular loops in days 2 and 6 (see Fig.\,\ref{fig: Halphaplots}, lower panels), those covering both the rise and decay of each optical flare. The timescales of these shorter loops are $\sim 30-60\, \rm min$ (when EW increases from $50$ to $200${~\AA}), and they are observed simultaneously with the deepest detected P-Cyg profiles in the reference line \ion{He}{\sc i}--$\lambda$5876 throughout the outburst (see Sec. \ref{outflows}).

\bigskip

\textbf{FWHM evolution}

\bigskip

\begin{figure*}
\includegraphics[width=\textwidth]{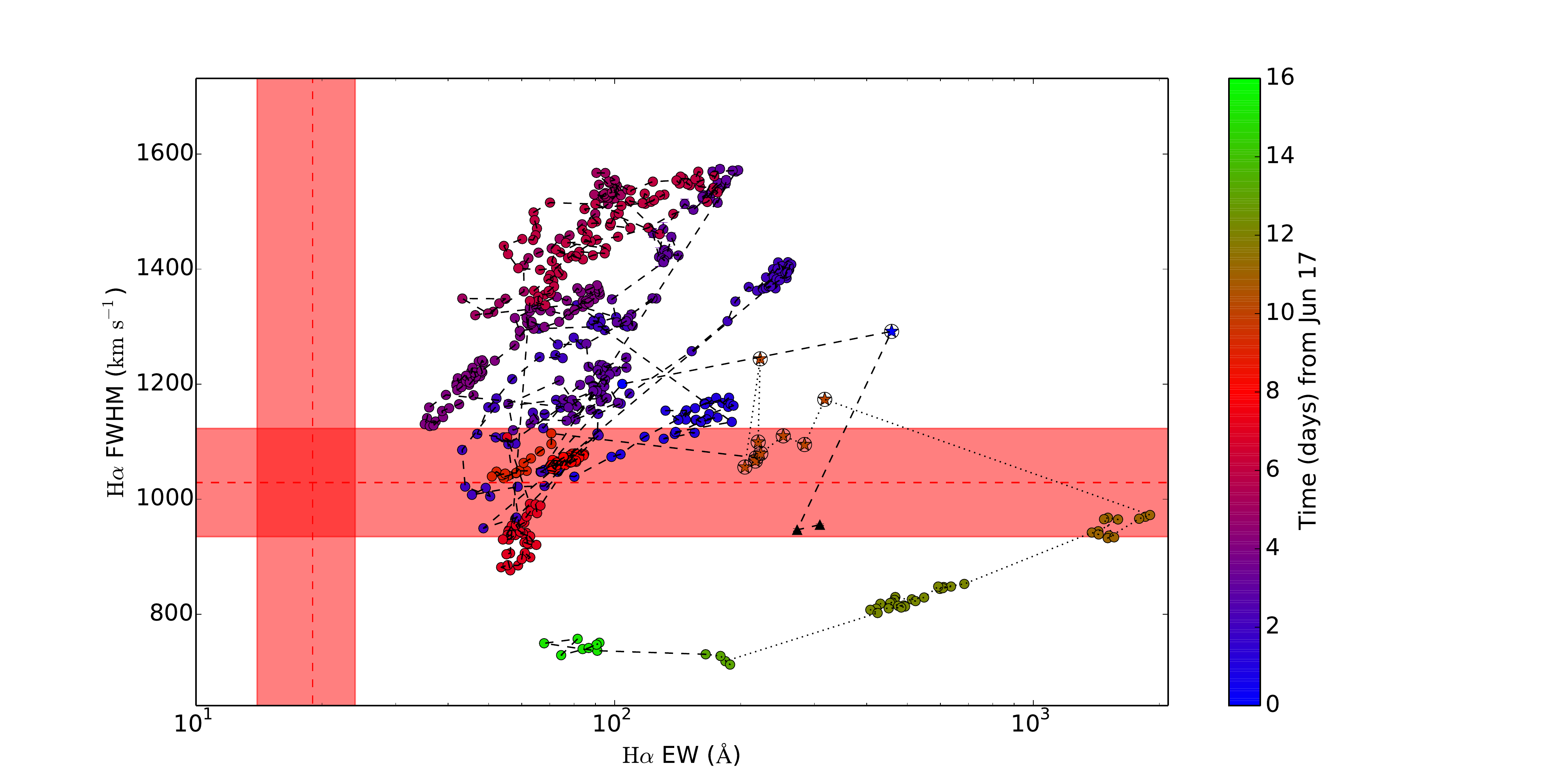}
    \caption{EW versus FWHM for $\rm H\alpha$. The colour bar refers to the observation date in days. Only observations up to day 15 are depicted. As in Fig.\,\ref{fig: rvsHalpha_flux}, encircled, star-shaped points refer to those where $\rm H\alpha$ saturated the detector. Black triangles refer to observations before the first X-ray detection of the outburst (day -1.8). The quiescent EW and FWHM values are depicted as red-dashed lines, with shaded regions for the uncertainty ($\rm EW=19.0 \pm 5.2$, $\rm FWHM=1029 \pm 94 \, km \, s^{-1}$, see \citealt{Casares2015}).}
    \label{fig: Halphafwhm}
\end{figure*}

The outer radius of an accretion disc in a LMXB can be estimated from the double peak separation  of emission lines \citep{Orosz1994}. This separation has been observed to decrease during the outburst state when compared to quiescent observations. The classic picture of LMXBs considers that the distance between peaks of an emission line profile decreases during outburst as the disc expands and reaches areas of lower Keplerian velocities. However, V404 Cyg (as well as some other LMXBs) does not retain a double-peaked profile during the most active phase of its outburst.

Incidentally, \citet{Casares2015} found that the $\rm H\alpha$ FWHM of quiescent LMXBs is linearly correlated with the projected orbital velocity of the companion star. He concluded that this was a consequence of the FWHM tracing the disc velocity at a certain radius ($\sim 42 \%$ of $\rm R_{L1}$). Therefore, the same disc expansion responsible for a decrease in the double-peak separation ought to produce narrower FWHMs. Indeed, this pattern has been observed in many LMXBs from outburst to quiescence (e.g. Swift J1357.2-0933, see \citealt{Corral-Santana2013}, \citealt{Torres2015} and \citealt{MataSanchez2015b}), including V404 Cyg during its previous 1989 event (see Fig. 2 in \citealt{Casares2015}, also Sec. \ref{Halpha1989and2015D}).

During 2015J, a Gaussian fit to $\rm H\alpha$ reveals clear changes in its FWHM. However, during the initial and most active phases of the outburst, the FWHM of V404 Cyg is systematically larger than the quiescent value $\rm FWHM=1029 \pm 94 \, km \, s^{-1}$ \citep{Casares2015}, as we show in Fig.\,\ref{fig: Halphafwhm}. This unexpected increase in the FWHM suggests that the nebular phase most likely contributes to the emission line throughout the outburst. Only at the end of the nebular phase, and for 9 days (from day 12 to 21) the FWHM drops below the quiescent level. However, the line flux remains high even if the continuum drops significantly, so this cannot be a fully disc-dominated state, but still contaminated by the expanding nebula.

\begin{figure*}
\includegraphics[keepaspectratio, width=0.6\textwidth]{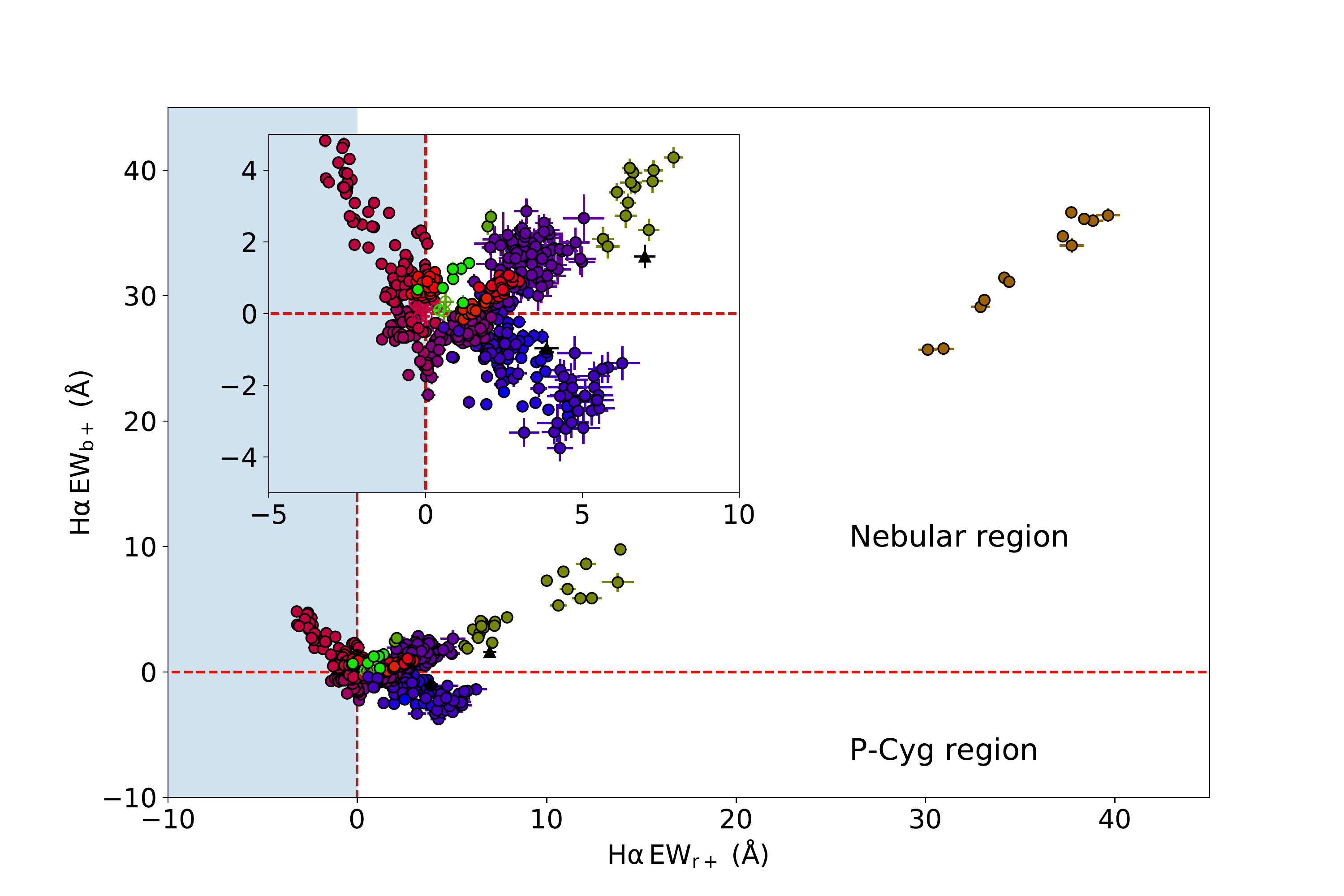}\includegraphics[keepaspectratio, width=0.4\textwidth]{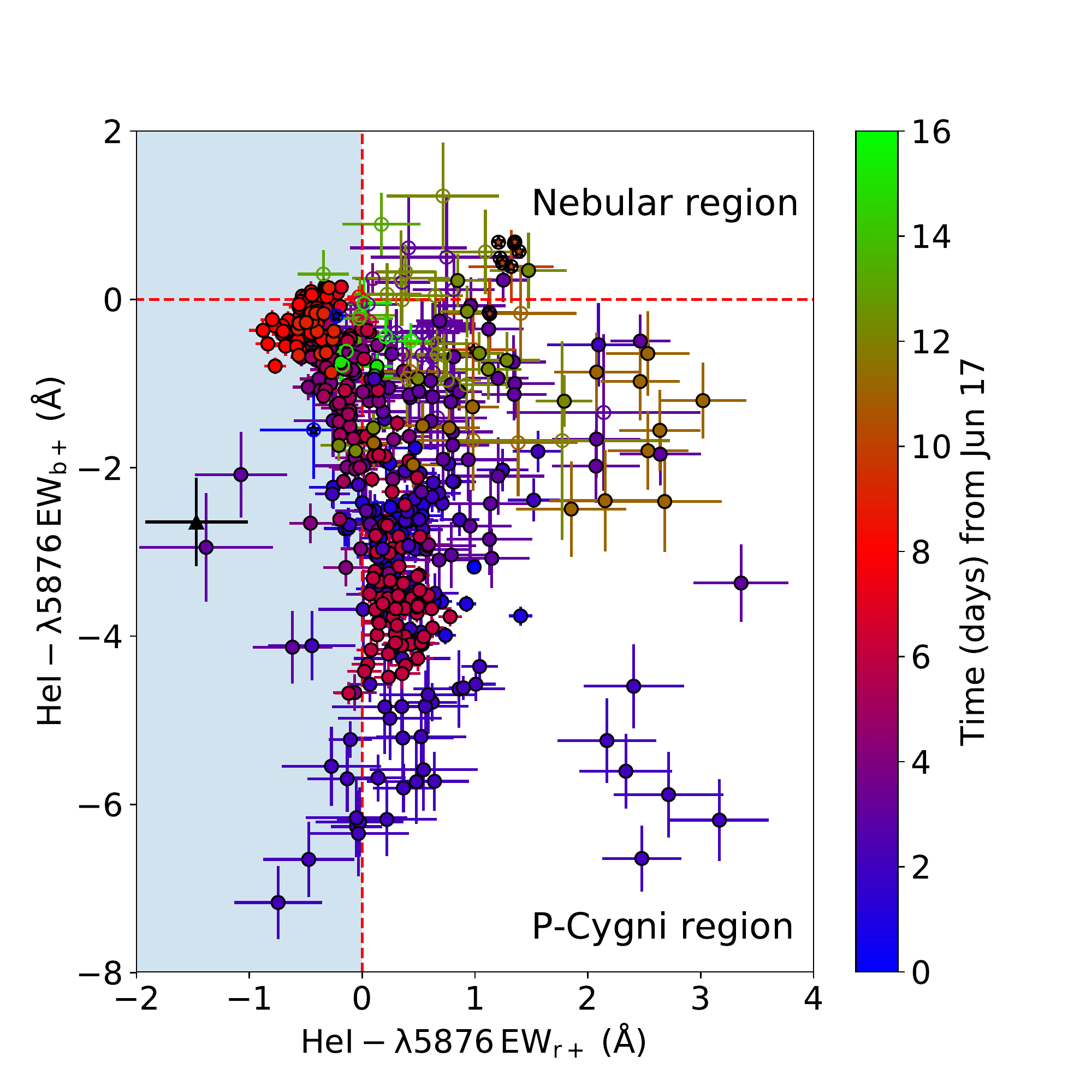}
    \caption{Left panel: EW excess (red vs blue halves) of the $\rm H\alpha$ line residual after a Gaussian subtraction. The inset depicts a zoom into the origin. Those spectra where $\rm H\alpha$ saturated the detector have been omitted. Black triangles refer to observations before the first X-ray detection of the outburst (day -1.8). Right panel: analogous diagram for the \ion{He}{\sc i}--$\lambda$5876 line residuals. Only observations up to day 15 are shown. The colour bar refers to the observing date. The nebular and P-Cyg regions are marked, while forbidden regions (e.g., contaminated by other nearby line transitions) are indicated with a blue shading.}
    \label{fig: Halpharedvsblue}
\end{figure*}
 
 \bigskip

\subsubsection{Search for outflows}
\label{outflows}
\begin{figure*}
\includegraphics[keepaspectratio, width=\textwidth]{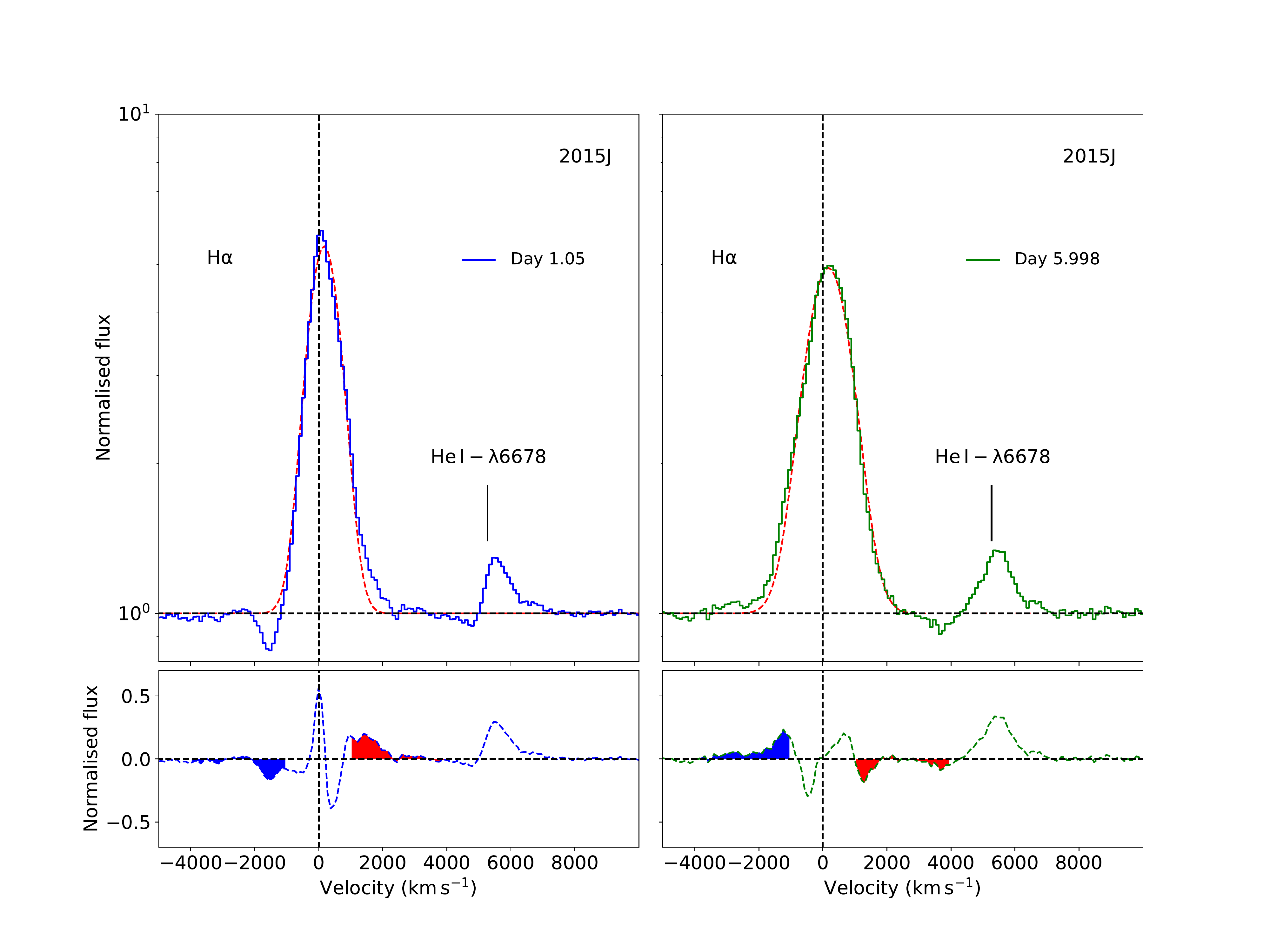}
    \caption{Left: individual spectrum (blue, solid line) obtained in day 1. The Gaussian fit to the $\rm H\alpha$ line is shown as a red, dashed line. Below, we present the resulting residual after subtracting the fit from the spectra. The blue and red filled regions correspond to the mask defined to obtain $\rm EW_{b+}$ and $\rm EW_{r+}$, respectively. Note that the P-Cyg in \ion{He}{\sc i}--$\lambda$6678 has a relatively low terminal velocity and does not affect the red wing of $\rm H\alpha$. Right: individual spectra obtained in day 6 (green, solid line), during the deepest P-Cyg profiles observed in \ion{He}{\sc i} lines. Both the Gaussian fit and residual are defined as in the left panel. The skewed profile of $\rm H\alpha$, as well as the presence of a P-Cyg absorption in \ion{He}{\sc i}--$\lambda$6678, affect the measurement of the residual excess in the profile wings.}
    \label{fig: specd6}
\end{figure*}

\bigskip

In order to search for the presence of outflows, we analysed the two main emission lines that show the most prominent outflow features: $\rm H\alpha$ (nebular phase) and \ion{He}{\sc i}--$\lambda$5876 (P-Cyg).

We subtracted the initial Gaussian fit to $\rm H\alpha$ from every spectrum and studied the residuals. The EW excess of the two velocity halves, as defined in Sec. \ref{methods} (blue and red, hereafter), can be used to classify the spectra in different groups (Fig.\,\ref{fig: Halpharedvsblue}, left panel). A blue absorption (i.e., negative $\rm EW_{b+}$) combined with a red excess (i.e., positive $\rm EW_{r+}$) reveals the presence of P-Cyg profiles (e.g., day 1; see Fig.\,\ref{fig: specd6}, left panel), while excess in both the blue and red halves are characteristic of the nebular phase (e.g., days 10--13). We note that the upper quadrant of the blue-shaded (forbidden) region of the diagram (blue excess and red absorption) is only populated by day 6 spectra. While the blue excess could mark the presence of the nebular phase, the red absorption is intriguing. After visual inspection of these spectra (see Fig. \ref{fig: specd6}, right panel), we found the origin of this red absorption in two factors: i) the intense P-Cyg profiles of the nearby \ion{He}{\sc i}--$\lambda$6678 line (see Fig.\,\ref{fig: specd6}, right panel), which achieves on this day the maximum terminal velocity observed ($ \rm  3000\,   km\,  s^{-1}$) and contaminates the red excess of $\rm H\alpha$, and ii) the skewed $\rm H\alpha$ blue half, probably due to the presence of a broad P-Cyg with similar terminal velocities to that of \ion{He}{\sc i}--$\lambda$6678, affecting the Gaussian fit. The rest of the spectra are compatible with null excess in both the blue and red halves of $\rm H\alpha$ (therefore, Gaussian profiles).

The same analysis was applied to \ion{He}{\sc i}--$\lambda$5876, which exhibits the most prominent P-Cyg profiles throughout the outburst. We note here that the mask defining the red ($\rm 1250$ to $4000 \,  \rm km\,  s^{-1}$) and blue ($ \rm-500$ to $ -4000\,  \rm  km\,  s^{-1}$) halves is different from that used for $\rm H\alpha$. This is due to both its narrower profile and the presence of interstellar absorption (Na {\sc i} $5890-5896${~\AA}). This interstellar feature was masked when fitting the Gaussian. The resulting EW excess diagram (Fig.\,\ref{fig: Halpharedvsblue}, right panel) reveals the deep blue absorptions (up to $7${~\AA}) of P-Cyg profiles. The nebular phase region (blue and red excess) is less populated than in the $\rm H\alpha$ line diagram, with the remarkable exception of day 10. We note that the red excess is always underestimated (due to the interstellar feature mentioned before) and produces the misplacement of some values in the blue-shaded region (red absorptions). The remaining spectra are compatible with null excess in both the blue and red halves of \ion{He}{\sc i}--$\lambda$5876 (therefore, Gaussian profiles).

\bigskip

\subsection{December 2015 and 1989 outbursts}

Our team also obtained optical spectroscopy during both the 2015D \citep{MunozDarias2017} and 1989 \citep{Casares1991} outbursts. In this section, we will extend the diagnostic diagrams and techniques defined for the 2015J event to all three outbursts. Given that the spectra corresponding to these events are not flux-calibrated, we will only inspect diagrams produced from normalised spectra.

\subsubsection{Evolution of $\rm H\alpha$}
\label{Halpha1989and2015D}

\begin{figure}
\includegraphics[width=0.5\textwidth]{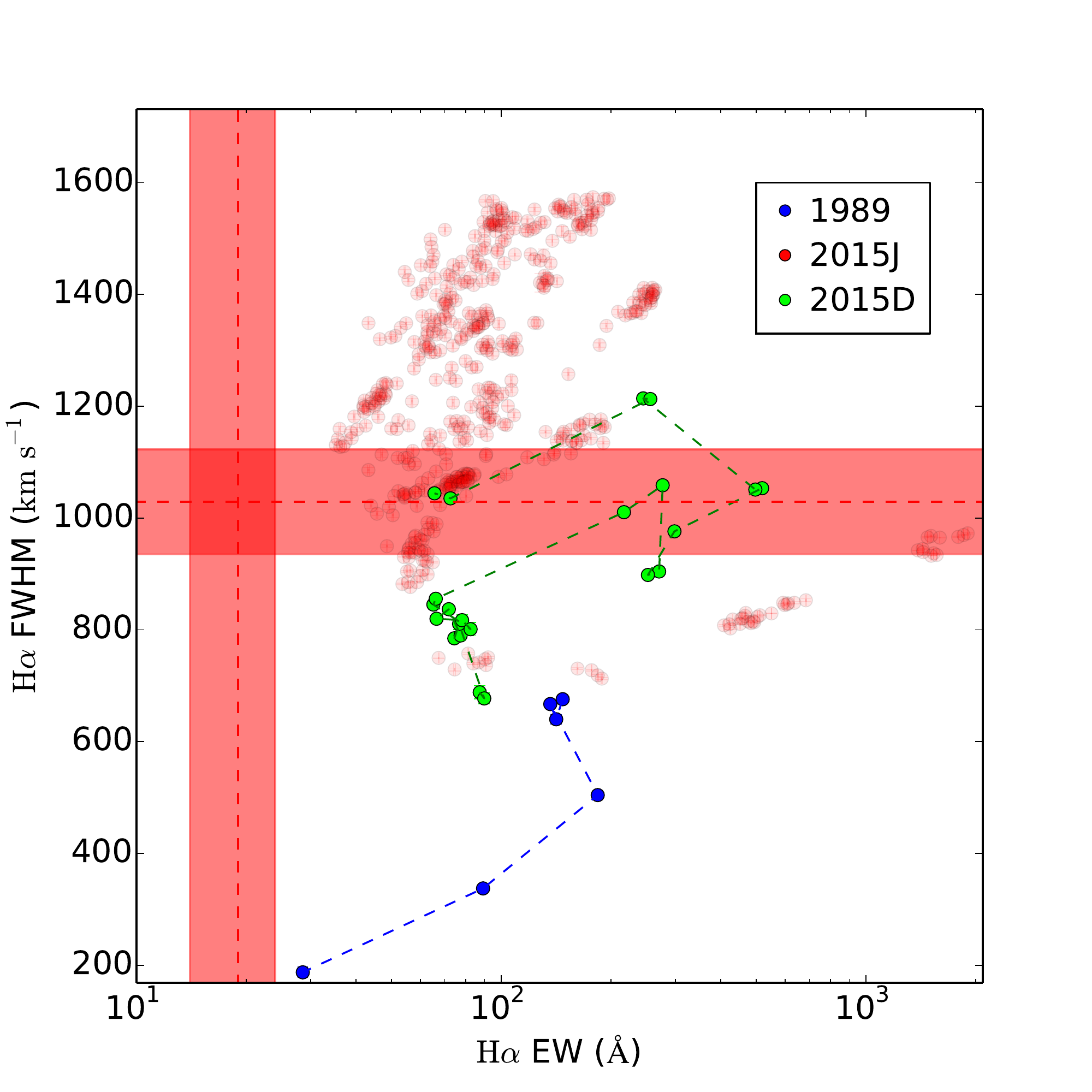}
    \caption{EW versus FWHM for $\rm H\alpha$. Colour coding of the points from the 3 outbursts is indicated in the legend. The quiescent values are marked with red-shaded bands for reference.}
    \label{fig: Halpha_EWvsFWHM_ALL}
\end{figure}

The EW of $\rm H\alpha$ and the FWHM of the Gaussian fit can be measured in the spectra of all three outbursts, as they do not require flux-calibrated spectra. The diagram produced by these parameters is shown in Fig. \ref{fig: Halpha_EWvsFWHM_ALL}.

Inspecting this diagram reveals that the 2015D outburst follows a nebular loop similar to that observed in 2015J, but reaching lower maximum values of EWs (it varies between $\rm EW=35-520\, ${\AA}). The FWHM of the Gaussian fit sometimes reaches values higher than the quiescent level ($\rm FWHM=1029 \pm 94 \, km \, s^{-1}$), but most of the time remains at equal or lower levels. 

On the other hand, the 1989 outburst exhibits a different behaviour. We first note that the $\rm H\alpha$ FWHM always remains below any measurement obtained in either the 2015J or 2015D outbursts. Indeed, the lowest value of $\rm FWHM=187 \pm  4 \, km \, s^{-1}$ is observed simultaneously with the lowest $\rm EW=28.6\pm 0.8\, ${\AA} ever measured in our dataset (close to the quiescent value $\rm EW=19.0 \pm 5.2\, ${\AA}). The $\rm H\alpha$ spectra obtained during the 1989 event only covers four days (MJD = 47678.129 -- 47682.194); it starts at relatively high values of FWHM ($\rm\sim 650 \, km \, s^{-1}$), then decays to the lowest one, always at $\rm EW \lesssim 200${~\AA}. FWHMs lower than the quiescent value are consistent with the classic picture of an expanding disc also observed in cataclysmic variables (CVs, see e.g., \citealt{Smak1984}; \citealt{Baptista2001}). 
 
\subsubsection{Search for outflows}

\begin{figure*}
\includegraphics[keepaspectratio, width=0.5\textwidth]{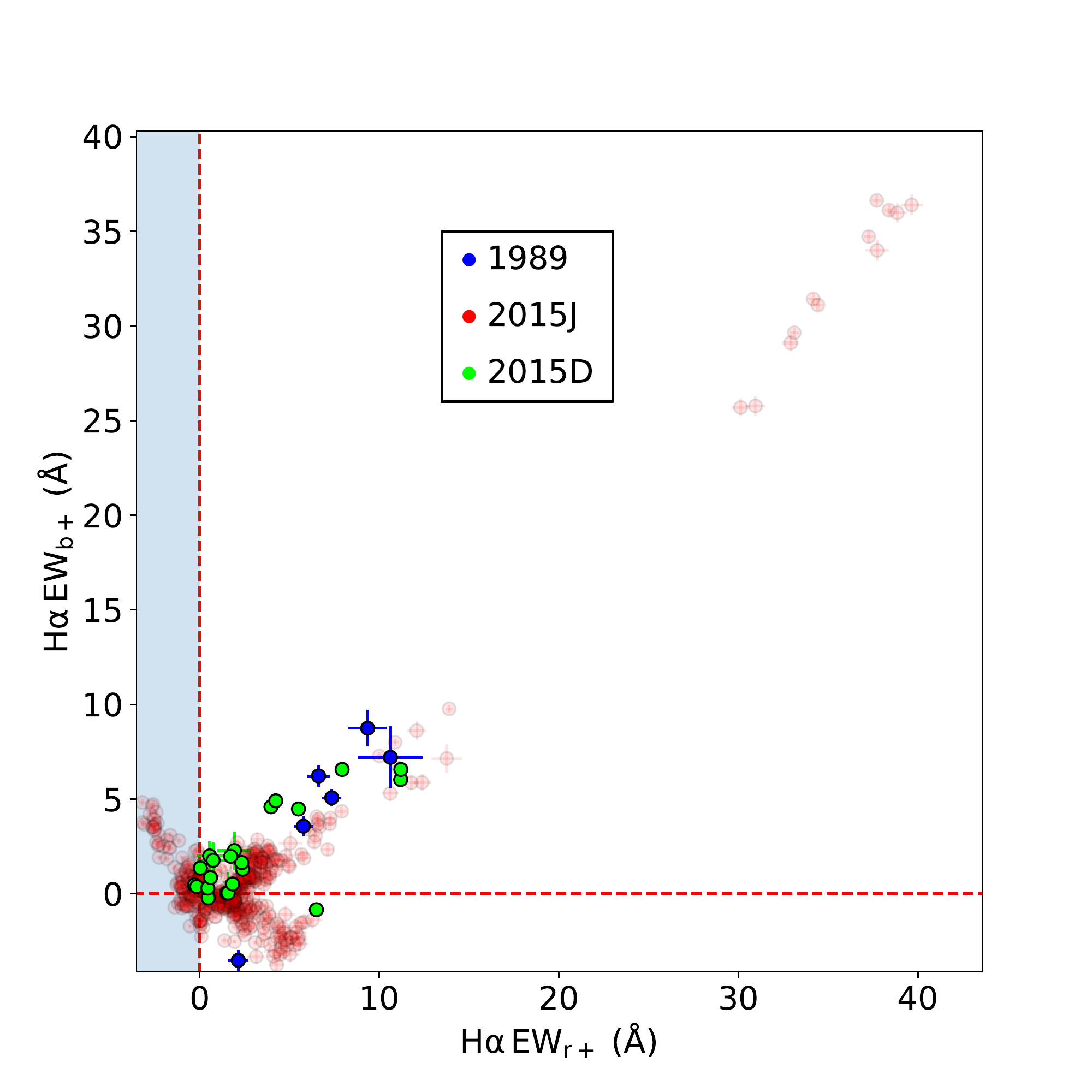}\includegraphics[keepaspectratio, width=0.5\textwidth]{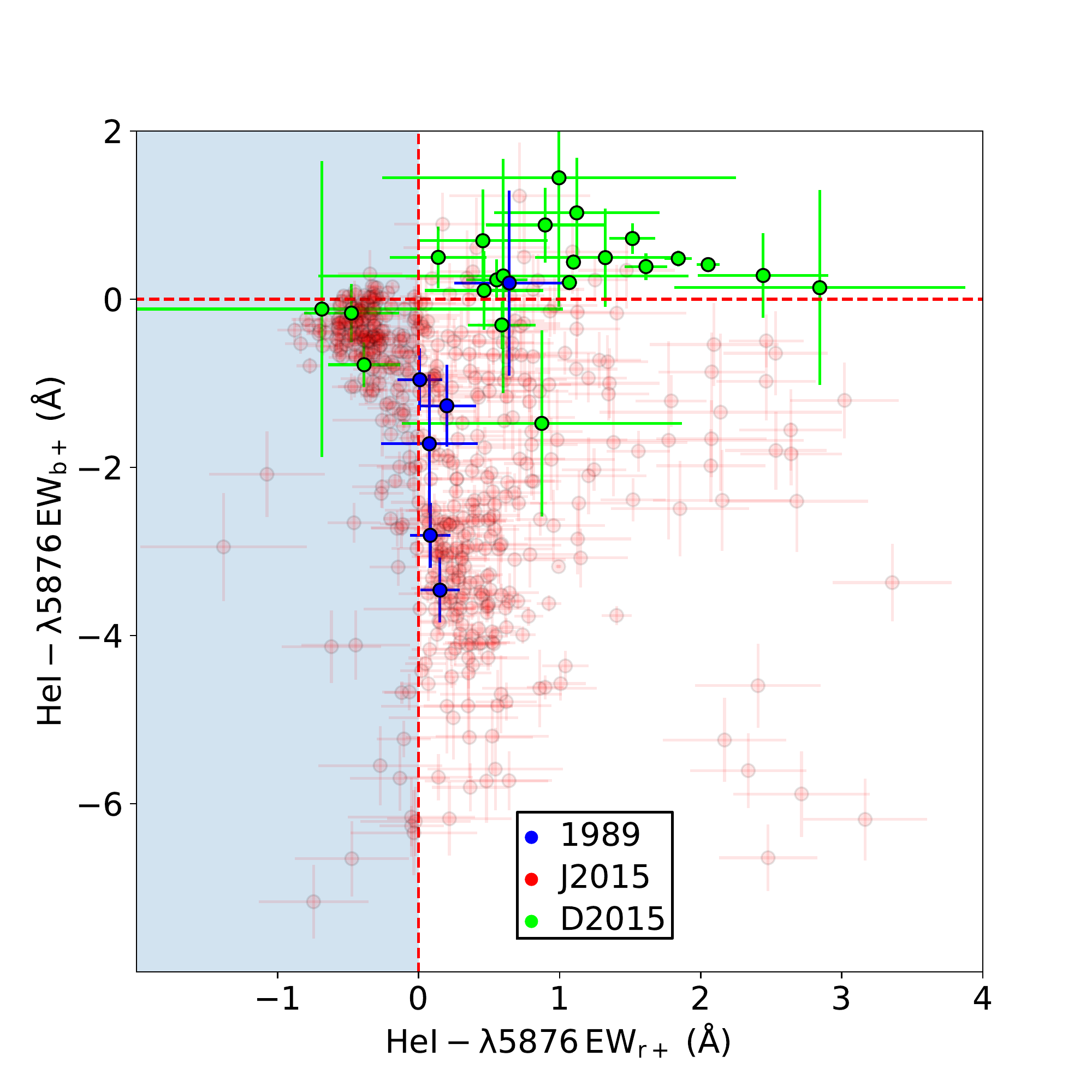}
    \caption{Left panel: EW excess (red vs blue halves) of the $\rm H\alpha$ line residual after a Gaussian subtraction. Those spectra where $\rm H\alpha$ saturated the detector have been omitted. Right panel: analogous diagram for the \ion{He}{\sc i}--$\lambda$5876 line residuals. The colour code is the same than that in Fig. \ref{fig: Halpha_EWvsFWHM_ALL}. Blue-shaded areas indicate the forbidden regions of the diagram.}
    \label{fig: Halpha_redblue_ALL}
\end{figure*}

We searched for outflows in the 2015D and 1989 events using the same technique as in the 2015J dataset (see Sec. \ref{outflows}). The EW excesses after a Gaussian subtraction (red vs blue halves) of the $\rm H\alpha$ line are depicted in Fig. \ref{fig: Halpha_redblue_ALL} left panel. Given both the narrowness of the lines and lower wind terminal velocities in the 1989 event, we use a different mask when calculating the residual EW ($\rm 50$ to $\rm 4000 \, km\, s^{-1}$ and $\rm -4000$ to $\rm -50 \, km\, s^{-1}$). We find that for both the 2015D and 1989 outbursts, the EW excesses (both red and blue halves) fall in the regions already covered by the 2015J event. Both the 2015D and 1989 outburst measurements are mainly distributed in the nebular region rather than in the P-Cyg one. Indeed, only two P-Cyg events are observed in $\rm H\alpha$: a P-Cyg with terminal velocity of $\rm \sim 1500 \, km\, s^{-1}$ in 2015D day 6, just before a nebular phase starts; and a narrow (terminal velocity of $\rm \sim 900 \, km\, s^{-1}$) but deep P-Cyg in day 15 of 1989 (with EW blue absorption similar to those observed in 2015J). The analogous diagram but for \ion{He}{\sc i}--$\lambda$5876 is depicted in Fig. \ref{fig: Halpha_redblue_ALL} right panel. During the 2015D outburst few and shallow P-Cyg profiles are found. On the other hand, the 1989 spectrum showing the deepest $\rm H\alpha$ absorption feature exhibits a shallower absorption in \ion{He}{\sc i}--$\lambda$5876, while the deepest absorption in the latter transition is found by day 11 (see Fig. 6 in \citealt{Casares1991}). During the remaining days, either nebular phase or gaussian profiles are observed.

\section{Discussion}

\subsection{The overall spectral evolution of the June 2015 outburst}
\label{Disc_spec_evol}

The detailed spectroscopic follow up of this outburst allowed us to observe the evolution of several spectral parameters in the optical range during the event. The V404 Cyg spectrum is highly variable on day-to-day timescales, but also within each observing block (with timescales of seconds to minutes, see \citealt{Kimura2016}).

\begin{figure*}
\begin{center}
\includegraphics[keepaspectratio, width=\textwidth]{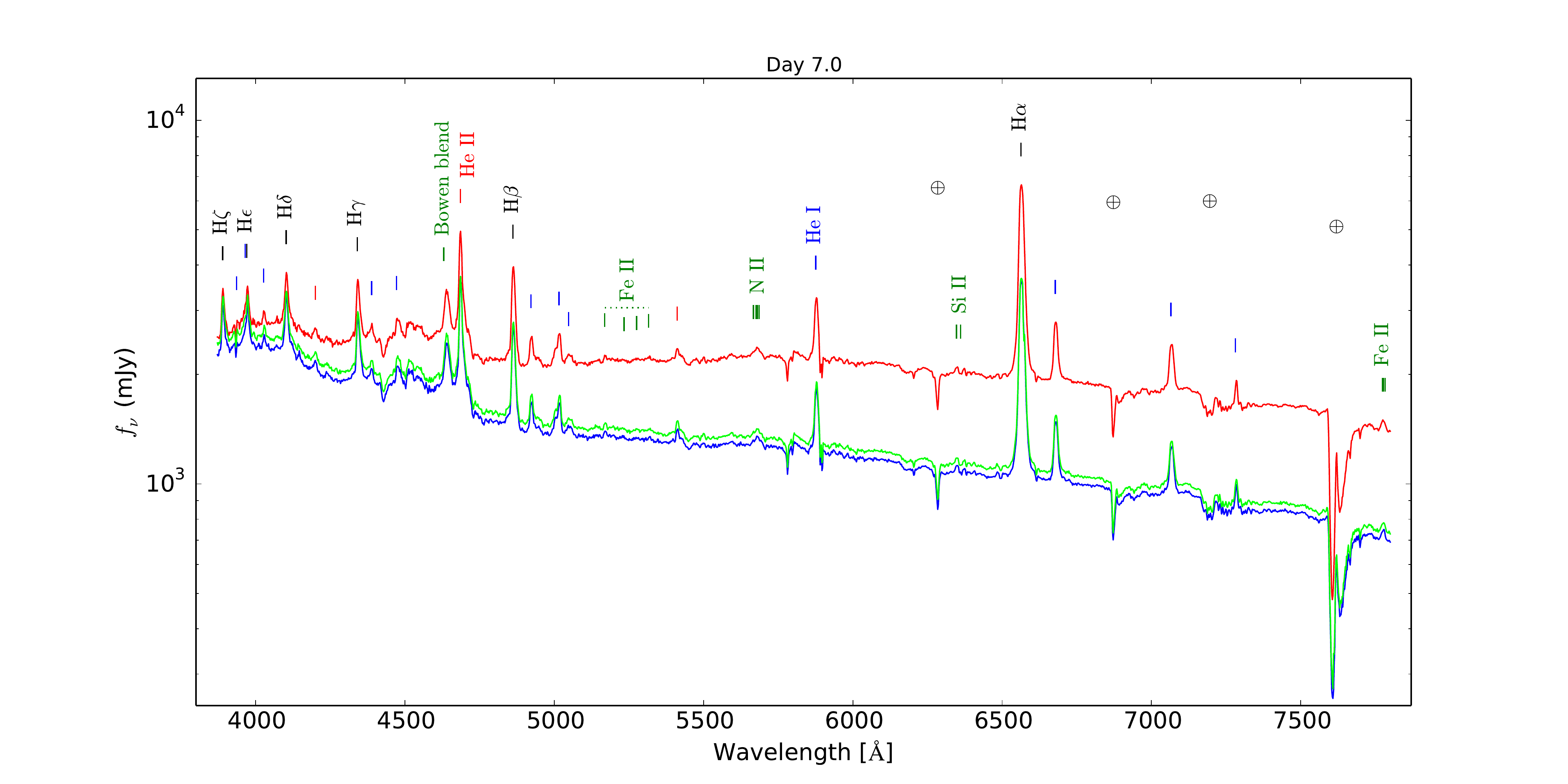}
\end{center}
    \caption{Day 7 spectral evolution during a rapid flare. The spectra correspond to before (blue), at the peak (red) and after the flare (green) respectively. Each spectrum has been acquired only 84 s after the previous one. Emission line identifications are also colour encoded: black for H, blue for \ion{He}{\sc i}, red for He {\sc ii} and green for other species (Si {\sc ii}, N {\sc ii}, Fe {\sc ii} and the Bowen blend). Telluric absorption lines are marked as $\oplus$.}
    \label{fig: flares7}
\end{figure*}

The spectral continuum varies both in brightness and shape as the outburst evolves, as seen in Fig. \ref{fig: spectra_fluxtot}. Among these variations, we highlight those occurring during day 7. The fast ($\rm < 5 \, min$), low amplitude flares (7.9--8.9) observed on this day (see Tab. \ref{tab:summatory}) produce a drastic change in the spectral continuum shape, which varies from negative step values to be almost flat in the peak of each flare, thus implying that the flare flux is red (see Fig. \ref{fig: flares7}). This is particularly interesting when compared with the regular, longer flares observed during the outburst (e.g., days 2 and 6, flare timescales of $\rm \sim  1\, hour$), where the continuum shape does not change that dramatically through the flaring event and, in fact, becomes bluer at the flare peaks.

Rapid (sub-second) red flares in $r'$ have been observed in V404 Cyg by \citet{Gandhi2016} on MJD 57199.20 (day 9.2 using our zero time). They propose that this variability arises from the base of the jet. This scenario has been recently supported by the coordinated X-ray, optical and radio observations presented in \citet{Gandhi2017}. In such work, they found a transition on day 8.18 involving a clear change in the timing correlation between the optical and X-rays, as subsecond delays appear. The transition is also associated with a short and fast (sub-second) flaring episode in their $r'$ light curve, plus a sharp rise in flux and spectral index of the radio emission, properties which support a jet origin for the fast optical variability. 

Although we have observations on days 8 and 9, they correspond to 8.00-8.03 and 9.04-9.06 respectively and therefore, the previously described events were not covered by our spectroscopic database. Nevertheless, the fast ($\rm < 5 \, min$, as we cannot resolve them with our lower temporal resolution), red flares found on day 7 might also have been produced by the variable contribution of the base of the jet.
 
\bigskip

Another peculiarity of the spectral evolution is the presence of double-peaked emission lines on day 13 after a faint re-brightening, while fainter spectra obtained on day 12 show single-peak profiles. The evolution from double-peaked to single-peaked during the brightening and \textit{vice versa} during the decay, has been observed in many outbursting LMXBs, but the physical origin of this phenomenon is still under debate. The observed change in the emission line profiles from single-peaked on day 12 ($r'\sim 13$) to double-peaked on day 13 ($r'\sim 12.4$) implies that, even if low continuum fluxes are required for the double peak to reappear (no double peaks are observed at $r' < 12.4$) there must be other factors involved.

The strong influence in the emission line profiles of the disc wind has been observed in many CVs, as well as variations of these line profiles on timescales of minutes (e.g., \citealt{Ringwald1998}). Indeed, the disc wind has also been proposed as the responsible for the double peak filling in high-luminosity non-magnetic CVs (e.g., \citealt{Murray1996}). While theoretical modelling of CV disc winds has been usually focused on the UV regime, recent works based on Monte Carlo simulations have extended this analysis to optical wavelengths, supporting the presence of wind-related single-peaked profiles in the optical emission lines \citep{Matthews2015}. 

These results might be consistent with an outflow contribution filling in the double peak in our V404 Cyg spectra. This would mean that from day 13, when the system is fainter than $r' \gtrsim 12.4$, the excitation of the surrounding nebula is probably not able to emit enough photons (via recombination) to fill in the double peak. The fact that we do observe single peaks on day 12 at a lower brightness is probably due to the extreme irradiation suffered by the nebula in previous days (the outburst peak occurs on day 9). Indeed, the most extreme example of the nebular contribution to the emission lines can be found by comparing the averaged spectra of day 11 (main nebular phase) and 15 (double peak profiles). Despite sharing analogous continuum fluxes (see Fig. \ref{fig: spectra_day11} and \ref{fig: spectra_day15}) the strength of the $\rm H\alpha$ emission line in day 11 is able to increase $r'$ by 1 magnitude.

In this scenario, the fact that our pre-outburst spectra (day -1.8, obtained 13 hours before the X-ray trigger) exhibited single peaks at similar brightness to day 12 ($r'\sim 13.2\pm 0.2$) implies that we probably missed a flare before these spectra were obtained, in order for the nebular contribution to be already present. This is consistent with the measured $\rm H\alpha$ EW= $250-305${~\AA} (see Fig. \ref{fig: rvsHalpha_flux}, black triangles), as well as with blue and red EW excesses that place the spectra in either the nebular (the earlier) or the P-Cyg (the later) region of the outflow diagram (Fig. \ref{fig: Halpharedvsblue}).

\subsection{The properties of the wind in June 2015, December 2015 and 1989 outbursts}
\label{windprop}

Narrow, blue-shifted absorption lines produced by highly ionised material (e.g., Fe {\sc xxv} and Fe {\sc xxvi}) have been observed in the X-ray spectra of some active LMXBs (e.g., GRS 1915+105, \citealt{Neilsen2009}). This hot wind likely has an equatorial geometry (given that it is only seen in edge-on systems, \citealt{Ponti2012}) and it is preferentially observed during the soft state when the radio jet is not present (see \citealt{Fender2016}, \citealt{Ponti2016}; see also \citealt{Homan2016}, \citealt{Bianchi2017}). During relatively high flux phases of the V404 Cyg 2015J outburst, a highly ionised wind was detected in the X-ray spectra taken by {\it Chandra} \citep{King2015} with terminal velocities between $\rm 1000\, km\,s^{-1}$ (Mg {\sc xii}) and $\rm 4000\, km\,s^{-1}$ (Fe {\sc xxvi}).

Thanks to the multiwavelength campaign on V404 Cyg during its 2015J outburst (MD16) we confirmed, for the first time, the detection of P-Cyg profiles in the optical simultaneous with radio jet emission. In the following, we will analyse the main properties of this cold wind, which might help us to understand the launching mechanism, the constraints that allow its detection in the optical spectra of LMXBs and discuss the possibility of a common origin with the hot, X-ray wind.

\subsubsection{June 2015}

The spectra show outflow-related signatures throughout the outburst, either as nebular phases or P-Cyg profiles. Observations of some nova events of CVs have already exhibited both similarly high BD values as well as critical changes in the emission lines, includying broadened and skewed profiles (see e.g., \citealt{Iijima2003}). We note that their spectra also exhibit emission lines from other atomic species such as Fe {\sc ii} and Si {\sc ii}, which are present in our nebular spectra as well. On the other hand, \citet{Southwell1996} discovered a bipolar outflow in the recurrent supersoft X-ray transient RX J0513.9-6951 in the form of blue- and red-shifted ``jet" components of He {\sc ii}--$\lambda$4686 and H$\beta$ at velocities comparable to the WD escape velocity ($\rm \sim 3800\, km\, s^{-1}$). Observations of other supersoft X-ray sources have revealed similarly high velocity P-Cyg profiles and emission line wings associated with such outflows, reaching terminal velocities ($\rm \lesssim 4000\, km\, s^{-1}$; e.g., \citealt{Crampton1996}) similar to those found in V404 Cyg ($\rm \lesssim 3000\, km\, s^{-1}$, MD16). Hot post-AGB candidates (AGB stars about to form a planetary nebula) also exhibit periodic mass ejections with similar P-Cyg and broad emission lines. However, these events are less dramatic than those occurring in nova events and, indeed, they exhibit lower terminal velocities ($\rm \sim 100 \, km\, s^{-1}$; e.g., \citealt{Smith1994}). More violent ejection events are observed in massive stars surrounded by low excitation nebulosities (e.g., the canonical luminous blue variable AG Carinae, \citealt{Thackeray1977}, or the P-Cygni star itself). They have generally lower expansion velocities than those of V404 Cyg (e.g., P-Cyg $\rm \sim 200 \, km\, s^{-1}$, \citealt{Smith2006}), but there have been reports of outflows as fast as $\rm 2000\, km\, s^{-1}$ (see \citealt{Thone2017}). All these similarities led MD16 to describe this phase of the V404 Cyg outburst as the nebular phase.

\begin{figure*}
\begin{center}
\includegraphics[keepaspectratio, width=\textwidth]{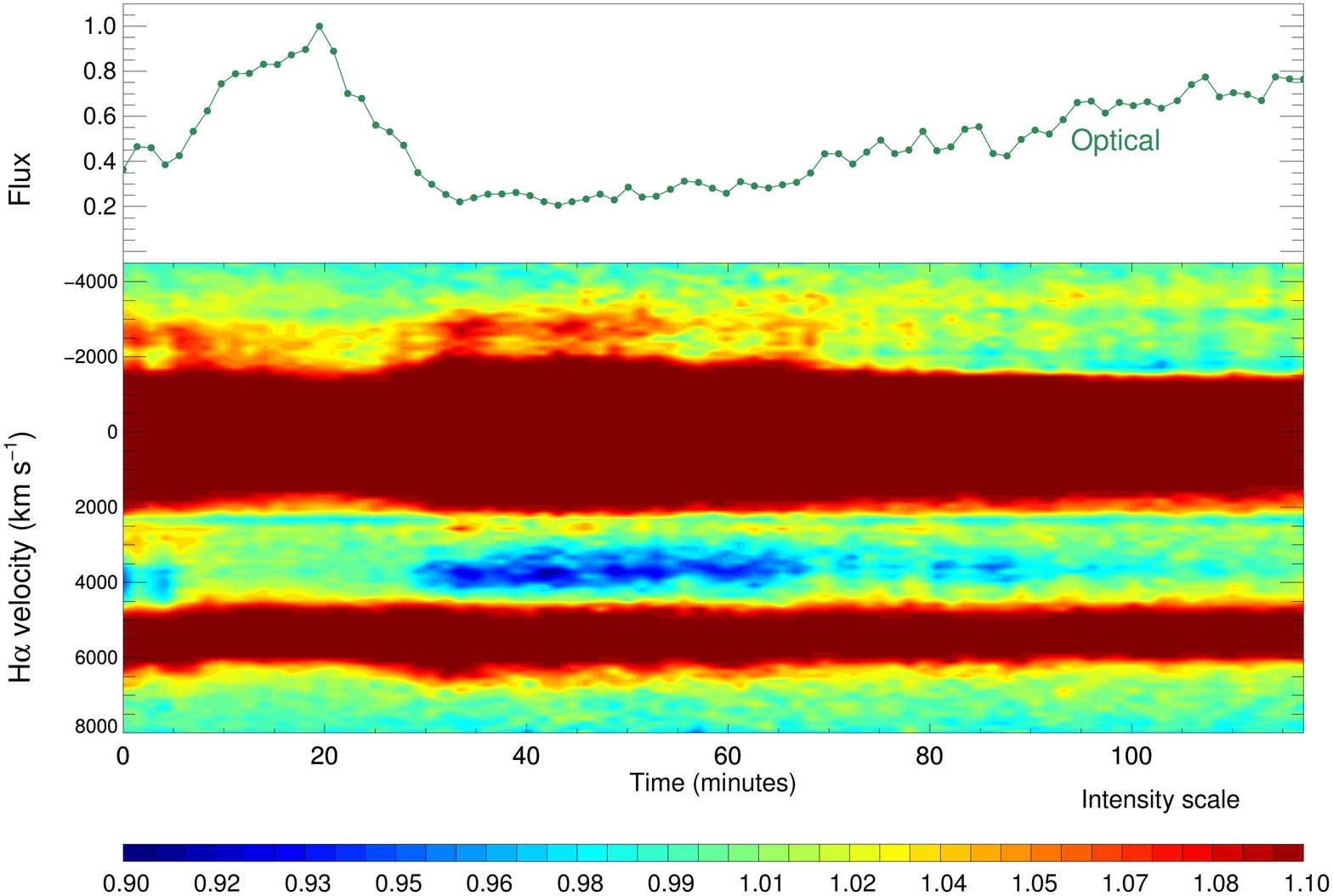}
\end{center}
    \caption{Trailed spectrum corresponding to data from day 6 of the 2015J outburst, centered on $\rm H\alpha$ and including \ion{He}{\sc i}--$\lambda$6678. The trail (bottom panel) covers 119 min with 85 spectra. Time corresponds to minutes from MJD 57195.96. The normalised intensity scale is such that absorptions are represented in blue, while emissions are plotted in red. The simultaneous optical ($r'$) light curve is shown normalised as green dots in the top panel. The presence of simultaneous optically thin and optically thick features (broad wings in $\rm H\alpha$, P-Cyg in \ion{He}{\sc i}--$\lambda$6678) is evident from min 30 to 70.}
    \label{fig: Trail-6}
\end{figure*}

Since the onset of the outburst (day -1.8, see Sec. \ref{Disc_spec_evol}) broad wings were present in $\rm H\alpha$. On the other hand, P-Cyg profiles in up to a dozen lines are found at almost every phase of the outburst, exhibiting different depths and shapes. These two signatures are, indeed, detected simultaneously in the spectra, as shown in the trailed spectrum of day 6 (see Fig. \ref{fig: Trail-6}). The flaring event covered by this day's observations shows a strong P-Cyg profile in \ion{He}{\sc i}--$\lambda$6678 as the system becomes fainter. Simultaneously, $\rm H\alpha$ wings become broader as the nebular contribution increases. On the other hand, at the flare peaks both features are much shallower (sometimes not even detected). All this argues in favour of a common origin for both signatures, that is, the expelled material. There are several possible scenarios for this. For instance, the outermost, optically thin layers of the ejected nebula would produce the characteristic excess in the $\rm H\alpha$ emission, while \ion{He}{\sc i}--$\lambda$6678 P-Cyg profiles would be produced in inner, denser and optically thick layers. Alternatively and depending on the plasma temperature, the different behaviour could be explained by He being highly recombined and optically thick (producing P-Cyg absorptions), while H remains mostly ionised and transparent.

\begin{figure*}
\begin{center}
\includegraphics[keepaspectratio, width=\textwidth]{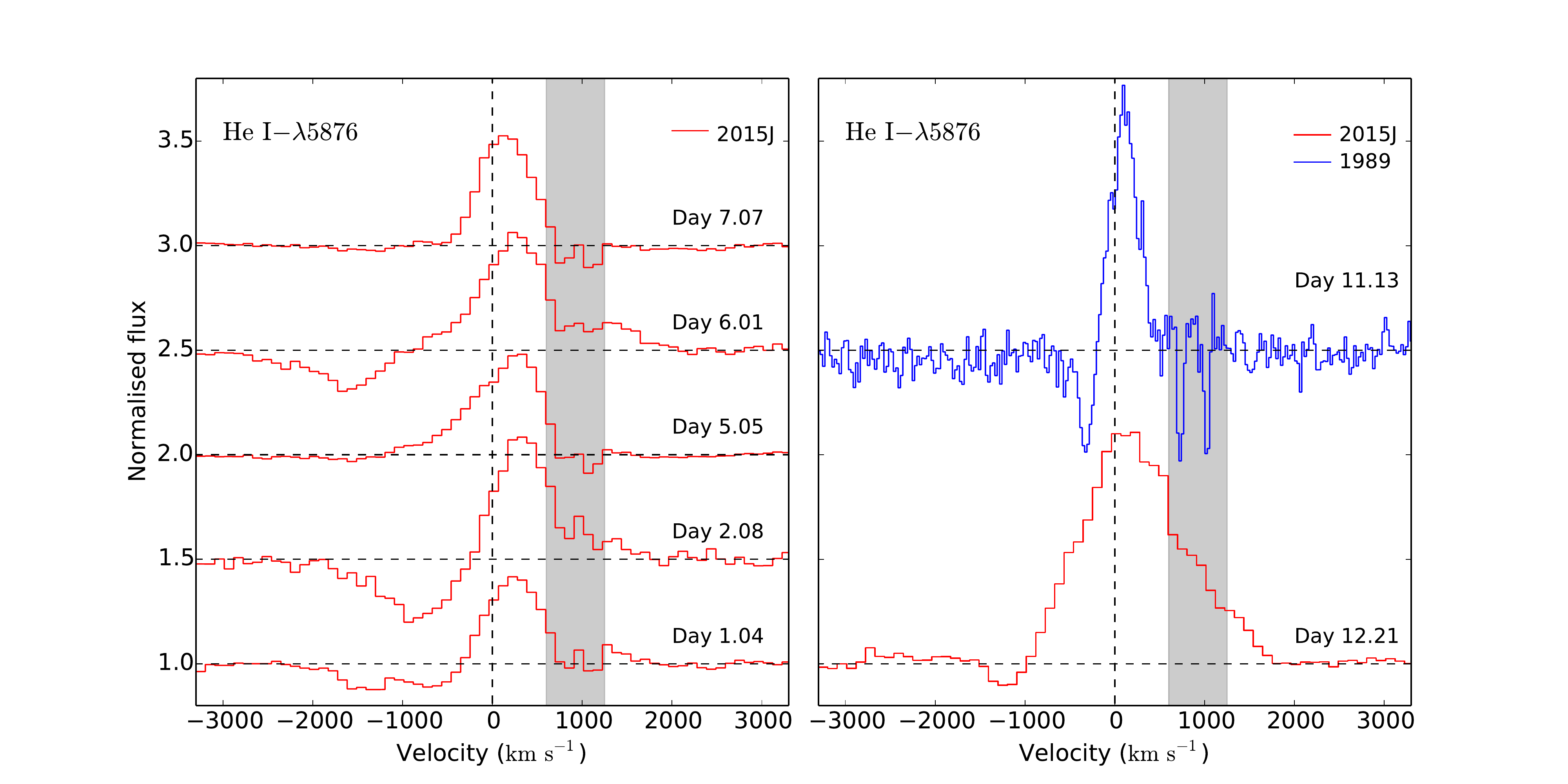}
\end{center}
    \caption{\ion{He}{\sc i}--$\lambda$5876 line profile in different days of the 2015J outburst, as well as one spectrum of 1989 outburst for comparison. Grey shading indicates regions contaminated by the interstellar Na {\sc i}--5890 doublet absorption.}
    \label{fig: outflows1}
\end{figure*}

Fig. \ref{fig: outflows1} shows \ion{He}{\sc i}--$\lambda$5876 through the first 7 days of the 2015J outburst (left panel), as well as in day 12 (right panel). The visual comparison between profiles shows that the blue absorption defining the P-Cyg has terminal velocities ($v_{\rm T}$) ranging from $\sim 1500\,  \rm km\, s^{-1}$ (day 12) to $\sim 3000\,  \rm km\, s^{-1}$ (day 6). We also note that the shape of this blue absorption is not always the same; for example, days 1 and 2 have similar absorption depths and terminal velocities, but the first shows a double absorption while the second exhibits only a single component. In day 5 we find skewed profiles that, in day 6, are complemented with a high terminal velocity absorption. A variable, optically thick contribution to the emission line may be responsible for the different outflowing profiles. Certainly, when the system reaches higher luminosities the P-Cyg profiles become shallower (day 7). This is interpreted as a result of the wind ionisation.

\begin{figure*}
\begin{center}
\includegraphics[keepaspectratio, width=\textwidth]{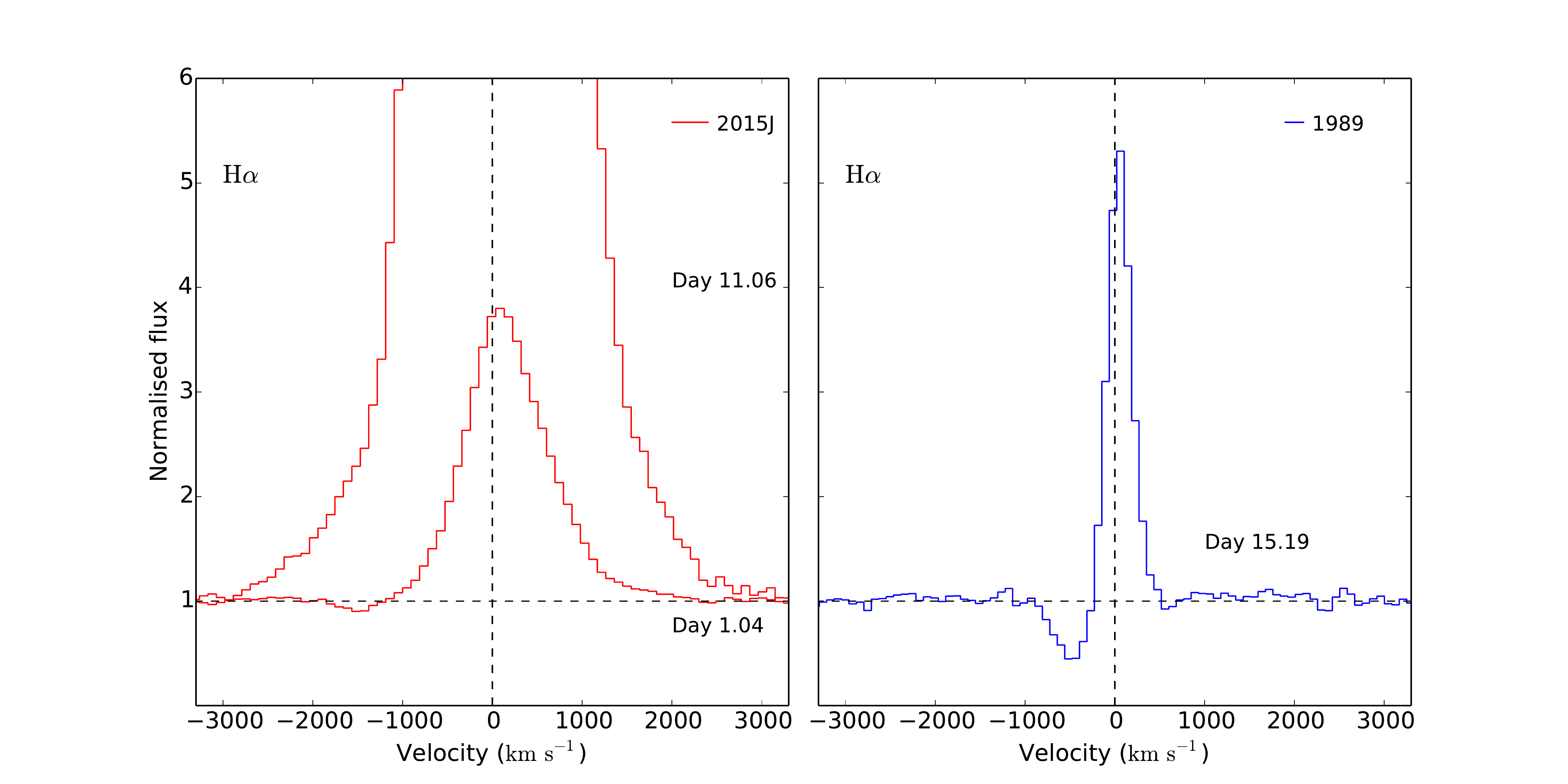}
\end{center}
    \caption{Same as Fig. \ref{fig: outflows1} for the $\rm H\alpha$ line.}
    \label{fig: outflows2}
\end{figure*}

In the left panel of Fig. \ref{fig: outflows2} we show $\rm H\alpha$ profiles on two different days. During day 1, a P-Cyg absorption is present in $\rm H\alpha$ with terminal velocities similar to those observed the same day in \ion{He}{\sc i}--$\lambda$5876 ($v_{\rm T}\sim 2000\,  \rm km\, s^{-1}$). On the other hand, the day 11 profile shows the prototypical nebular spectrum where $\rm H\alpha$ sits in broad wings reaching $v_{\rm T}\sim 3000\,  \rm km\, s^{-1}$, similar to the terminal velocities observed in \ion{He}{\sc i}--$\lambda$5876 during day 6. The comparison with \ion{He}{\sc i}--$\lambda$5876 on day 11 is not that straightforward because the P-Cyg is much shallower. In general, the P-Cyg blue absorptions observed in $\rm H\alpha$ are typically shallower than those observed in \ion{He}{\sc i}--$\lambda$5876, but on the other hand, the optically thin contribution (nebula) is more prominent in the former line.

The highest velocities of the wind ($v_{\rm T} \sim 3000\, {\rm km\, s^{-1}}$) are observed after a bright flare (day 6, $r' \sim 8.1-9.7$), while the lowest velocities ($v_{\rm T} \sim 1500\, {\rm km\, s^{-1}}$) are found at much lower brightness (day 12, $r' \sim 12.8$). However, during the peak of the outburst (day 9, $r' \sim  7.7$) shallow P-Cyg profiles are observed at $v_{\rm T} \sim 2000\, {\rm km\, s^{-1}}$, similar to those observed at day 2 ($r' \sim 9.0-11.3$). Therefore, we conclude that there is no obvious connection between the system brightness and the wind terminal velocity.



\subsubsection{December 2015}

The 2015D event is significantly fainter ($V_{\rm peak}\sim 13$, \citealt{MunozDarias2017}) than 2015J ($V_{\rm peak}\sim 11$, \citealt{Kimura2016}) but both share similar phenomenology. The duration of both events is similar, exhibiting $\sim 2-3$ weeks of strong variability. Compared with other LMXBs and taking into account the longer orbital period of V404 Cyg (i. e. large size of the accretion disc) these timescales are extremely short. This, combined with the fact that both events are separated by only 6 months led MD16 to propose that during the 2015J accretion event only the innermost part of the disc was accreted. 

2015D can be seen as a fainter and less extreme version of the 2015J event. The $\rm H\alpha$ line EW is found to vary between $60-600${~\AA}, while 2015J covered a wider range of $30-2000${~\AA}. The FWHM varies in the range $\rm 600-1200 \, km \, s^{-1}$ during 2015D compared with $\rm 700-1600 \, km \, s^{-1}$ for 2015J. The spectroscopic coverage during 2015D is less intensive than during 2015J, but there are spectra exhibiting outflow features such as P-Cyg (day 10, the peak of the outburst) and a nebular phase (day 8). The fact that the nebular phase is observed before the outburst peak is not troubling; given the limited observing window (due to the small  angular distance to the Sun), some phenomenology (e.g., flares) might have been missed.

A remarkable feature of 2015D is that through the event, the brightness of the system frequently drops to values comparable with the quiescent level ($V \sim 18.5$). The best example of this is day 8, the very day that the strongest nebular phase is observed. During this day, the system brightness reaches $V\sim 18$ (see \citealt{MunozDarias2017}) but the emission lines are broad, with $\rm H\alpha$ reaching $\rm EW \sim 600${~\AA}. On the other hand, day 10 shows one of the few clear detections of a P-Cyg profile during this event, with terminal velocities of $\rm \sim 2500  \, km \, s^{-1}$, similar to those observed in 2015J.

\subsubsection{1989}

The comparison with the 1989 event is not straightforward. The 1989 spectra were obtained during a bright phase of the outburst and there is no spectroscopic coverage during the initial rise (see \citealt{Kimura2016}). The photometric coverage is scarce and not simultaneous with the spectroscopic data. Indeed, the first detection of the 1989 outburst was reported on 1989 May 21 \citep{Makino1989}, while our spectra were obtained on days 11--21 (taking May 21st 00:00 UT as the reference time; see Tab. \ref{tab:1989}). 

The brightness of the system during 1989 (see e.g., \citealt{Wagner1991}) peaks at $\rm V\sim 11.5$ during day 9, (\citealt{Casares1991}; corresponding to a de-reddened $\rm r'\sim 7$) then fades down to $\rm V\sim 14.5$ ($\rm r'\sim 10$) by day $12$. However, instead of a full return to quiescence, the system re-brightens to $\rm V\sim 13.5$ ($\rm r'\sim 9$) by day 16, and after at least $\sim 2$ more weeks of strong variability, slowly fades to quiescence in $\sim 400$ days (see \citealt{Han1992}). This might be comparable with the 2015J outburst, which peaks at $\rm r'\sim 7.5$ on day 9, drops in brightness to $\rm r\sim 13$ by day $12$, and returns to quiescence between $40-120$ days after the initial X-ray detection ($\sim 2-3$ weeks of strong variability). The 2015J event exhibits a much sharper decay than the 1989 event and, while it lacks of a rebrightnening phase, instead shows a sequel outburst only 6 months later (2015D).

The spectroscopic data obtained during 1989 event reveals a different phenomenology than that observed in 2015J and 2015D. The spectra exhibit narrower profiles that populate a different region in the $\rm H\alpha$ FWHM-EW diagram ($\rm FWHM<700 \, km \, s^{-1}$, $\rm EW=30-200${~\AA}) than the spectra obtained in both 2015 events (see Fig. \ref{fig: Halpha_EWvsFWHM_ALL}). This might point to a systematically smaller nebular contribution to the emission lines during 1989.

On the other hand, the 1989 spectrum depicted in the right panel of Fig. \ref{fig: outflows1} (1989 day 11) reveals a blue absorption in \ion{He}{\sc i}--$\lambda$5876 with terminal velocity $v_{\rm T}\sim 700\,  \rm km\, s^{-1}$, lower than any observed in either the 2015J (day 12, depicted in same panel) or the 2015D events. Furthermore, the deepest P-Cyg absorption in $\rm H\alpha$ ever detected is found in day 15 of the 1989 outburst and also has a low terminal velocity ($\sim 900\,  \rm km\, s^{-1}$). We note that these deep, low velocity P-Cyg profiles are found during days with system brightness as high as $\rm V \sim 13$, comparable to that of the 2015D peak or even to that of 2015J day 6 (the day with the highest velocity P-Cyg). These results favour the interpretation of the P-Cyg terminal velocity not being directly correlated with the system brightness. Nevertheless, we must note that the non-simultaneity of the photometric to spectroscopic data, combined with the extreme variability already observed in the 2015J and the 2015D events (drops of $\rm \Delta V \sim 3$ in $\rm \sim 30 \, min$), could place the spectra of 1989 at much lower brightness. Likewise, the observed luminosity might be, during some epochs, different from the intrinsic (see \citealt{Motta2017a,Motta2017b}).

\begin{figure}
\includegraphics[width=0.5\textwidth]{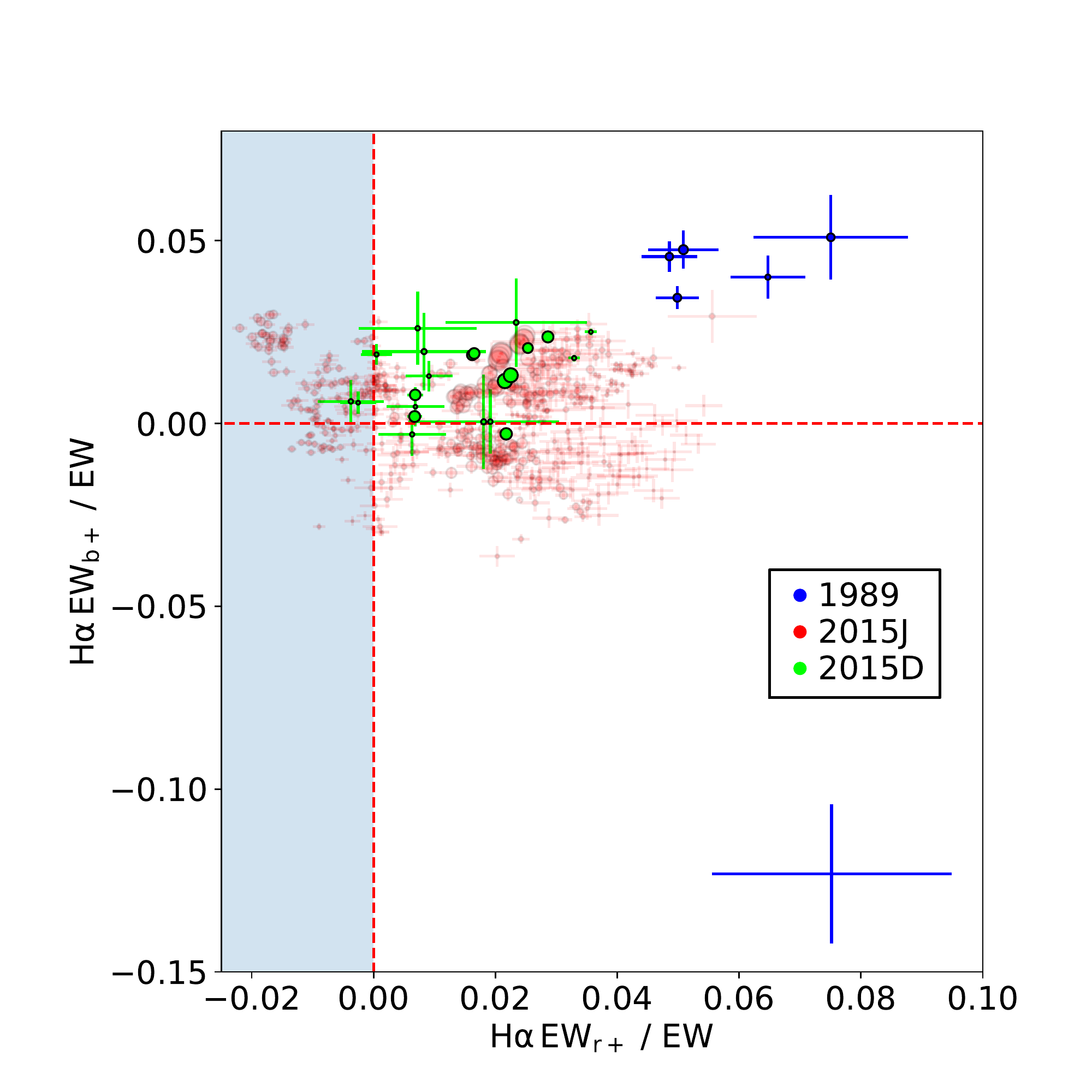}
    \caption{Ratio of the EW excess (red vs blue halves) of $\rm H\alpha$ to its EW. Those spectra where $\rm H\alpha$ saturated the detector have been omitted. The colour code is the same as that in Fig. \ref{fig: Halpha_redblue_ALL}. The size of each dot is proportional to the $\rm H\alpha$ EW. Blue-shaded areas indicate the forbidden regions of the diagram.}
    \label{fig:ouflows_percentaje}
\end{figure}

We finally note that both blue and red EW excesses of $\rm H\alpha$ populate similar regions of the diagram during the three outbursts. However, if we plot the ratio of the EW excesses normalised by the total EW of the line (see Fig. \ref{fig:ouflows_percentaje}), we find that 2015J and 2015D spectra have blue and red excesses always below $6\%$ of the total EW, while the 1989 spectra show higher ratios ($4-12\%$ for the blue absorption of the P-Cyg). Even if these measurements could be affected by the narrower mask employed to analyse the 1989 spectra, the combination of low EW from $\rm H\alpha$ ($\rm EW=30-200${~\AA}) and the narrow but deep absorptions (see Fig. \ref{fig: outflows2}) suggest these results remain secure. We note that the higher ratios found in 1989 do not necessarily imply larger outflows. The lower $\rm H\alpha$ EW and FWHM suggest a small nebular contribution, while the lower terminal velocities might suggest that the wind originates in outer regions of the disc compared to those of 2015J.

\subsubsection{The origin of the wind}

Winds signatures have been observed in accreting compact objects across a wide range of masses, including CVs, LMXBs and Active Galactic Nuclei (AGN). Depending on the compact object mass and the mass accretion rate, different wind launching mechanisms can be at play. The main candidates to launch winds in LMXBs are: 

\begin{itemize}
\item[i)] Thermal: they are driven via Compton heating. The launching region roughly corresponds to the accretion disc radius where the thermal velocity of the plasma matches the escape velocity (e.g., \citealt{Begelman1983a,Begelman1983b}).
\item[ii)] Magnetic: field lines might leave the disc surface and extend to large distances. This might produce a wind for certain magnetic field configuration. (e.g., \citealt{Blandford1982}). 
\item[iii)] Radiation pressure: when the system approaches the Eddington luminosity the radiation pressure is able to not only prevent the accretion of material but also to expel it, thereby launching a wind. However, we note that many of our wind observations were recorded at very low luminosities (e.g., the 2015D outburst) when this mechanism should not be ocurring. The best example might be day -1.8 spectra (obtained half day before the X-ray trigger) were clear outflows signatures are found (see Tab. \ref{tab:summatory}).
\item[iv)] Line-driven: in CVs, certain resonance lines in the UV range are able to drive a wind that is observed as P-Cyg features (e.g., \citealt{Proga2002a}). This mechanism is not expected to be efficient in LMXBs due to over-ionisation of atoms in the presence of a much higher hard X-ray irradiation (see e.g., \citealt{Proga2002b}).

\end{itemize}

The main launching mechanism of hot, X-ray winds in LMXBs is thought to be either thermal (e.g., \citealt{Begelman1983a,Begelman1983b}; \citealt{Ponti2012}) or magnetic (e.g., \citealt{Blandford1982}; \citealt{Miller2006}). 
While a magnetic origin for the wind cannot be ruled out, the X-ray wind terminal velocities found in the V404 Cyg 2015J outburst ($\rm 1000-4000\, km\, s^{-1}$, \citealt{King2015}) are consistent with thermal winds at a launching radius of $R_l=0.8-13.3\cdot 10^5 \,{\rm  km}$ (assuming a $M_{\rm BH} = 10\, M_{\odot}$). \citet{King2015} found that the disc radius for the formation of Si {\sc xiii} emission lines (which they assume are formed in the wind) was $40-300\cdot 10^{5} \, {\rm  km}$. 

The similar terminal velocities observed in the optical wind during both 2015J and 2015D ($\rm 1500-3000\, km\, s^{-1}$), lead us to consider a thermal origin for this wind. Such velocities are consistent with a launching radius of $R_l=1.5-6.0\cdot 10^5 \, {\rm  km}$. On the other hand, the lowest terminal velocity observed in the 1989 event ($\rm 700\, km\, s^{-1}$) would point to a radius as far as $R_l=27\cdot 10^5 \, {\rm  km}$. 

We note that, by applying a standard accretion disc model with the assumptions introduced in MD16 (which does not take into account any irradiation effects), we obtained surface temperatures of $T_{\rm D}\sim 5000-20000\, \rm K$ for the derived optical wind launching radius. On the other hand, the surface temperature at the X-ray launching radius would be $T_{\rm D}\sim 2500-35000\, \rm K$, and $T_{\rm D}< 1500\, \rm K$ for the Si {\sc xiii} (too low for this line formation). This emphasises the importance of taking into account the disc irradiation and photo-ionisation effects. At this point it is interesting to consider the scenario presented in \citet{Motta2017b}. They propose a particular geometry during the outburst, with an inner (inflated) slim disc that shields a large fraction of the X-ray irradiation from the central source. In this case, the X-ray irradiation of the outer disc might be much weaker. It is also interesting to note that, given that the X-ray irradiation of the outer disc is lower, an optical line-driven wind similar to that observed in CVs could be launched. However, this would be different from that observed in X-rays. 

\subsubsection{On the impact of the wind in the outburst evolution}

Accretion disc winds are thought to have an important impact on the accretion process. On theoretical grounds, \citet{Luketic2010} presented simulations (for thermally driven winds) supporting mass outflow rates several times larger than the accretion rate down to luminosities of a few per cent the Eddington limit. This seems to be confirmed by several observations from which mass outflow rates comfortably exceeding the contemporaneous accretion rate have been inferred (\citealt*{Neilsen2011}; \citealt{Ponti2012}). Translating these observables into the full outburst mass/energy balance,  \citet*{Fender2016} found, under relatively simple assumptions, that $>50$~\% of the total mass \textit{consumed} during the outburst could go into the wind. This mass balance considers that the hot X-ray wind is only present during soft states. 
In V404 Cyg we have found that the above picture could be even more extreme. In the first place, the cold optical wind is simultaneous to the radio jet (MD16), which indicates that it is present in the hard and/or the hard-intermediate states. Secondly, there is strong (albeit indirect) evidence suggesting that the mass carried by the wind is (at least) comparable to both the accreted mass (i.e. transformed into luminosity) and that transferred from the donor during the 26 yr of quiescence. This, together with the unusual outburst evolution, with just two weeks of strong activity abruptly finished by a sharp luminosity drop, led to MD16 to put forward a wind instability scenario. We proposed a toy model in which the outflow depletes the outer accretion disc, and thus only the inner regions are the ones eventually accreted. In this context, the 1989 outburst share many of the phenomenology observed in 2015J. However, the wind has a significantly lower terminal velocity (see Sec. \ref{windprop}), which would imply that the disc depletion occurred at larger radii. This is consistent with the longer outburst duration, as compared with 2015J.  This wind instability scenario has been proposed to explain the outbursts of the ultraluminous X-ray binary HLX-1 \citep{Soria2017}.

As noted in MD16, extreme wind phenomenology seem to be a common property of systems with long orbital periods and therefore large accretion discs. This is the case of the three BH transients with the longest periods (>2.5 d):  GRS~1915+105 \citep*{Neilsen2009}, V404 Cyg (MD16) and GRO~J1655-40 (e.g. \citealt{Miller2008}). In addition, the fourth in this list would be V4641 Sgr, for which \citet*{MunozDarias2018} have recently found conclusive evidence for optical winds with very similar properties to that of V404 Cyg (see also \citealt{Chaty2003}; \citealt{Lindstrom2005}).  
Finally, we note that an alike accretion/outflow coupling, including winds, is likely present in neutron star systems (e.g. \citealt*{Ponti2014}; \citealt{MunozDarias2014}; \citealt{Homan2016}; \citealt*{DiazTrigo2016}).

\subsection{The properties of the plasma}
\label{BP}

The broad spectral range of our spectra allow us to analyse the evolution of the Balmer lines. The analysis of their relative intensity has been previously used in AGN studies to determine the physical conditions of the emitting plasma (e.g., \citealt{Ilic2012}). In particular, we will use the so-called Boltzmann plot. 

Assuming that i) the emitting plasma is in local thermodynamic equilibrium (LTE), ii) the lines are optically thin, and iii) the Balmer series originates in the same region, one can derive the following equation:

$$\log{(F_{\rm n})}=B-A\,  E_{\rm u};\qquad F_{\rm n}=\frac{F_{\rm ul}\lambda}{g_{\rm u}A_{\rm ul}}$$

$F_{\rm ul}$ is the line flux of a transition from the upper ($\rm u$) to the lower ($\rm l$) level, $\lambda$ is the transition wavelength, $A_{\rm ul}$ is the spontaneous transition probability and $g_{\rm u}$ is the statistical weight of the upper level. $F_{\rm n}$ is the line flux normalized by atomic constants, $E_{\rm u}$ is the energy of the upper level, and both A and B are constant parameters. In particular, A is called the temperature parameter and is defined in terms of the average excitation temperature:

$$A = \frac{\log{(e)}}{kT_{\rm exc}} \approx \frac{5040}{T_{\rm exc}}$$
 
Therefore, we can obtain the excitation temperature of the emitting plasma through a linear fitting in the Boltzmann plot, that is, $\log{(F_{\rm n})}$ versus $ E_{\rm u}$. 

We applied this method to every flux-calibrated spectrum of the 2015J outburst where transitions from $\rm H\alpha$ to $\rm H\epsilon$ were present. The resulting diagrams revealed that the vast majority of the data follow positive trends  (i.e., $A<0$; see Fig. \ref{fig: BP} left panel). These would correspond to negative excitation temperatures and therefore, they are not allowed. This indicates that the plasma does not fulfil the aforementioned conditions. Plausible explanations include different emission regions (e.g., accretion disc and ejecta) or a strong deviation from LTE. 

Only during the main nebular phase that follows the sharp drop in luminosity (and thus in the flux of the ionising continuum), observed at the end of the 2015J outburst (days 11--13), negative slopes are obtained. Albeit the fits are not good (see Fig. \ref{fig: BP}), we note that the general trend is in agreement with the presence of a hot expanding plasma where the gas excitation is regulated by collisions with free electrons released during the bright phases of the outburst. This results in $T_{\rm exc}$ in the range of $\rm 7000-12000\, K$; values fully compatible with the presence of H recombining.

\section{Conclusions}

We have presented a very detailed analysis of the optical spectroscopic campaign performed during the June 2015 outburst of V404 Cyni. A day-by-day summary of the outburst reveals the fast variability and dramatic changes taking place in the spectral lines. Among them, outflow signatures in the form of P-Cyg profiles and nebular lines are particularly remarkable. We studied the system brightness, the main properties of the emission lines, and other derived parameters, such as the Balmer decrement and ionisation ratio.

\begin{itemize}
\item We find an overall correlation between the X-ray and optical fluxes, but note that day-by-day observations do not always follow the trend.

\item We have constructed several diagnostic diagrams which allow us to identify these wind-related features and follow their evolution. We find that outflows are present even before the X-ray trigger and also during the decay phase, with P-Cyg features at fluxes as low as $r' \sim 12.8$.

\item Besides the main nebular phase event witnessed at the end of the outburst, it is worth noting the short-lived ($\rm \sim 2\, h$) nebular loops found on days 2 and 6 (each covering a flare event) suggesting the continuous presence of these nebular cycles.
\item The $\rm H\alpha$ line exhibits broad wings as the brightness of the system decays while, at the same time, P-Cyg absorptions become deeper in \ion{He}{\sc i}--$\lambda$5876. The simultaneous presence of both features, combined with their similar velocities, strongly suggests that they correspond to different phases of the same outflow.

\item Thanks to our extended spectroscopic database, we also analyse the transition of the emission lines from double-peaked to single-peaked profiles. The results favour the interpretation of a nebular contribution filling the double peak throughout the outburst. 

\item We finally investigate the properties of the plasma through Boltzmann plots of the Balmer series and derived realistic plasma temperatures during the nebular phase ($T_{\rm exc} \sim \rm 7000-12000\, K$).

\end{itemize}

We extended this analysis to the spectra obtained during the December 2015 and 1989 events. The 2015D observations reveal a similar behaviour to that of 2015J, but at a lower brightness. On the other hand, the 1989 outburst is characterised by systematically narrower emission lines and lower terminal velocities in the P-Cyg profiles. 
These results will serve as a reference for the extreme, rapid changes taking places during LMXB outbursts.

\section*{Acknowledgements}

DMS acknowledges Fundaci\'on La Caixa for the financial support received in the form of a PhD contract. TMD acknowledges support via a Ram\'{o}n y Cajal Fellowship (RYC-2015-18148) and the Spanish MINECO project AYA2017-83216-P. J.C. acknowledges financial support from the Spanish Ministry of Economy, Industry and Competitiveness (MINECO) under the 2015 Severo Ochoa Program MINECO SEV-2015-0548, and to the Leverhulme Trust through grant VP2-2015-046. PAC gratefully acknowledges the receipt of a Leverhulme Emeritus Fellowship from the Leverhulme Trust. MAP's research is funded under the Juan de la Cierva Fellowship Programme (IJCI-2016-30867) from MINECO. ML is supported by EU's Horizon 2020 programme through a Marie Sklodowska-Curie Fellowship (grant nr. 702638). MAPT acknowledges support via a Ramón y Cajal Fellowship (RYC-2015-17854). A. W. S. is supported by an NSERC Discovery Grant and a Discovery Accelerator Supplement. AJCT acknowledges support from the Spanish Ministry MINECO Project AYA 2015-71718-R (including FEDER funds).

We thank Javier M\'endez (head of the ING Service Program), Chris Benn (head of Astronomy), Marc Balcells (director of ING) and all the ING SAs for their help in monitoring this unique event during ING Service and Discretionary time. We also thank Romano Corradi (director of GTC), Antonio Cabrera (head of science operations) and the GTC support team for the quick response during the multiples triggers of our program. This research has been carried out with telescope time awarded by the CCI International Time Programme at the Canary Islands observatories (program ITP13-8). We also thank the GTC DDT committee (coordinated by the IAC director, Rafael Rebolo) for the awarded time to our observing program.
This paper is partially based on observations with \textit{INTEGRAL}, an ESA project with instruments and science data centre funded by ESA member states (especially the PI countries: Denmark, France, Germany, Italy, Switzerland, Spain), and Poland, and with the participation of Russia and the USA. MOLLY software developed by T. R. Marsh is gratefully acknowledged.




\bibliographystyle{mnras}
\bibliography{MiBiblio}



\appendix
\section{Supplementary tables and figures}

\begin{figure*}
\includegraphics[keepaspectratio, trim=2cm 0cm 2cm 0cm, clip=true, width=\textwidth]{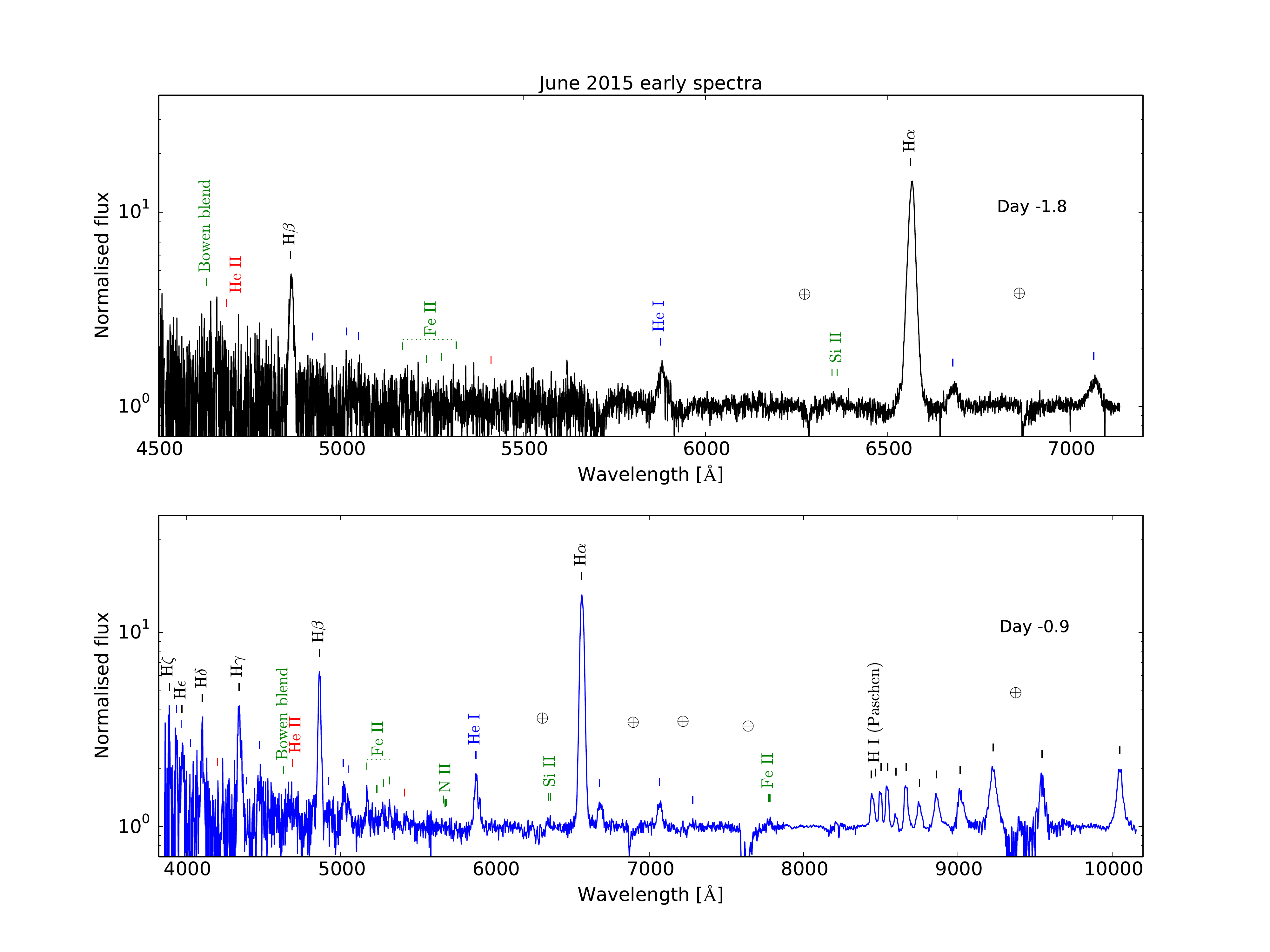} \caption{Normalised, averaged spectrum obtained during day -1.8 with WHT/ISIS, resulting from the combination of the red and blue arm. Bottom panel: Normalised, averaged spectrum obtained during day -0.9 with GTC/OSIRIS, where consecutive R1000B and R2500I spectra are shown together to cover the wider wavelength range among the database. Emission line identifications are colour encoded: black for H, blue for \ion{He}{\sc i}, red for He {\sc ii} and green for other species (Si {\sc ii}, N {\sc ii}, Fe {\sc ii} and the Bowen blend). Telluric absorption lines are marked as $\oplus$.}
    \label{fig: spectra_special}
\end{figure*}

\begin{figure*}
\center
\includegraphics[keepaspectratio, width=0.5\textwidth]{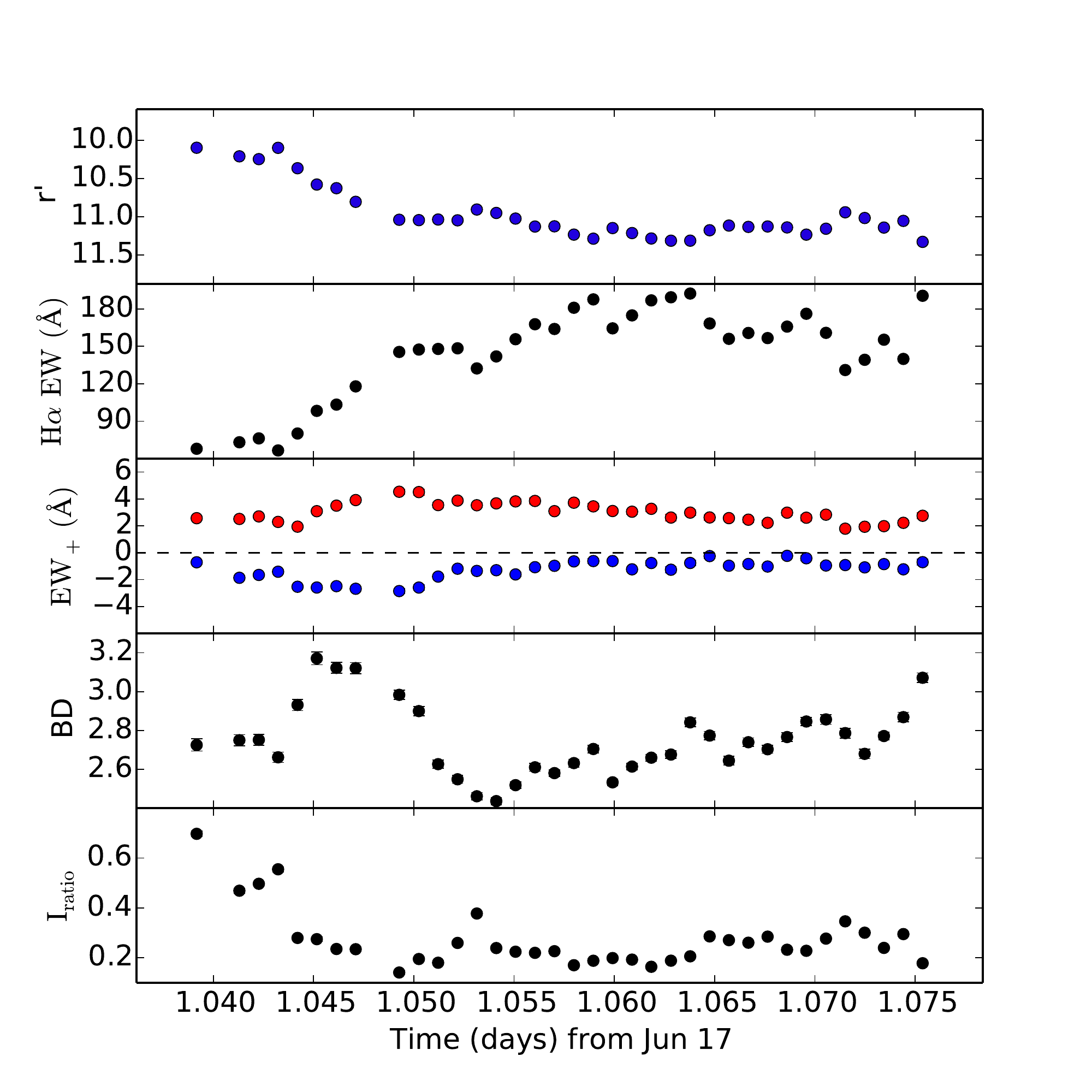}\includegraphics[keepaspectratio, width=0.5\textwidth]{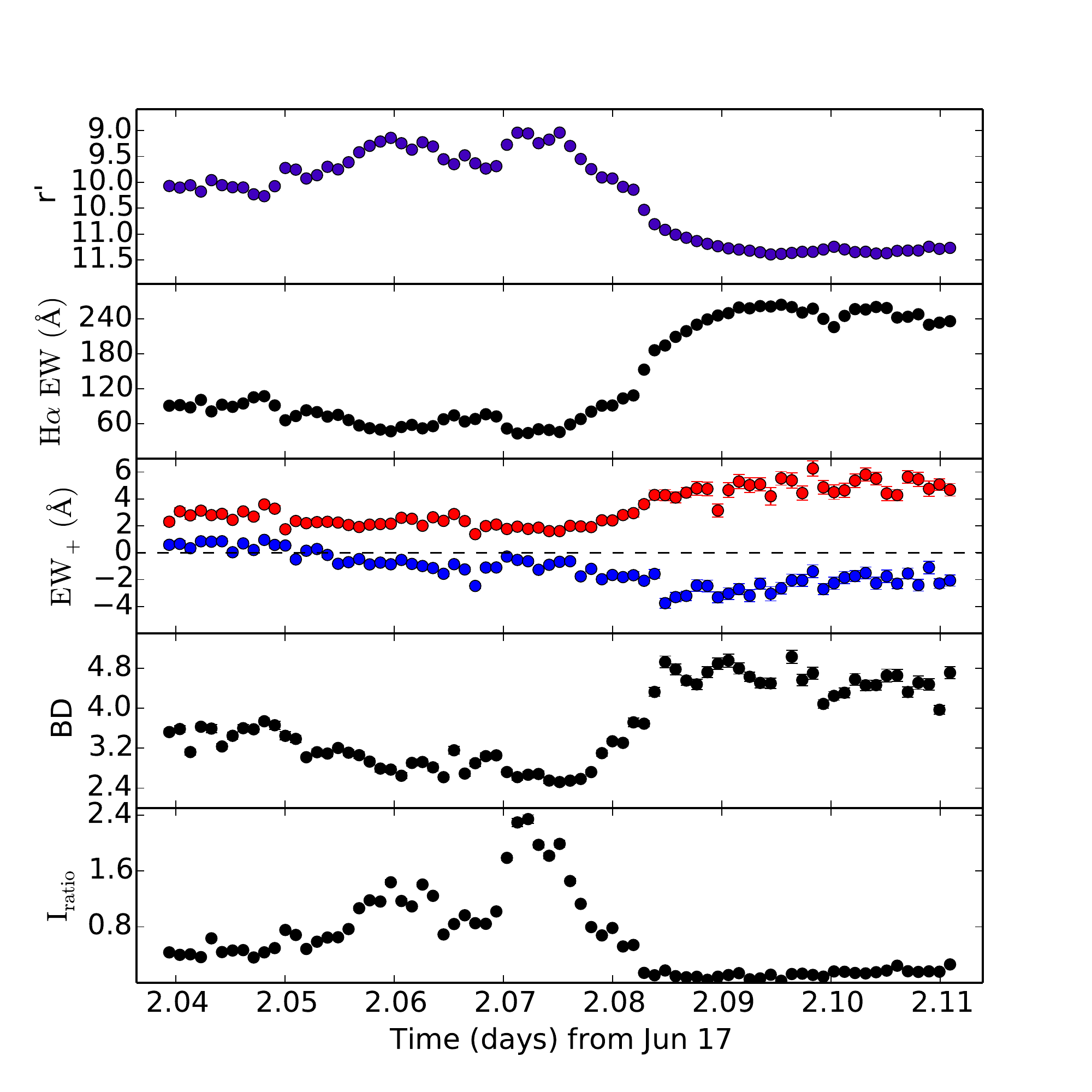}
    \caption{Temporal evolution along days 1 (left) and 2 (right) of the following parameters: $r'$, $\rm H\alpha$ $\rm EW_{+}$ (blue and red points correspond to $\rm EW_{b+}$ and $\rm EW_{r+}$ respectively), BD and $I_{\rm ratio}$.}
    \label{fig: paramd1-2}
\end{figure*}

\begin{figure*}
\center
\includegraphics[keepaspectratio, width=0.5\textwidth]{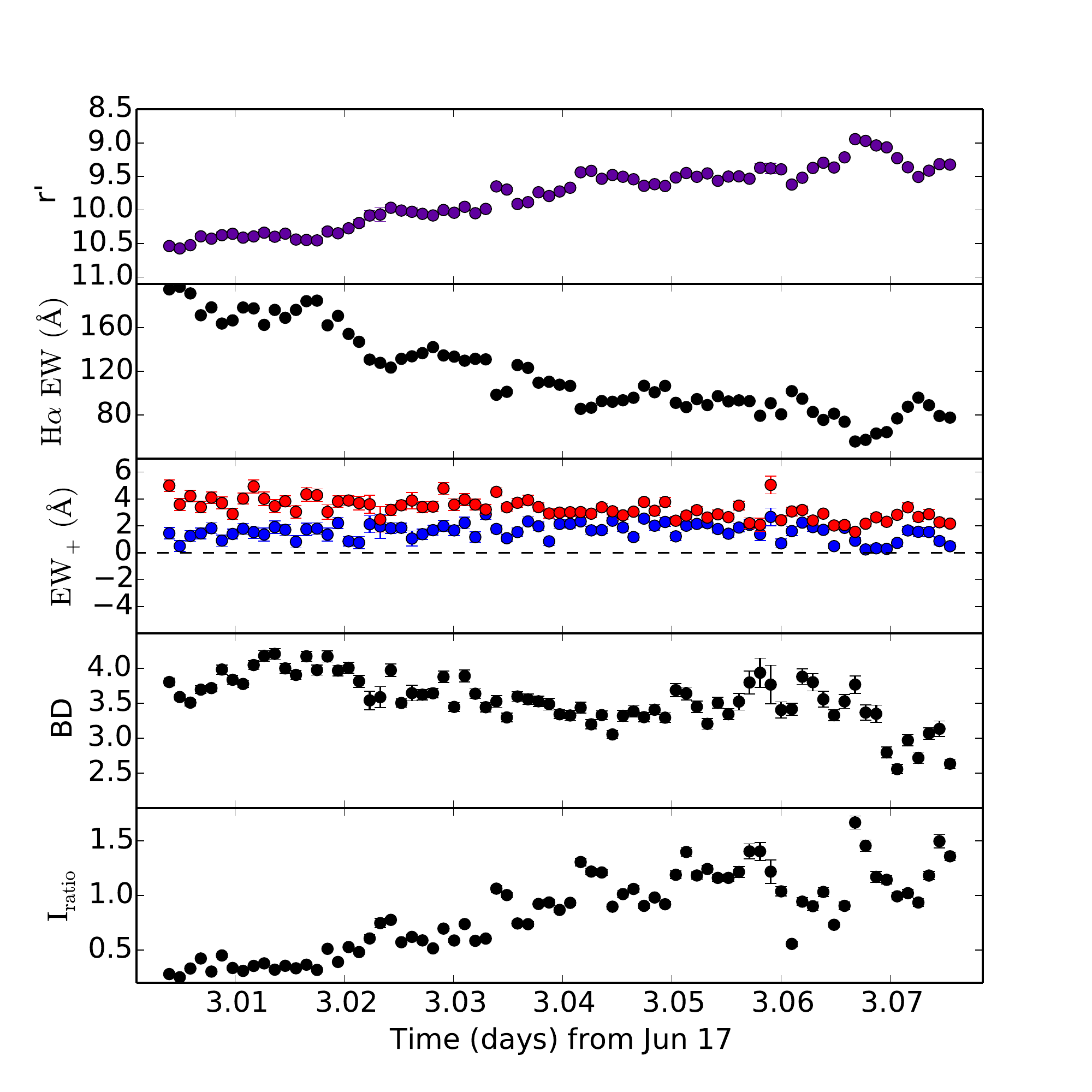}\includegraphics[keepaspectratio, width=0.5\textwidth]{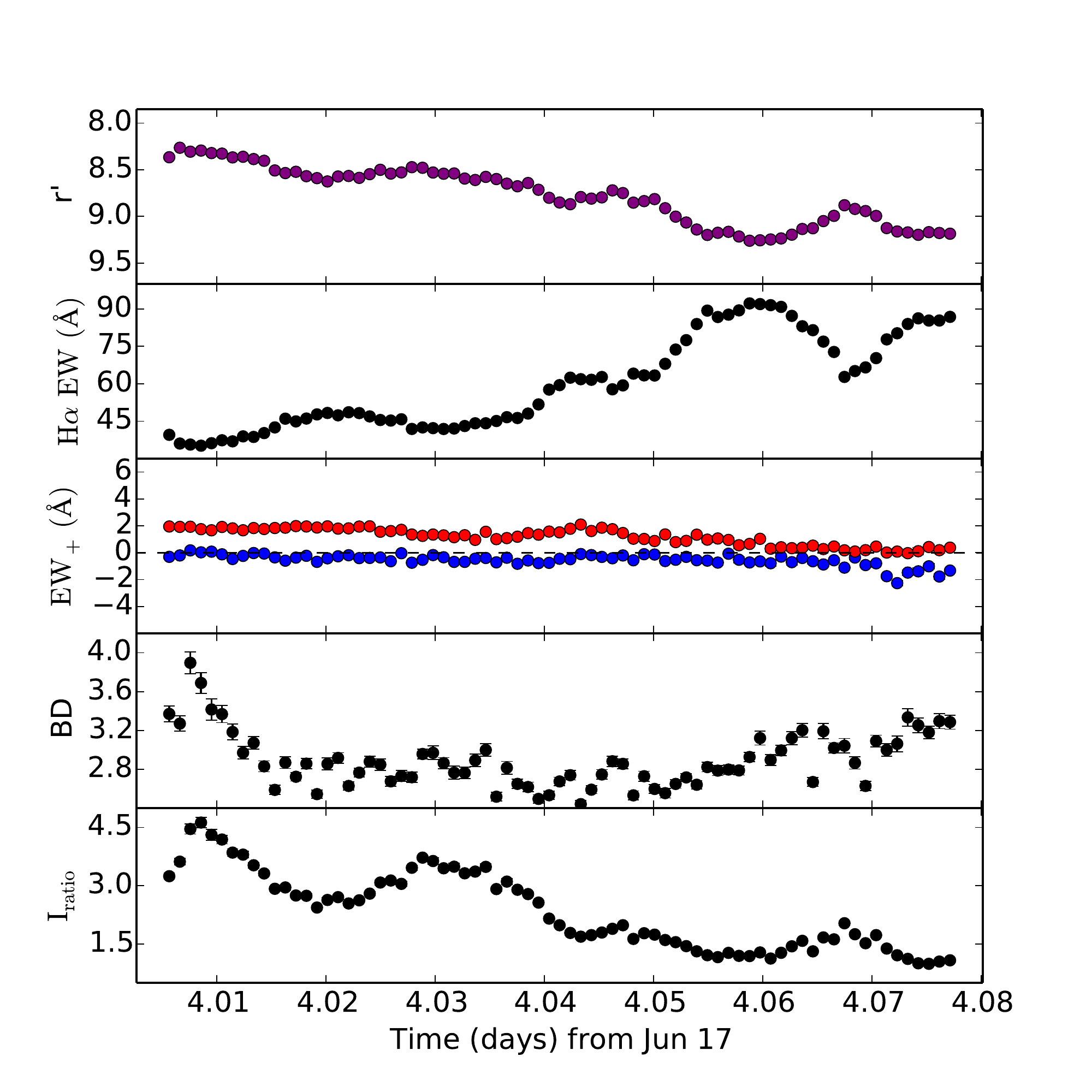}
    \caption{Temporal evolution along days 3 (left) and 4 (right) of the following parameters: $r'$, $\rm H\alpha$ $\rm EW_{+}$ (blue and red points correspond to $\rm EW_{b+}$ and $\rm EW_{r+}$ respectively), BD and $\rm I_{ratio}$.}
    \label{fig: paramd3-4}
\end{figure*}

\begin{figure*}
\center
\includegraphics[keepaspectratio, width=0.5\textwidth]{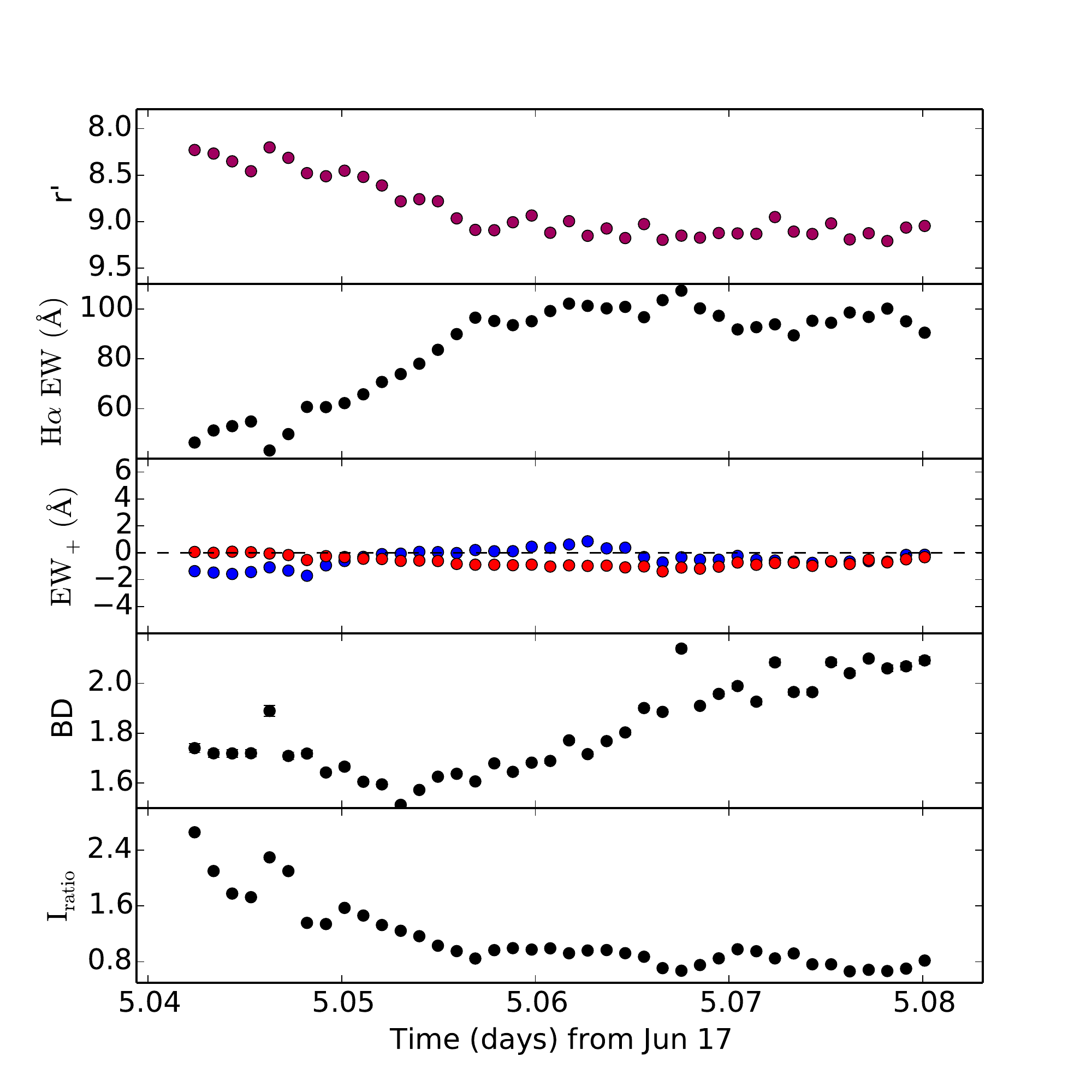}\includegraphics[keepaspectratio, width=0.5\textwidth]{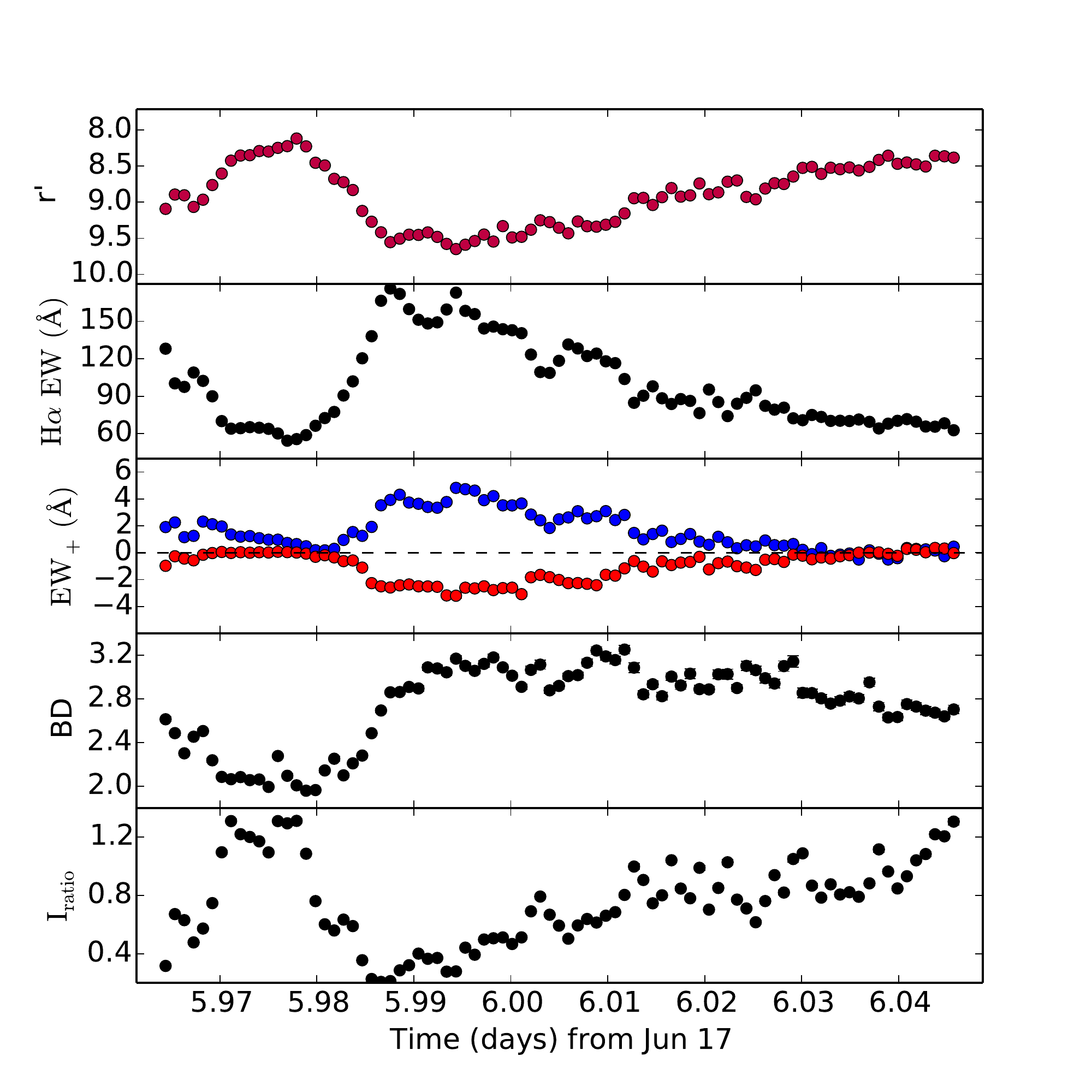}
    \caption{Temporal evolution along days 5 (left) and 6 (right) of the following parameters: $r'$, $\rm H\alpha$ $\rm EW_{+}$ (blue and red points correspond to $\rm EW_{b+}$ and $\rm EW_{r+}$ respectively), BD and $\rm I_{ratio}$.}
    \label{fig: paramd5-6}
\end{figure*}

\begin{figure*}
\center
\includegraphics[keepaspectratio, width=0.5\textwidth]{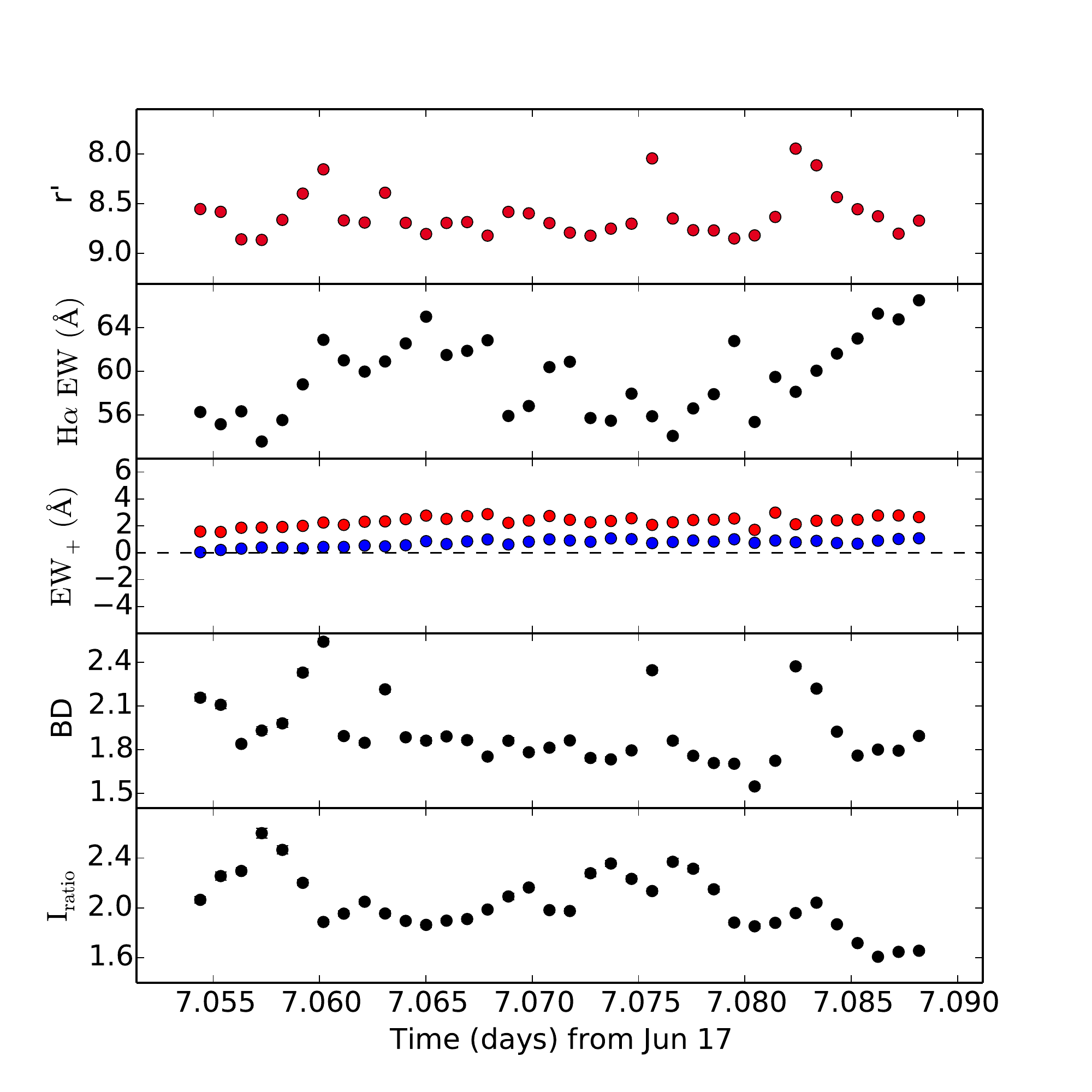}\includegraphics[keepaspectratio, width=0.5\textwidth]{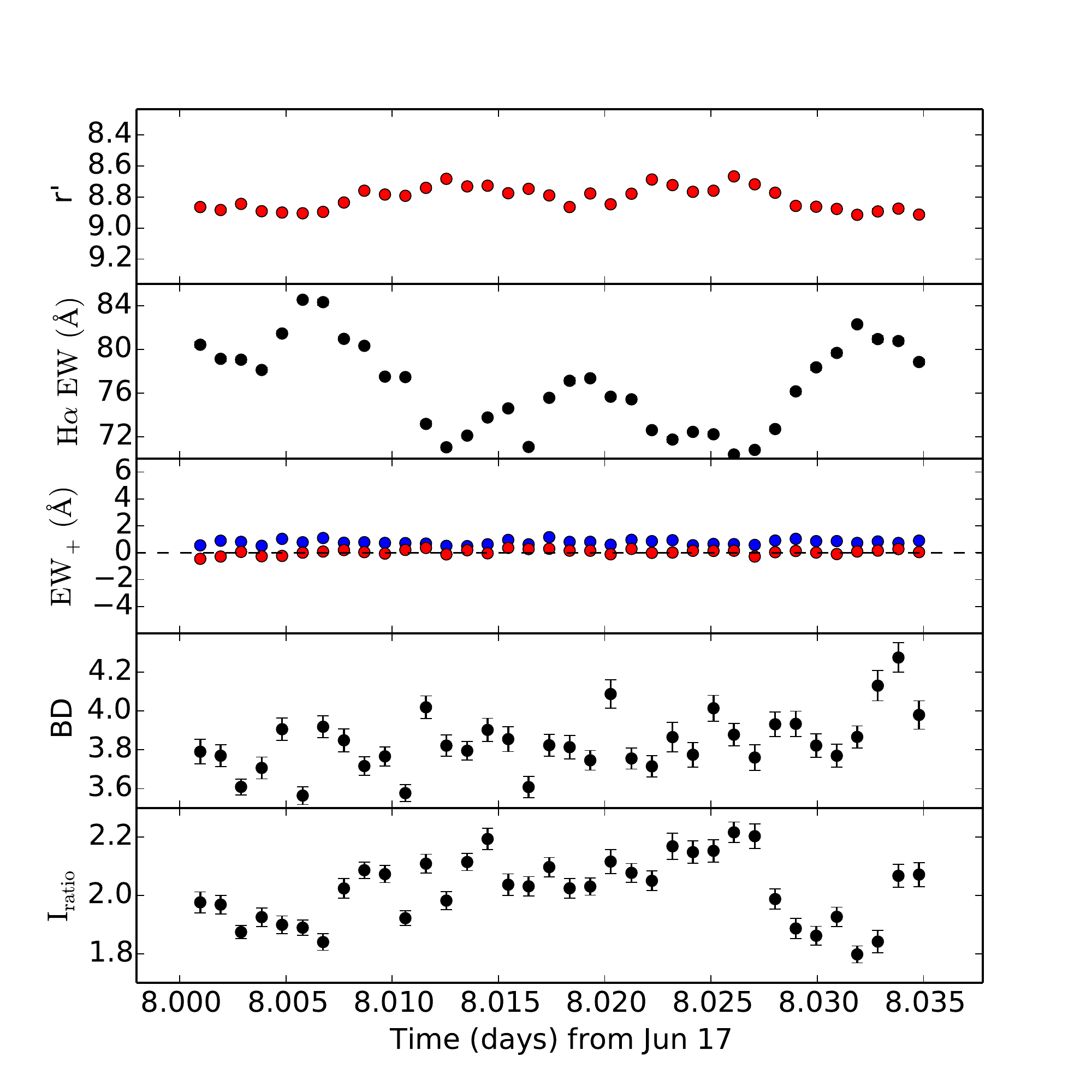}
    \caption{Temporal evolution along days 7 (left) and 8 (right) of the following parameters: $r'$, $\rm H\alpha$ $\rm EW_{+}$ (blue and red points correspond to $\rm EW_{b+}$ and $\rm EW_{r+}$ respectively), BD and $\rm I_{ratio}$.}
    \label{fig: paramd7-8}
\end{figure*}

\begin{figure*}
\center
\includegraphics[keepaspectratio, width=0.5\textwidth]{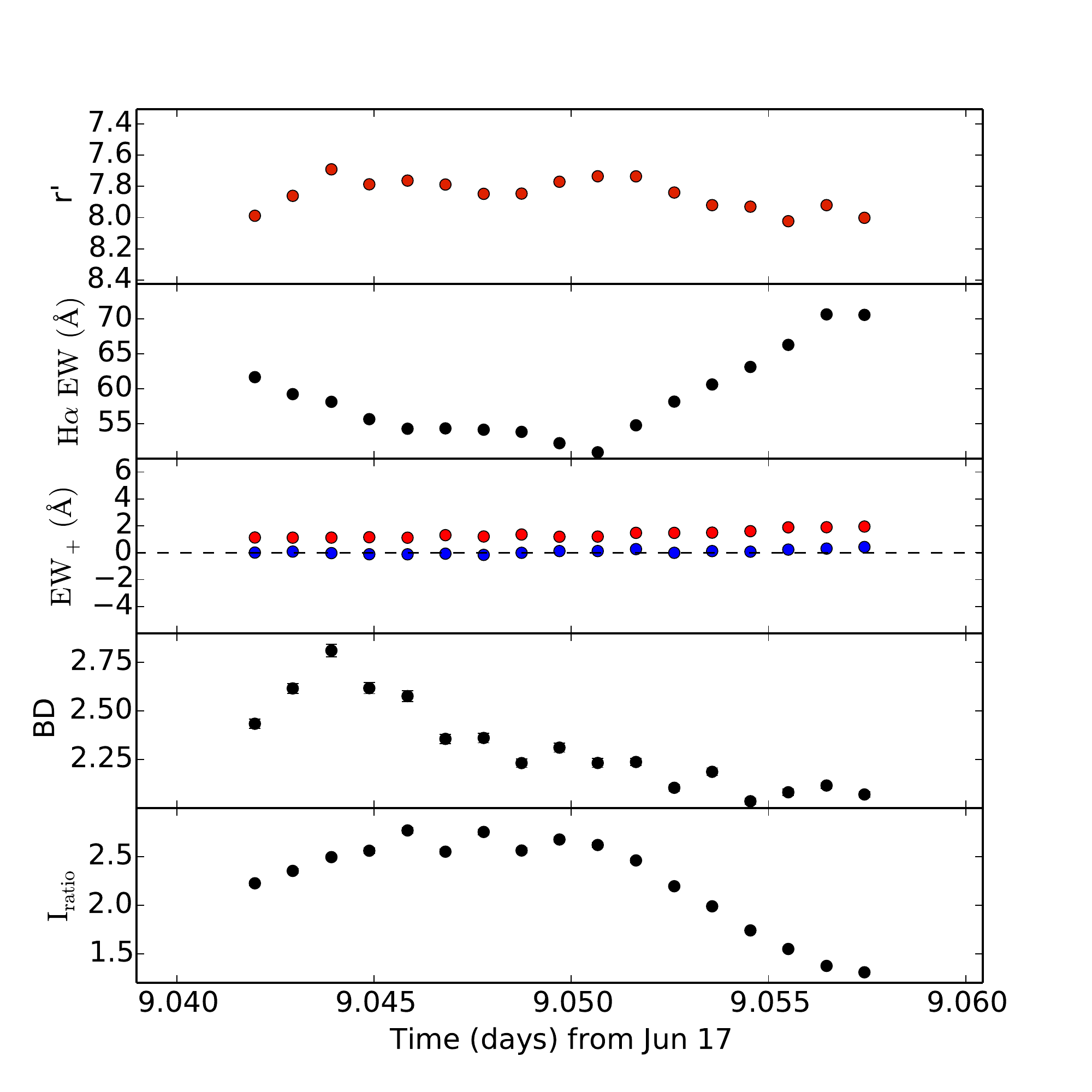}
    \caption{Temporal evolution along day 9 of the following parameters: $r'$, $\rm H\alpha$ EW, $\rm H\alpha$ $\rm EW_{+}$ (blue and red points correspond to $\rm EW_{b+}$ and $\rm EW_{r+}$ respectively), BD and $\rm I_{ratio}$.}
    \label{fig: paramd9}
\end{figure*}

\begin{figure*}
\center
\includegraphics[keepaspectratio, width=0.5\textwidth]{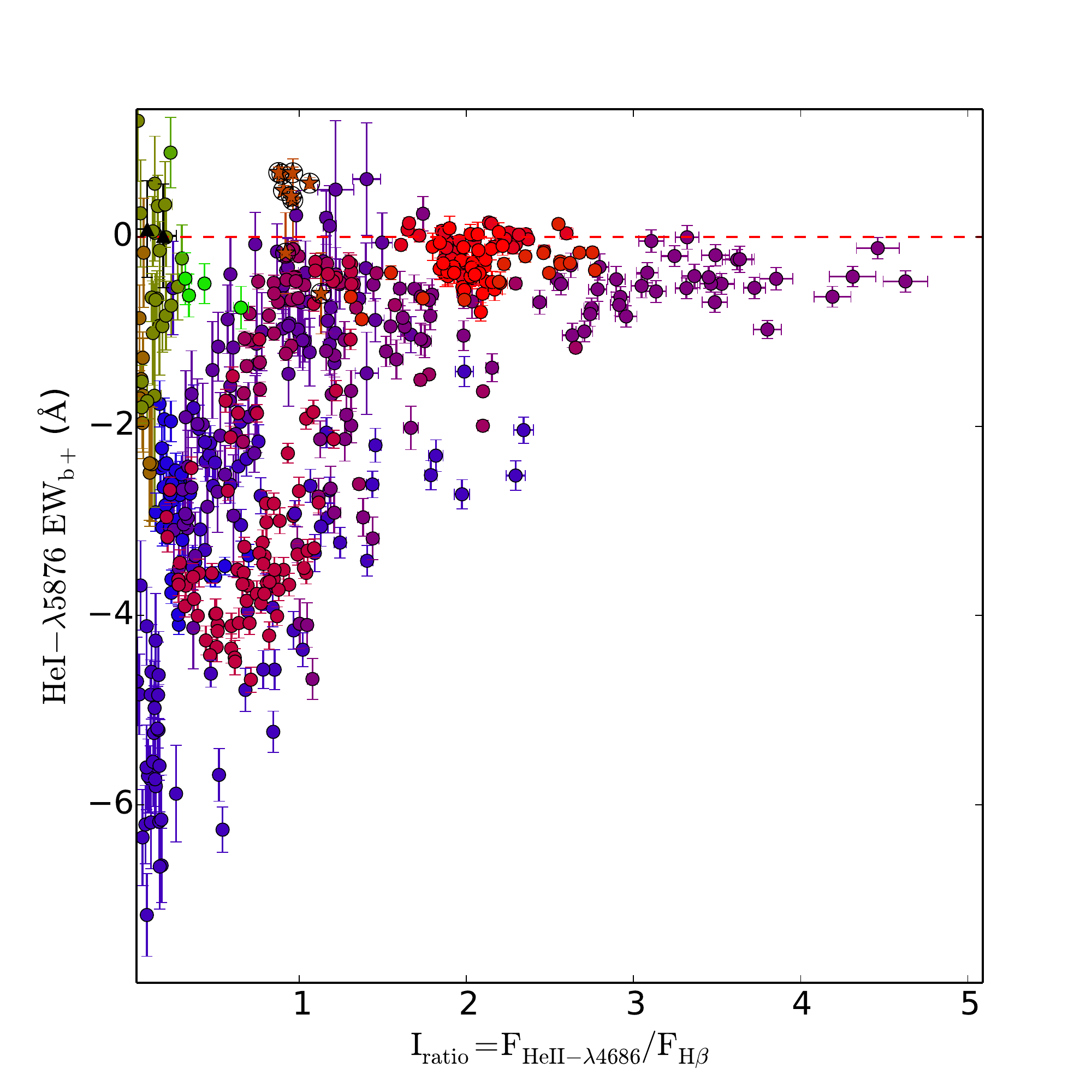}\includegraphics[keepaspectratio, width=0.5\textwidth]{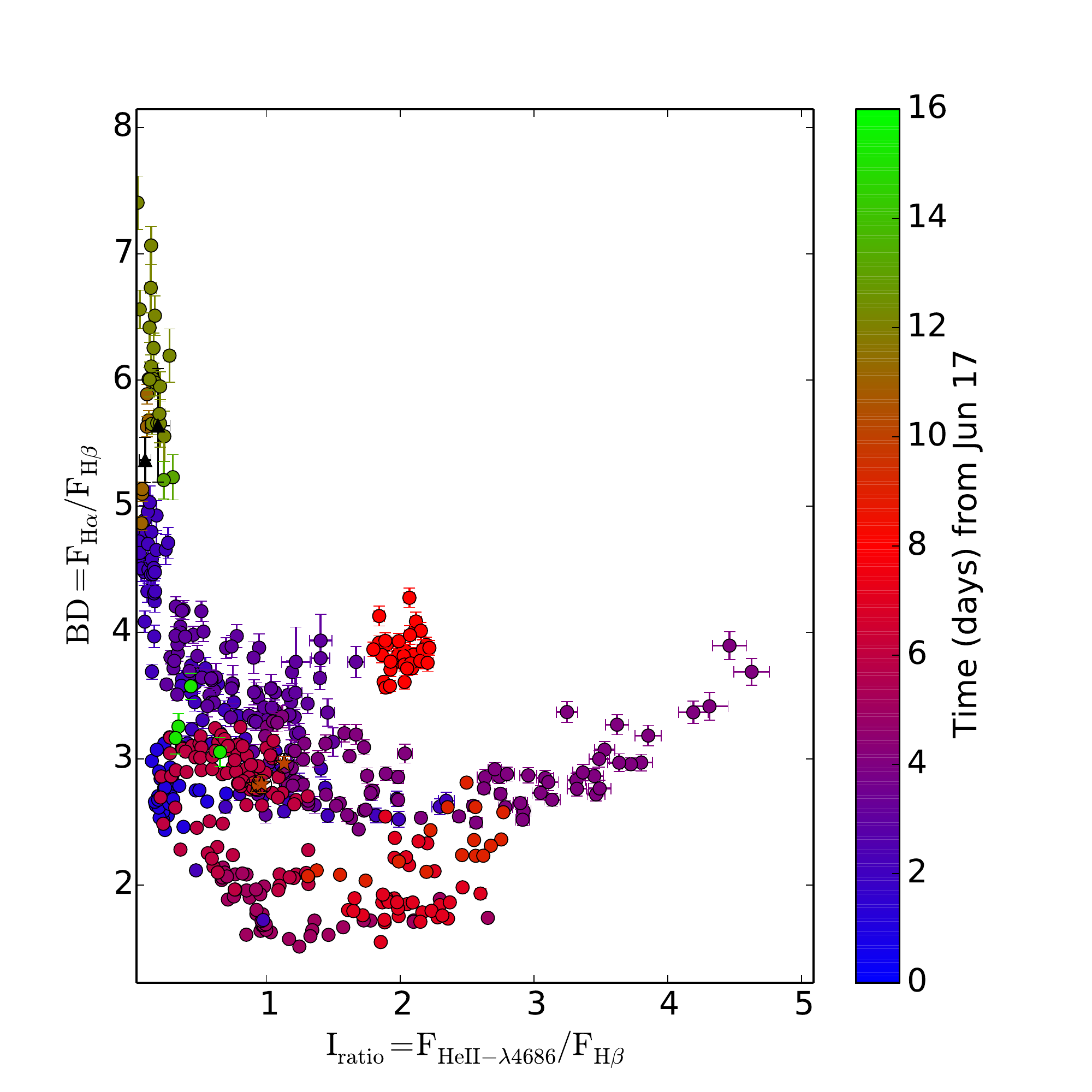}
    \caption{Left panel: $I_{\rm ratio}$ vs EW excess (blue half) of the \ion{He}{\sc i}--$\lambda$5876 line. Right panel: BD vs $I_{\rm ratio}$. The colour bar defines the observation date, and are only depicted up to day 15. Encircled, star-shaped points refer to those where $\rm H\alpha$ saturated the detector. }
    \label{fig: IratiovsBD}
\end{figure*}

\begin{figure*}
\center
\includegraphics[keepaspectratio, width=0.5\textwidth]{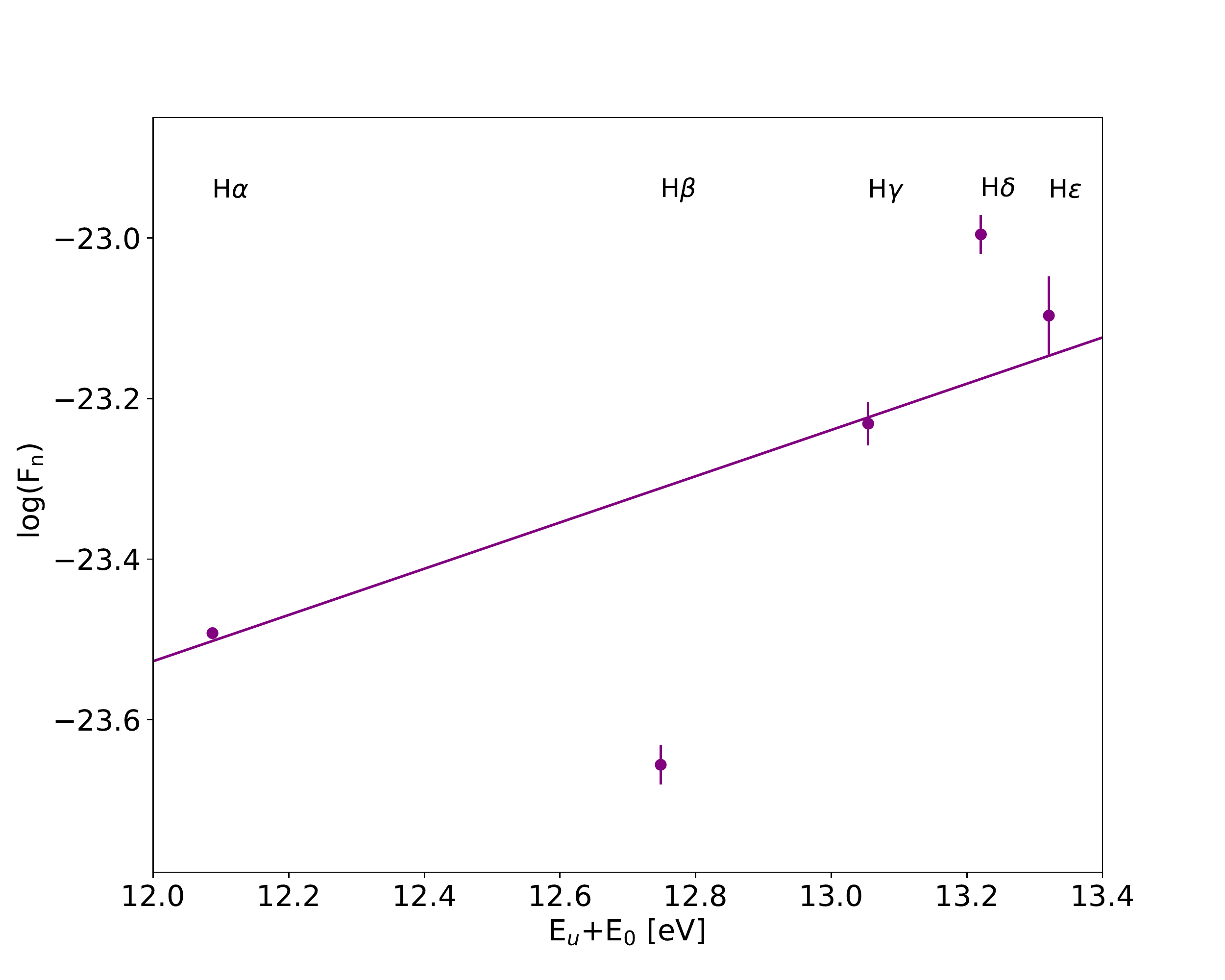}\includegraphics[keepaspectratio, width=0.5\textwidth]{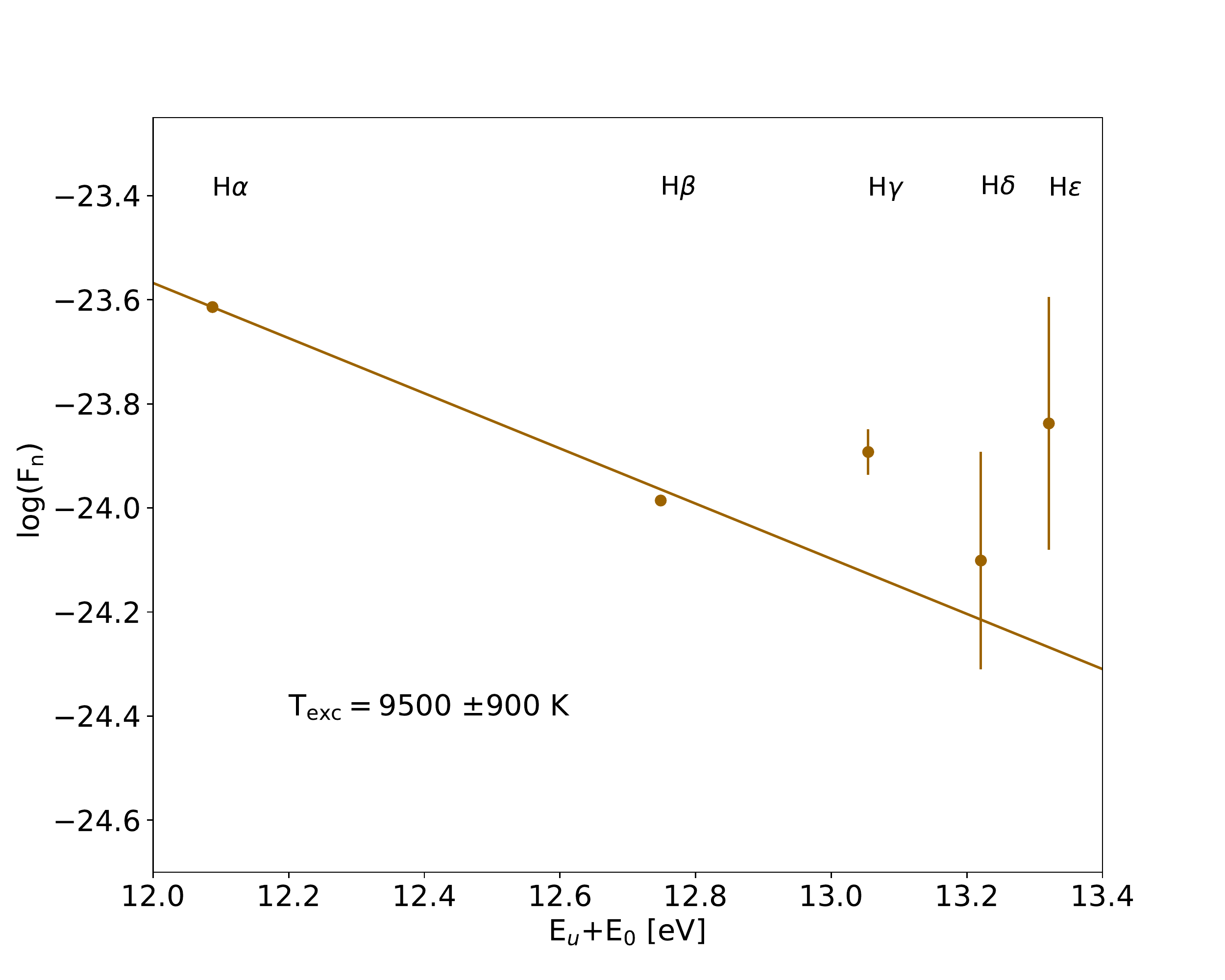}
    \caption{Boltzmann plots using the strongest five transitions of the Balmer series ($\rm H\alpha$ to $\rm H\epsilon$) corresponding to day 4 (left) and day 11 (right). The linear fit defining the trend is depicted as a solid line. The excitation temperature ($T_{\rm exc}$) derived from the fit is shown for the right panel.}
    \label{fig: BP}
\end{figure*}

We present here the observation logs and a summary table with the main characteristics per day of the outburst. We also present here extra figures, including particular spectra (Fig. \ref{fig: spectra_special}), but also showing the temporal evolution of different parameters during those days were we obtain a sufficient number of spectra (Fig. \ref{fig: paramd1-2}, \ref{fig: paramd3-4}, \ref{fig: paramd5-6}, \ref{fig: paramd7-8} and \ref{fig: paramd9}). The evolution of the ionisation ratio ($I_{\rm ratio}$) and the Balmer decrement ($\rm BD$) along the outburst is presented in Fig. \ref{fig: IratiovsBD}. Fig. \ref{fig: BP} shows the Boltzmann plots corresponding to two spectra obtained in days 4 and 11, respectively.

\begin{table*}
\caption{Observation log of the 2015J V404 Cyg outburst.} 

\bigskip

\centering
\begin{threeparttable}
\begin{tabular}{l l l l l l l}
\hline
Night / Time since trigger & Tel./Inst.$^{\dagger}$          & Setup & Spec. res. $\,(\rm{km\,s^{-1}})$ &      $N_{\rm {SPEC}}$  & $ T_{\rm{EXP}} \,(\rm{s})$ & Flux calibration \\
\hline
2015 Jun 14 / Day -2  &	WHT/ISIS	& R600B	&	120 &		2		&		600	 & Y$^{a}$ \\ 
	 &	& R1200R	 &	35	&	2		&		600	 & Y$^{a}$  \\ 
2015 Jun 15 / Day -1	 &	GTC/OSIRIS  & R1000B	&	250	&	1		&			600	  & N \\ 
 & &  R2500V & 100 &	1		&			600	  & N \\
	 &	  &  R2500I & 150 &	1		&			600	  & N \\

2015 Jun 16 / Day 0		&	NOT/FIES		&  &	10	&	1				&	1800  & N \\
2015 Jun 17 / Day 1			&GTC/OSIRIS & R1000B & 250 		&		36	&			60	  & Y \\
2015 Jun 18 / Day 2	&	GTC/OSIRIS & R1000B	 & 250	&		75	&			60	 & Y \\
				&	WHT/ACAM		&	CLEAR+V400  & 640 &	2		&		120	& Y  \\
				&	NOT/ALFOSC	&	Grism 7 	 & 200 &	2			&	600		& N  \\
2015 Jun 19 / Day 3		&		GTC/OSIRIS & R1000B		 & 250 &				75			&			60		& Y  \\
2015 Jun 20 / Day 4		&		GTC/OSIRIS & R1000B	 & 250	&				75		&				60				& Y  \\ 
2015 Jun 21 / Day 5		&		GTC/OSIRIS & R1000B	 & 250	&				40		&				60			& Y 	 \\
2015 Jun 22 / Day 6			&	GTC/OSIRIS & R1000B	 & 250	&				85		&				60				& Y 	 \\
2015 Jun 23 / Day 7			&	GTC/OSIRIS & R1000B	 & 250	&				36		&				60				& Y 	 \\
2015 Jun 24 / Day 8		&	GTC/OSIRIS & R1000B		 & 250	&			36		&				60				& Y 	 \\
2015 Jun 25 / Day 9		&	GTC/OSIRIS & R1000B	 & 250		&			17			&			60				& Y 	 \\
2015 Jun 26 / Day 10			&	GTC/OSIRIS & R1000B	 & 250		&			3				&		40			& Y 	 \\
				&		 & R2500V		 & 100	&			6			&			70				& Y 	 \\
				&		 & R2500R		 & 100	&			6			&			70					& Y 	 \\
2015 Jun 27 / Day 11		&		GTC/OSIRIS &R1000B	 & 250			&	6			&		40							& Y 	 \\	
					&	 &R2500V	 & 100			&		6			&			70								& Y 	 \\	
				&	 &R2500R		 & 100	&			6			&			70								& Y 	 \\	
2015 Jun 28 / Day 12		&		GTC/OSIRIS &R1000B	 & 250	&				19			&			20-120$^{b}$				& Y 	 \\
					&	 &R2500V	 & 100		&			3			&			360					& Y 		 \\
					&	&R2500R		 & 100	&			3				&		120							& Y 		 \\
2015 Jun 29 / Day 13			&	GTC/OSIRIS &R1000B	 & 250		&			2			&			120								& Y 	 \\
					&	 &R2500R		 & 100		&		2				&		120							& Y 		 \\
\hline
\end{tabular}
\begin{tablenotes}
\item[a]{No comparison star was present in the slit during the observations. Therefore, these spectra were flux-calibrated using a standard star observed with the same instrumental configuration.}
\item[b]{The exposure time for each spectrum varies within this range.}
\item[$\dagger$]{Gran Telescopio Canarias (GTC), Optical System for Imaging and low-Intermediate-Resolution Integrated Spectroscopy (OSIRIS); William Herschel Telescope (WHT), Intermediate dispersion Spectrograph and Imaging System (ISIS); Nordic Optical Telescope (NOT), Andalucia Faint Object Spectrograph and Camera (ALFOSC), FIbre-fed Echelle Spectrograph (FIES); Isaac Newton Telescope (INT), Intermediate Dispersion Spectrograph (IDS).}
\end{tablenotes}
\label{tab:summer1}
\end{threeparttable}
\end{table*}

\newpage

\begin{table*}
\caption{Continuation of table \ref{tab:summer1}.} 

\bigskip

\centering
\begin{threeparttable}
\begin{tabular}{l l l l l l l}
\hline
Night / Time since trigger  &  Tel./Inst.          & Setup  & Spec. res. $\,(\rm{km\,s^{-1}})$ &   $N_{\rm {SPEC}}$  & $ T_{\rm{EXP}} \,(\rm{s})$ & Flux calibration \\
\hline
2015 Jul 01 / Day 15	&		GTC/OSIRIS &R1000B		& 250 	&			4			&			60-120$^{a}$								& Y 	 \\
				&		 &R2500R		& 100		&		4			&			60-120	$^{a}$							& Y 			 \\
2015 Jul 04 / Day 18		&		GTC/OSIRIS & R2500R		& 100	&			2			&			130							& Y 			 \\
2015 Jul 07 / Day 21	&		WHT/ISIS & R300B	& 230	&	3				&		1375-1800$^{a}$				& Y 			 \\
 & & R600R & 70 &	9				&		430-900$^{a}$	 & Y\\		
					&	NOT/ALFOSC		&	Grism 7	 & 200 &		1				&		900					& N 					 \\
2015 Jul 12 / Day 26		&		WHT/ISIS & R300B	& 230 &		5			&			900				& Y 						 \\
		&		 & R600R	& 70 &		10			&			900		 & Y							 \\
2015 Jul 14 / Day 28		&		NOT/ALFOSC		&	Grism 7 & 200	&			2		&				900					& N				 \\
2015 Jul 15 / Day 29		&		NOT/ALFOSC		&	Grism 7 & 200		&		1		&				900						& N				 \\
2015 Jul 18 / Day 32		&	WHT/ISIS		&	R300B	& 230	&		3				&		900-1800$^{a}$				& Y$^{b}$ 			 \\
			&		&	R600R		& 70 &		4				&		900-1800$^{a}$	 & Y							 \\
2015 Jul 19 / Day 33		&	WHT/ISIS		&	R300B	& 230	&		10			&			900-1800$^{a}$		& Y 						 \\
			&			&	R600R	& 70	&		10			&			900-1800$^{a}$			 & Y					 \\
2015 Jul 20 / Day 34	&	INT/IDS & R632R		& 100	&	6				&		1800	& Y 									 \\
2015 Jul 21 / Day 35		&		WHT/ISIS		&	R300B	& 230	&		5				&		1800					& N 				 \\
	&		&	R600R	& 70	&		6				&		1800		 & Y							 \\
2015 Jul 27 / Day 41 &		WHT/ACAM	&	GG395+V400	& 370	&			2			&			900			& Y 								 \\
2015 Jul 31 / Day 45		&		GTC/OSIRIS & R2500R	& 100		&			2			&			139					& Y 					 \\
2015 Aug 06 / Day 51		&		WHT/ISIS & R1200R		& 35 &			6			&			1800		& N							 \\
				&		NOT/ALFOSC		&	Grism 7 	& 200	&		2			&			900			& N							 \\			
2015 Sep 01 / Day 77		&		WHT/ISIS				&		R300B	& 230	& 1				&		1800				& Y 						 \\
	&						&		R600R& 70&		2				&		900		 & Y								 \\
2015 Oct 28 / Day 134		&		WHT/ACAM			&		GG395+V400& 370	&		2				&		900			& Y 							 \\

\hline
\end{tabular}
\begin{tablenotes}
\item[a]{The exposure time for each spectrum varies within this range.}
\item[b]{Some exposures had the field star spectra at the edge of the chip, making flux calibration unreliable. }
\end{tablenotes}
\label{tab:summer2}
\end{threeparttable}
\end{table*}


\begin{table*}
\caption{Observation log of the 2015D V404 Cyg outburst.}

\bigskip

\centering
\begin{threeparttable}
\begin{tabular}{l l l l l l l}
\hline
Night / Time since trigger &  Tel./Inst.          & Setup & Spec. res. $\,(\rm{km\,s^{-1}})$ &      $N_{\rm {SPEC}}$  & $ T_{\rm{EXP}} \,(\rm{s})$ \\
\hline
23-12-2015 / Day 1 &	NOT/ALFOSC & Grism 7 & 200	 &		2		 &		600		 \\ 
24-12-2015 / Day 2		 &	GTC/OSIRIS	 &	R1000B	& 250	 &				4				 &		125-300$^{a}$			 \\  
				 &		INT/IDS	 &	R632R	& 100		 &		3				 &		1200					 \\ 
25-12-2015 / Day 3	 &	GTC/OSIRIS	 &	R1000B	& 250		 &			2			 &			125					 \\ 
27-12-2015 / Day 5 &	GTC/OSIRIS	 &	R1000B	& 250		 &			2			 &			360						 \\ 
28-12-2015	 / Day 6	 &		WHT/ISIS		 &		R300B  & 230 &				2			 &			900						 \\ 
		 &		WHT/ISIS		 &		R600R & 70 &				2			 &			900						 \\ 
29-12-2015	 / Day 7	 &		WHT/ISIS		 &	 R300B	 & 230&				1			 &			900						 \\ 
		 &		WHT/ISIS		 &		R600R & 70 &				1			 &			900						 \\ 
30-12-2015	 / Day 8		 &	GTC/OSIRIS	 &	R1000B& 250		 &				2			 &			125				 \\ 
31-12-2015	 / Day 9		 &	GTC/OSIRIS	 &	R1000B	& 250		 &			2			 &			125						 \\ 
01-01-2016	 / Day 10		 &	GTC/OSIRIS	 &	R1000B	& 250		 &			3			 &				60-125$^{a}$					 \\ 
03-01-2016		 / Day 12	 &	WHT/ISIS			 &		R300B	& 230	 &	1				 &		900							 \\ 
		 &		WHT/ISIS		 &		R600R  & 70 &				1			 &			900						 \\ 
06-01-2016 / Day 15	 &	GTC/OSIRIS	 &	R1000B & 250		 &			2			 &			125							 \\ 
07-01-2016 / Day 16	 &	GTC/OSIRIS	 &	R1000B	& 250		 &			2			 &			125							 \\ 
09-01-2016 / Day 18		 &		WHT/ACAM			 &	 CLEAR+V400	& 640 &			1				 &		900							 \\ 
10-01-2016 / Day 19		 &	GTC/OSIRIS	 &	R1000B	& 250		 &			4			 &			125							 \\ 
11-01-2016	 / Day 20	 &		GTC/OSIRIS	 &	R1000B	& 250		 &			4				 &		125							 \\ 
28-04-2016	 / Day 128	 &		NOT/ALFOSC		 &	Grism 7	 & 120 &				2			 &			600					 \\ 				
24-05-2016	 / Day 154	 &		WHT/ISIS			 &	 R600R	 & 70 &			1			 &			900							 \\ 
29-06-2016	 / Day 190	 &		WHT/ISIS				 &	R600R	& 70	 &		3			 &			713-1800$^{a}$					 \\ 
04-07-2016	 / Day 195	 &		WHT/ISIS				 &	R600R	& 70 &		1			 &			900							 \\ 
 \\

\hline
\end{tabular}
\begin{tablenotes}
\item[a]{The exposure time for each spectrum varies within this range.

No reliable flux calibration is available in this dataset. For a more detailed description of the GTC coverage in this outburst, see \citet{MunozDarias2017}. }
\end{tablenotes}
\label{tab:winter}
\end{threeparttable}
\end{table*}


\begin{table*}
\caption{Observation log of the 1989 V404 Cyg outburst.}

\bigskip

\centering
\begin{threeparttable}
\begin{tabular}{l l l l l l l}
\hline
Night / Time since trigger &  Tel./Inst.          & $\lambda$ [{\AA}] range & Spec. res. $\,(\rm{km\,s^{-1}})$  &      $N_{\rm {SPEC}}$  & $ T_{\rm{EXP}} \,[\rm{s}]$ \\
\hline
31-05-1989 / Day 11	 &	INT/IDS+IPCS &	4000-5000	 & 60 &		6		 &			600			 \\ 
				 &		INT/IDS+IPCS	 &	3500-4500		& 60 	 &	1				 &		1200				 \\ 	
				 &		INT/IDS+IPCS	 &	5800-6800		& 60 	 &	3			 &			100-1500$^{a}$				 \\ 		
01-06-1989 / Day 12		 &		INT/IDS+IPCS	 &	4000-5000	& 60 		 &	2			 &			600				 \\ 			
				 &		INT/IDS+IPCS	 &	3500-4500		& 60 	 &	1			 &			100					 \\ 
				 &		INT/IDS+IPCS	 &	5800-6800		& 60 	 &	1			 &			600					 \\ 
02-06-1989 / Day 13			 &	INT/IDS+IPCS &	3000-8000	& 230 	 &		1			 &			900				 \\ 			
03-06-1989 / Day 14		 &		INT/IDS+IPCS	 &	4000-5000	& 60 		 &	1			 &			600				 \\ 			
04-06-1989 / Day 15			 &	INT/IDS+IPCS	 &	3000-8000	& 230 	 &		1			 &			600				 \\ 			
05-06-1989 / Day 16		 &		INT/IDS+IPCS	 &	4000-5000	& 60 	 &		3			 &			700-1200$^{a}$			 \\ 					
06-06-1989	 / Day 17		 &	INT/IDS+IPCS	 &	4000-5000	& 60 		 &	8				 &		400-900$^{a}$				 \\ 				
07-06-1989 / Day 18			 &	INT/IDS+IPCS	 &	4000-5000	& 60 	 &		11			 &			600			 \\ 
10-06-1989	 / Day 21		 &	INT/IDS+IPCS	 &	4000-5000	& 60 		 &	18				 &		600			 \\

 \\

\hline
\end{tabular}
\begin{tablenotes}
\item[a]{The exposure time for each spectrum varies within this range.}

No reliable flux calibration is available in this dataset. For more details see  \citet{Charles1989} and \citet{Casares1991}. 
\end{tablenotes}
\label{tab:1989}
\end{threeparttable}
\end{table*}

\begin{table*}
\caption{Key Spectral Characteristics of V404 Cyg during the 2015J Outburst.} 

\bigskip

\centering
\begin{threeparttable}
\begin{tabular}{l l l l l l l}
\hline
Day &  $r'$ &  BD &  $I_{\rm ratio}$  & $\rm H\alpha$ EW ({\AA}) & $\rm  H\alpha\, FWHM \, (km \, s^{-1})$  & Notes\\
\hline
    & & & & & &     \\ 
-2	 &	$13.2$	& $5.3-5.7$ 	&	 $0.05-0.24$	&		$260-305$	&		$940-956$	 & Previous to the X-ray trigger.  \\ 
    & & & & & &     \\ 
1	 &	$10.2-11.4$	& $2.4-3.2$ 	&	 $0.1-0.7$	&		$65-195$	&		$1020-1080$ & Slow decay.  \\ 
    & & & &  & &     \\ 
2	 &	$9.0-11.3$	& $0.9-5.1$ 	&	 $0.03-2.5$		&		$40-270$	&		$945-1415$ & Full flare, nebular  \\      
   & & & & & & loop and deep  \\
      & & & & & & P-Cyg.  \\ 
          & & & & & &     \\ 
3	 &	$8.9-10.6$	& $2.5-4.3$ 	&	 $0.2-1.7$			&		$55-200$	&		$1110-1580$ & Slow rise.  \\ 
    & & & & & &     \\ 
4	 &	$8.2-9.3$	& $2.4-4.0$ 	&	 $1.0-4.7$			&		$35-92$	&		$1120-1375$ & Slow decay.  \\ 
    & & & & & &     \\ 
5	 &	$8.2-9.2$	& $1.5-2.2$ 	&	 $0.6-2.6$			&		$43-105$	&		$1314-1569$ & Decay and plateau.  \\ 
    & & & & & &    \\ 
6	 &	$8.1-9.7$	& $1.9-3.3$ 	&	 $0.2-1.3$			&		$53-177$	&		$1335-1573$  & Full flare, nebular  \\      
   & & & & & & loop and deep  \\
      & & & & & & P-Cyg. \\ 
          & & & & & &     \\ 
7	 &	$7.9-8.9$	& $1.5-2.6$ 	&	 $1.6-2.6$			&		$53-67$	&		$875-993$  & Fast, red flares. \\ 
    & & & & &     \\ 
8	 &	$8.6-8.9$	& $3.5-4.3$ 	&	 $1.8-2.2$			&		$71-84$	 & $1049-1082$  &Bright plateau. \\ 
    & & & & &     \\ 
9	 &	$7.7-8.0$	& $2.0-2.8$ 	&	 $1.3-2.8$			&		$50-71$	 & $1035-1116$ & Flux peak. \\     
& & & & & &     \\ 
10	 &	$ 9.1$	& $>2.5$ 	&	 $1.0-1.1$			&		$>205$	 &	 $<1320$ & Nebular phase \\
   & & & & & & starts. \\ 
    & & & & & &     \\ 
11	 &	$ 11.5-12.0$	& $4.8-5.9$ 	&	 $0.06-0.12$			&		$1472-1904$	 & $932-974$ & Nebular phase.  \\ 
    & & & &  & &     \\ 
12	 &	$12.8-13.1$	& $5.5-7.4$ 	&	 $0.03-0.3$			&		$408-685$	 & $801-856$ & Nebular phase. \\ 
    & & & & & &     \\ 
13	 &	$12.4$	& $5.3$ 	&	 $0.2$			&		$183-189$	 & $711-732$ &Nebular phase\\      
  & & & & & & ends, double-\\
   & & & & & &  peaked lines. \\ 
    & & & & & &     \\ 
15	 &	$12.5-12.8$	& $3.0-3.6$ 	&	 $0.3-0.7$			&		$74-91$	 &$726-759$ & Double-peaked  \\
   & & & & & & lines. \\ 
    & & & & & &     \\ 
\hline
\end{tabular}
\label{tab:summatory}
\end{threeparttable}
\end{table*}

\bsp	
\label{lastpage}
\end{document}